\documentclass[useAMS,usenatbib]{mn2e}
\usepackage{graphicx}
\usepackage{times}
\usepackage{wasysym}
\title[Type III migration; Outward migration]{Numerical simulations of
type III planetary migration:\\ III. Outward migration of massive planets}
\author[A. Pepli\'nski et al.]{A. Pepli\'nski,$^1$\thanks{E-mail:
adam@astro.su.se} P. Artymowicz$^2$ and G. Mellema$^1$\\ $^1$Stockholm
University, AlbaNova University Centre, SE-106 91 Stockholm, Sweden\\
$^2$University of Toronto at Scarborough, 1265 Military Trail, Toronto,
Ontario M1C 1A4, Canada}
\begin{document}
\voffset=-0.8in 
\date{Accepted 0000 . Received 0000 ; in original form 0000 Month}
\pagerange{\pageref{firstpage}--\pageref{lastpage}} \pubyear{0000}

\maketitle

\label{firstpage}
\begin{abstract}
We present a numerical study of rapid, so called type III migration
for Jupiter-sized planets embedded in a protoplanetary disc. We limit
ourselves to the case of outward migration, and study in detail its
evolution and physics, concentrating on the structure of the
co-rotation and circumplanetary regions, and processes for stopping
migration. We also consider the dependence of the migration behaviour
on several key parameters. We perform this study using global,
two-dimensional hydrodynamical simulations with adaptive mesh
refinement. We find that the outward directed type III migration can
be started if the initial conditions support $Z > 1$, that corresponds
to initial value $M_\rmn{\Delta} \ga 1.5$. Unlike the inward directed
migration, in the outward migration the migration rate increases due
to the growing of the volume of the co-orbital region. We find the
migration to be strongly dependent on the rate of the mass
accumulation in the circumplanetary disc, leading to two possible
regimes of migration, fast and slow. The structure of the co-orbital
region and the stopping mechanism differ between these two regimes.

\end{abstract}

%

\section{Introduction}

The discovery of extra-solar planetary systems
(\citealt{1995Natur.378..355M}, \citealt{2000prpl.conf.1285M},
\citealt{2002ApJ...568..352V}) has invoked a strong interest in the
theories of planetary system formation. The occurrence of so-called
`hot Jupiters' (objects with minimum masses $M>M_{\jupiter}$ and
semi-major axes $a< 0.1$~AU) has shown that planets can be found on
orbits very different from the formation sites of planetary
cores. Since the in situ formation of these objects is difficult both
in the core accretion scenario \citep{1996Icar..124...62P} and through
direct gravitational instability \citep{2001ApJ...563..367B}, the
global migration of planetary cores has become the subject of a large
number of studies. The inward migration of the planets due to
planet-disc gravitational interaction was first described by
\citet{1979ApJ...233..857G,1980ApJ...241..425G}, prior to the
discovery of extra-solar planetary systems. Later on it was worked out
in a large number of investigations summarised in the reviews by
\citet{1993prpl.conf..749L}, \citet{2000prpl.conf.1111L} and
\citet{2007prpl.conf..655P}.

In the standard picture the planet exchanges angular momentum with
the disc through the excitation of spiral density waves at the
Lindblad resonances and through the interaction at the corotation
resonance.  However until recently, the second mechanism was assumed
to be inefficient and was therefore neglected. The non-axisymmetric
pattern of the waves excited at Lindblad resonances acts back the
planet providing the torque driving its orbital migration. This leads
to two possible regimes of the migration: type I for planets embedded
in the disc and type II for planets opening the gap. In both cases the
direction of migration usually is inward. Note that type II is however
capable of moving the planet outward provided the disk has local
density gradients supporting outward viscous spreading
\citep{1993prpl.conf..749L,2003LPI....34.1736W,2007MNRAS.377.1324C}.

Recently the corotation resonance have been found to be important and
capable of slowing down and even reverting the direction of migration in
type I regime (\citealt{2006ApJ...652..730M},
\citealt{2006A&A...459L..17P}). However, it can also lead to a new and
very fast migration mode (called type III) that is not truly resonant, but
depends strongly on the gas flow in the planet vicinity and does not
have a predetermined direction (\citealt{2003ApJ...588..494M};
\citealt{2004ASPC..324...39A}, \citealt{2007prpl.conf..655P}). This type
of migration was studied numerically by \citet{2003ApJ...588..494M} who
performed global two-dimensional simulations of an inward and outward
directed migration. The inward directed migration was studied by
\citet{2005MNRAS.358..316D}. \citet{2005CeMDA..91...33P} considered
local shearing box simulations in 2-D, and \citet{Pawel_Miguel1} in
3-D. 

This paper is the third in a series devoted to a numerical
investigation of type III migration for the high-mass planets. In
\citet{PaperI} (henceforth Paper~I) we showed how the applied disc
model critically influences the outcome of the simulations. We found
two physically motivated corrections to the standard disc model, both
necessary to remove non-physical effects from the simulations. The
first one is a correction of the gas acceleration (due to
self-gravity) which ensures that the circumplanetary disc moves together
with the planet. The second is a modification of local-isothermal
approximation, which allows an increase of the temperature inside the
planet's Roche sphere and thus decreases the amount of gas accumulated in the
planet's vicinity. Paper~I contains the detailed description of the disc model,
the code and the convergence tests. 

This method was applied to the study of inward directed type III
migration in \citet{PaperII} (henceforth Paper~II). There we focussed
on the physical aspects of type III migration and described in detail
the mechanisms driving rapid migration.  We also discussed the
mechanisms for stopping type III migration (or rather transforming it to
the very much slower type II regime).

In the present paper we focus on outward directed type III
migration. Although the driving mechanism is the same for both the
inward and outward directed migration, the existence of the
differential Lindblad torque breaks the symmetry between both cases as
shown in \citet{PaperIV}. The analytical theory in \citet{PaperIV}
also shows that unlike inward migration, outward migration only is
allowed for a co-orbital mass deficit $M_\rmn{\Delta}$ larger than
one. Paper~II showed that inward migration will always slow down due
to the shrinking of the size of the co-orbital region. For outward
migration one would expect the reverse, i.e.\ a speed up of migration,
allowing the planet to travel long distances in the disc. 

The layout of the paper is as follows. Section~\ref{disc-model} gives
the numerical setup of the disc model. In Sections~\ref{outward_mig}
and~\ref{sect_dep_on_par} we describe the results of simulations of
outward migrating protoplanets and the dependency of the migration on
the various parameters. In Sections~\ref{stop_mig} and~\ref{sect_ecc_ev}
we analyse the stopping of type III migration and eccentricity
evolution. Finally in Section \ref{conclusions} we discuss their
implications for extrasolar planetary configurations.


\section{Overview of the numerical setup}

\label{disc-model}

The full description of the adopted disc model, the numerical method and
the boundary conditions was given in Paper~I. In this chapter we
will give only a short description of the adopted units and initial
conditions. 

For our simulations we use the {\it Flash} hydro code version 2.3
written by the {\it FLASH Code Group} from the Centre for
Astrophysical Thermonuclear Flashes at the University of
Chicago\footnote{http://www.flash.uchicago.edu} in 1997. {\it Flash}
is a modular, adaptive-mesh, parallel simulation code capable of
handling general compressible flow problems. It is designed to allow
users to configure initial and boundary conditions, change algorithms,
and add new physics modules. We use the code in the pure
hydrodynamical mode in two dimensions, and the adaptive mesh is used
to achieve high resolution around the planet (4 levels of
refinement). The simulations are performed on the Cartesian grid in
the inertial reference frame.

In the simulations we adopt non-dimensional units, where the sum of star
and planet mass $M_\rmn{S} + M_\rmn{P}$ represents the unit of mass. The
time unit and the length unit are chosen to make the gravitational
constant $G=1$. This makes the orbital period of Keplerian rotation at a
radius $a=1$ around a unit mass body equal to $2 \pi$. However, when it
is necessary to convert quantities into physical units, we use a
Solar-mass protostar $M_\rmn{S}=M_{\astrosun}$, a Jupiter-mass
protoplanet $M_\rmn{P}=M_{\jupiter}$, and a length unit of $5.2AU$. 
This makes the time unit equal to $11.8/2\pi$ years. 

In all simulations the grid extends from $-4.0$ to $4.0$ in both
directions around the star and planet mass centre. This corresponds to a
disc region with a physical size of $20.8$ AU. 

The initial surface density $\Sigma$ profile is given by a modified
power law: 
\begin{equation}
\Sigma_\rmn{init} = \psi(r_\rmn{c}) \Sigma_\rmn{0}  (r_\rmn{c}/r_\rmn{0})^{\alpha_\rmn{\Sigma}},
\end{equation}
where $r_\rmn{c} = |\bmath{r} - \bmath{r}_\rmn{C}|$ is the distance to
the mass centre of the planet-star system, $r_\rmn{0}$ is a unit
distance, and $\psi$ is a function that allows introducing a sharp
edges in the disc. To start the outward directed migration we
introduce a density jump at the initial planet position (see Paper~I
Fig.~2). The disc mass is characterised by the disc to the primary mass
ratio
\begin{equation}
\mu_\rmn{D} = {{\Sigma_\rmn{init}(r_\rmn{0}) \pi r_\rmn{0}^2} \over {M_\rmn{S}}} = 
{{\Sigma_\rmn{0} \pi r_\rmn{0}^2} \over {M_\rmn{S}}}.
\end{equation}
We investigate different density profiles by changing
$\alpha_\rmn{\Sigma}$ from $-1.5$ to $0.0$, and $\mu_\rmn{D}$ from
$0.005$ to $0.01$. For the Minimum Mass Solar Nebula (MMSN)
$\mu_\rmn{D}=0.00144$ and $\alpha_\rmn{\Sigma}=-3/2$. 

To enforce rapid migration the planet is introduced instantaneously on a
circular orbit of semi-major axis equal $0.8$. We performed the
simulations with $M_\rmn{P}/M_\rmn{S}$ equal $0.0007$, $0.001$ (i.e. one
Jupiter mass, $M_{\jupiter}$, for a one-solar-mass star) and $0.0013$.

The aspect ratio for the disc with respect to the star is fixed at
$h_\rmn{s} = 0.05$, whereas the circumplanetary disc aspect ratio
$h_\rmn{p}$ can range form $0.4$ to $0.6$. 

The smoothing length of the stellar potential, $r_\rmn{soft}$, is taken
to be $0.5$. For the planet this parameter was chosen be a fraction of
the planet's Hill radius $R_\rmn{H} =
a{[M_\rmn{P}/(3M_\rmn{S})]}^{(1/3)}$, and was set to $r_\rmn{soft} =
0.33 R_\rmn{H}$. The corresponding size of the envelope $r_\rmn{env}$
was set to $0.5 R_\rmn{H}$. 

{\it Flash} has the ability to track multiple fluids. We use this
feature to investigate the evolution of the fluid in corotation
region. In our simulations we introduce a tracer fluid that has a
value of 1 in the (initial) corotation region $(a_\rmn{init} -
2R_\rmn{H})< r<(a_\rmn{init}+2R_\rmn{H})$, and zero outside it. This
allows us to distinguish between the fluid captured by the planet in
the horseshoe region and the fluid flowing through the corotation
region.

\begin{figure*}
\includegraphics[width=84mm]{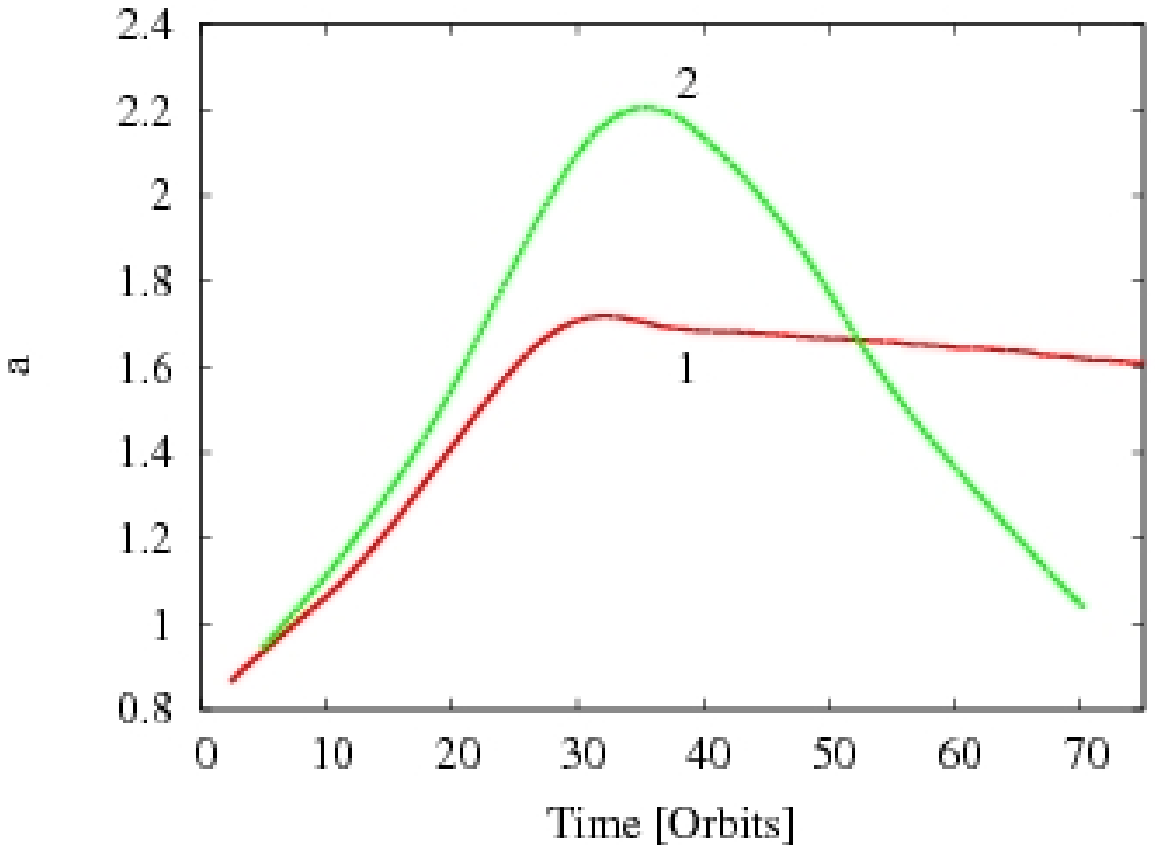}
\includegraphics[width=84mm]{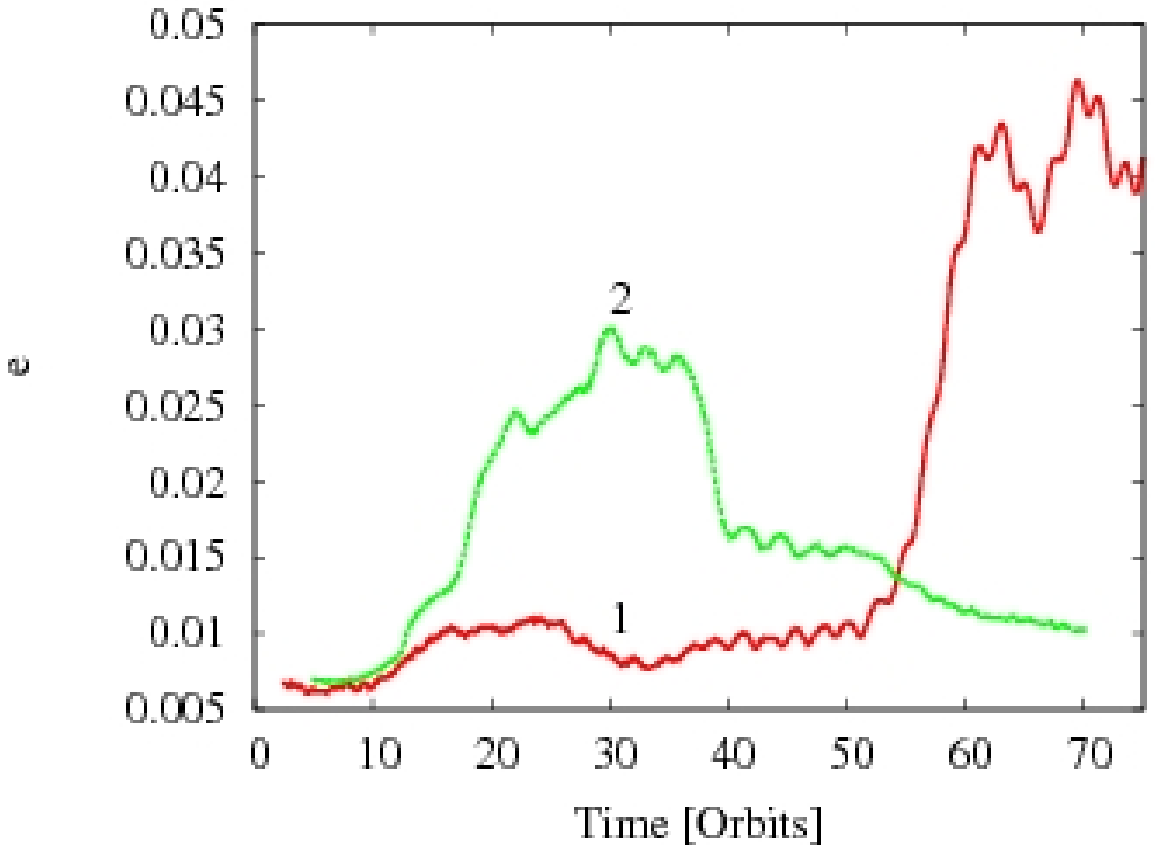}
\includegraphics[width=84mm]{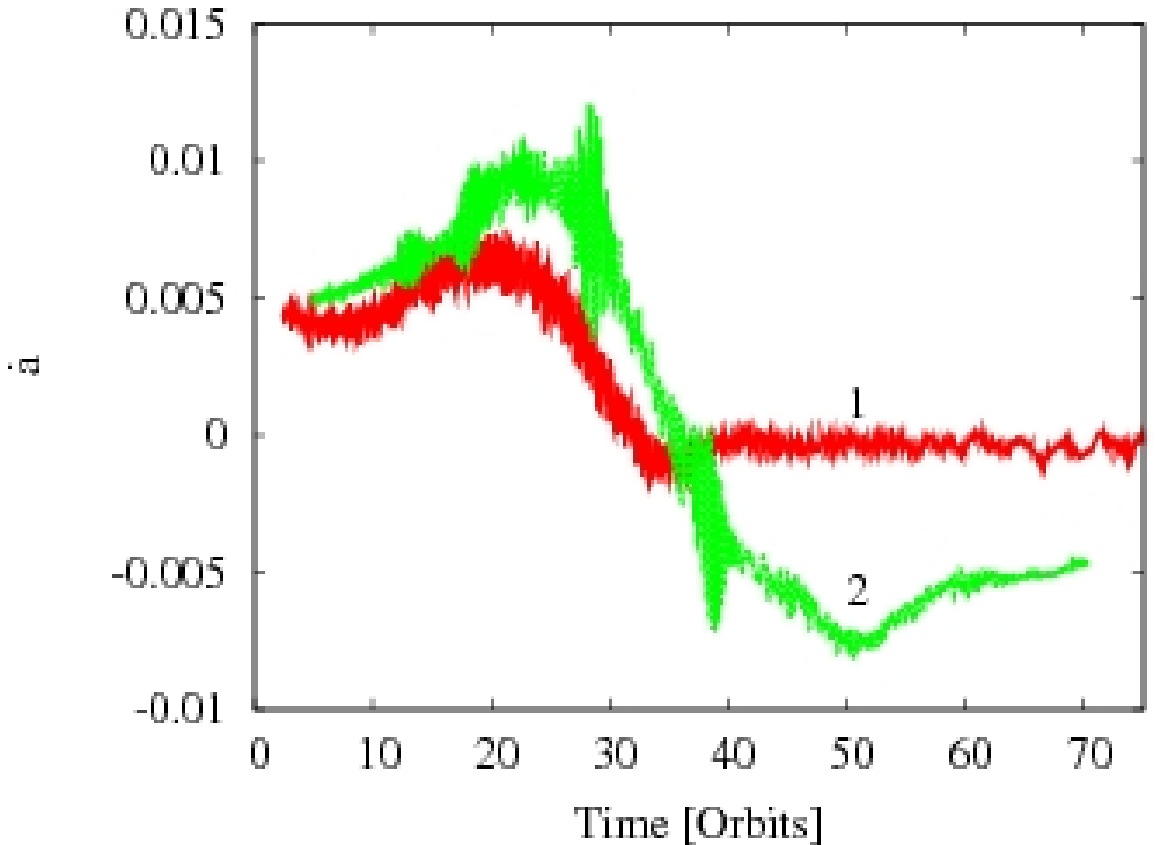}
\includegraphics[width=84mm]{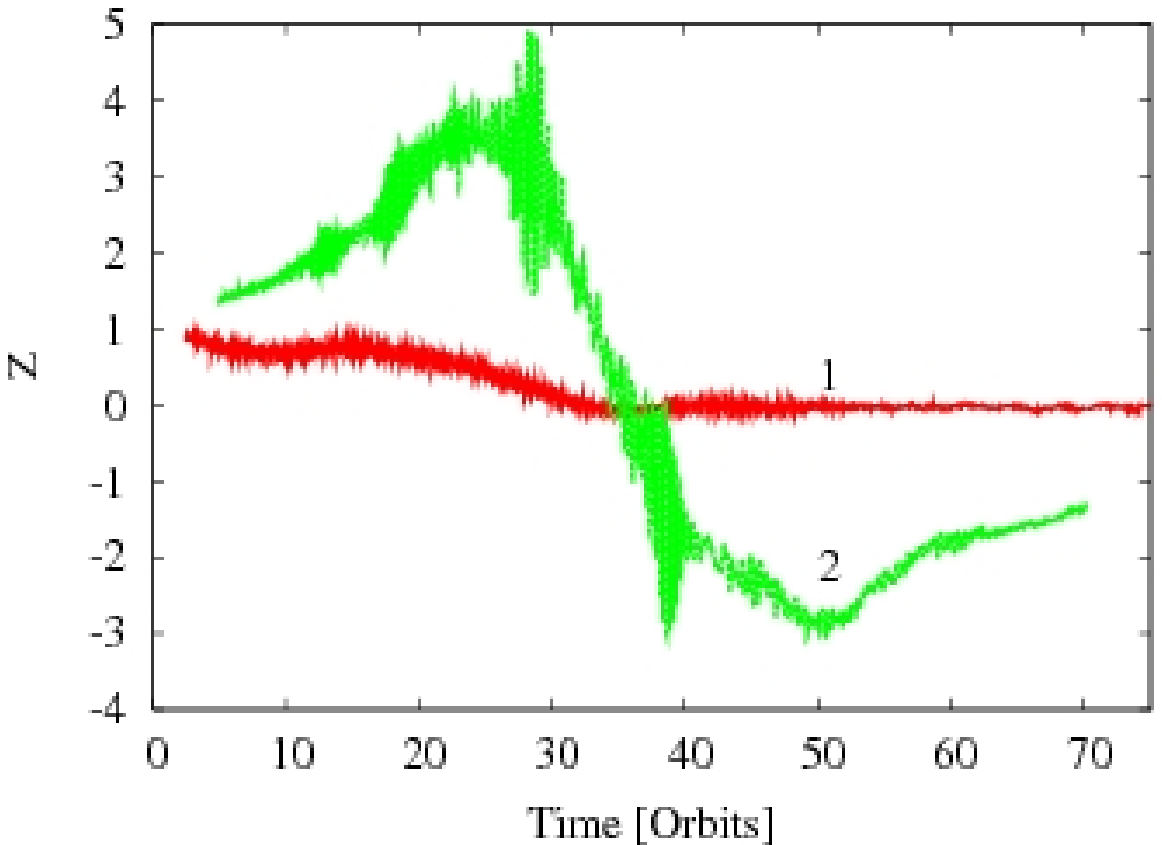}
\includegraphics[width=84mm]{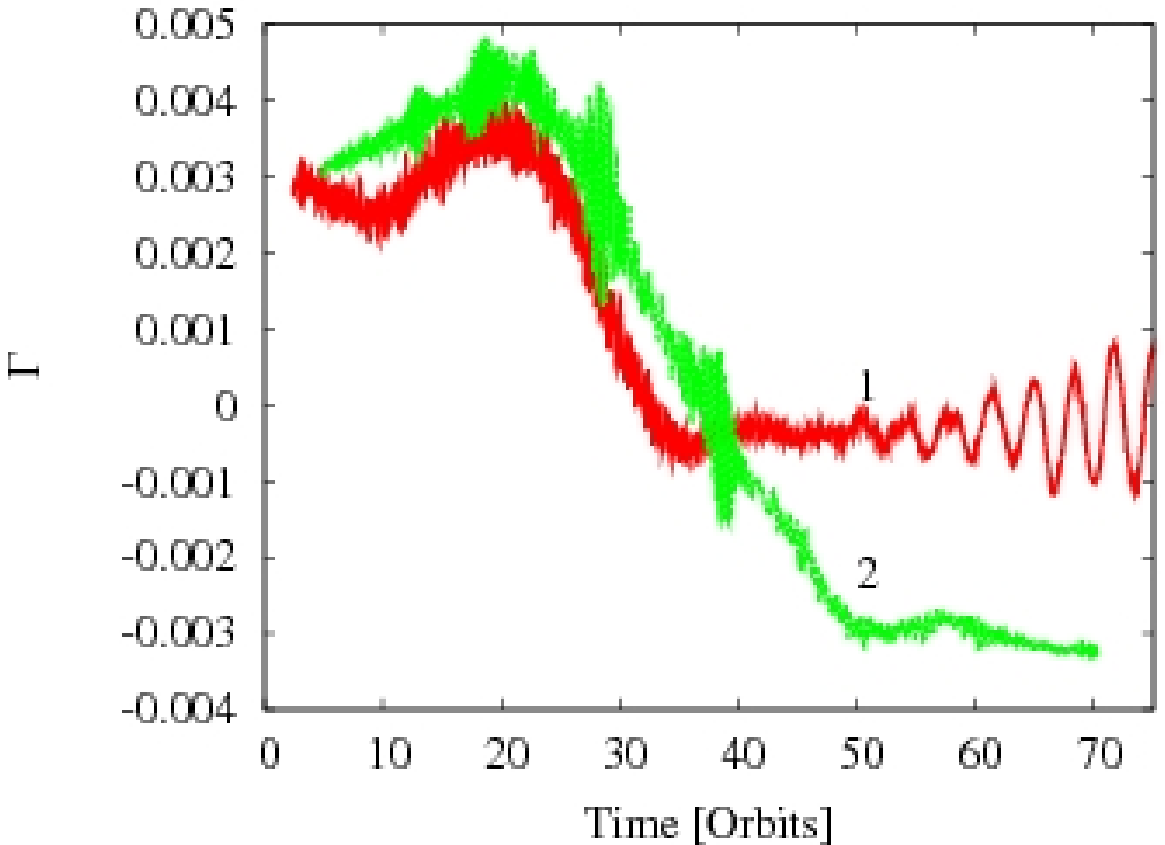}
\includegraphics[width=84mm]{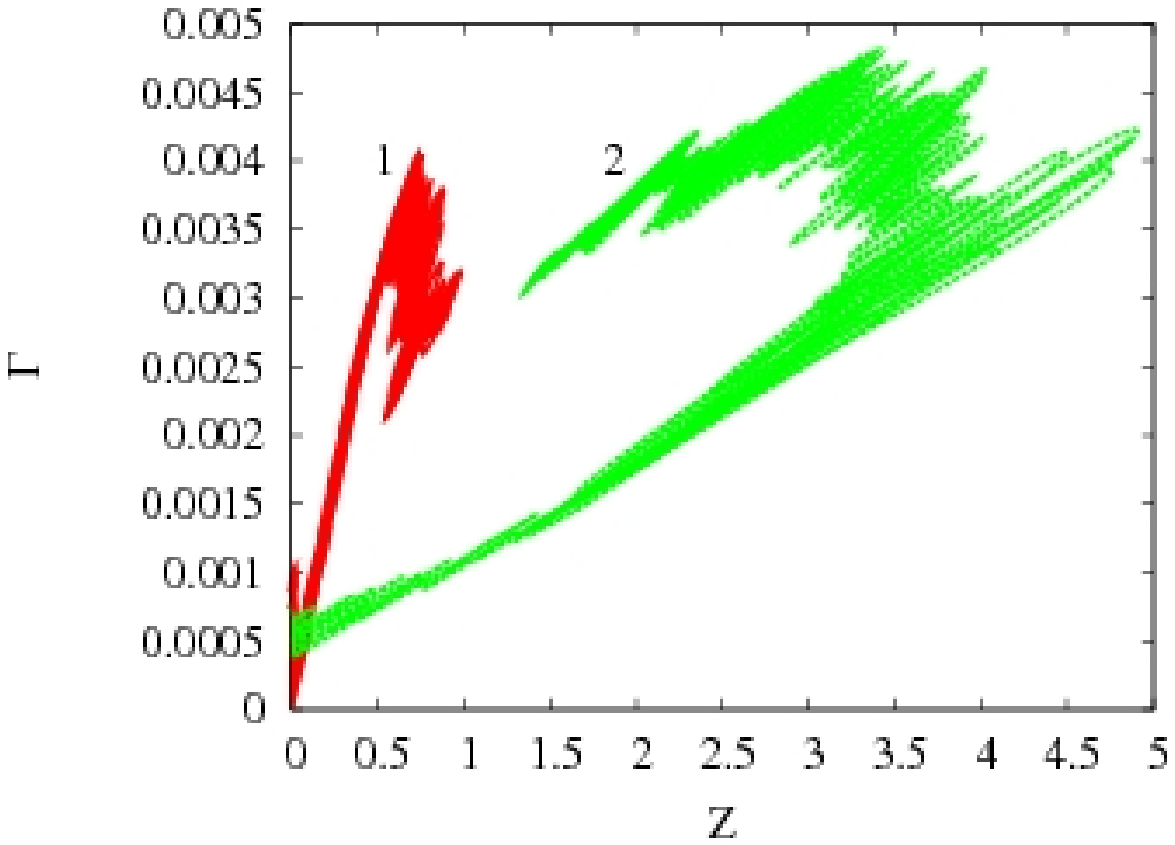}
\caption{Orbital evolution of the outward migrating planet. Upper left
 and upper right panels show the planet's semi-major axis $a$ and
 eccentricity $e$. The migration rate $\dot a$ and the non-dimensional
 migration rate $Z$ are presented in middle left and middle right
 panels. The bottom row shows the torque exerted by the gas on the
 planet $\Gamma$ as a function of time (left plot) and the
 non-dimensional migration rate (right plot). Curves 1 and 2 correspond
 to the models with a time-dependent
 ${M_\rmn{P}}^* =\widetilde M_\rmn{P} =M_\rmn{P}+ M_\rmn{soft}$ (model~M1)
 and constant 
 ${M_\rmn{P}}^* = M_\rmn{P} = 0.001$ planet's mass (model~M2)
 respectively. To remove strong short-period oscillations the data of
 M1 and M2 were averaged over 5 and 10 orbits respectively.} 
\label{fim1_a}
\end{figure*}

\begin{figure*}
\includegraphics[width=84mm]{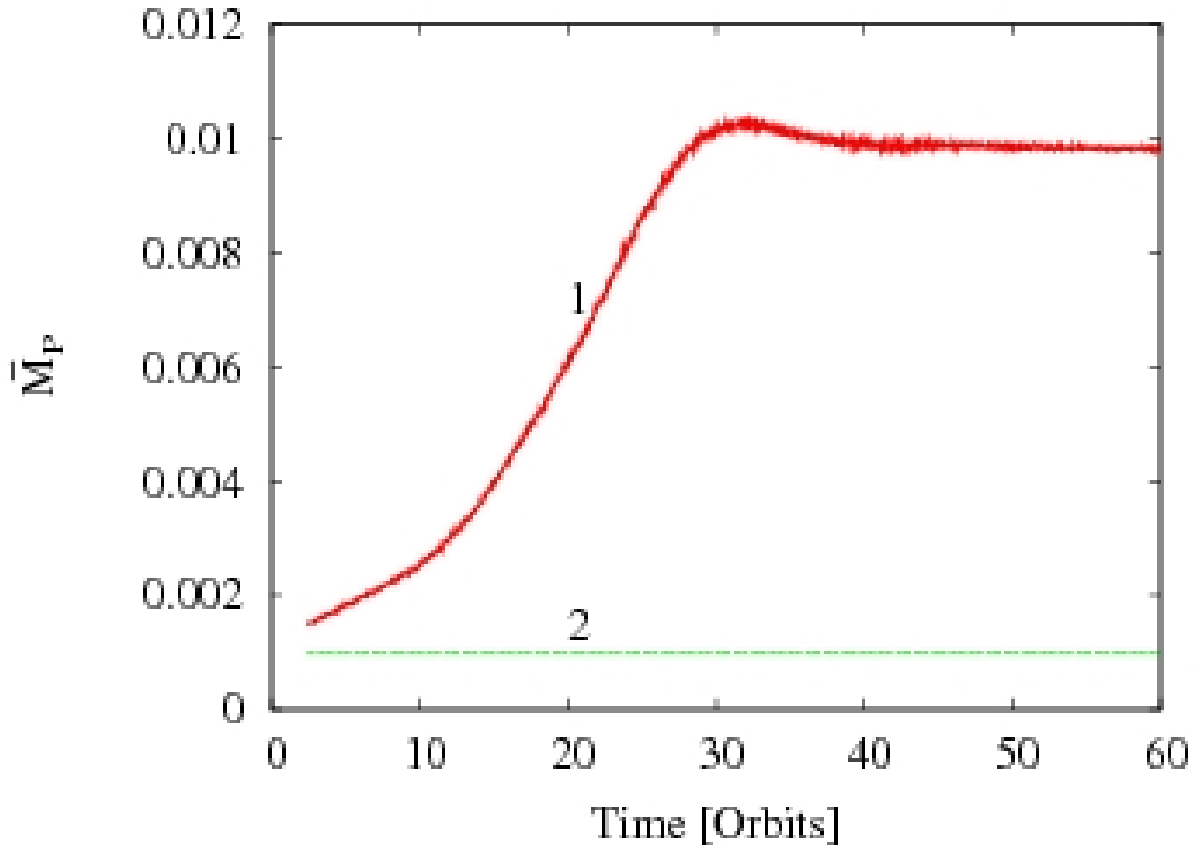}
\includegraphics[width=84mm]{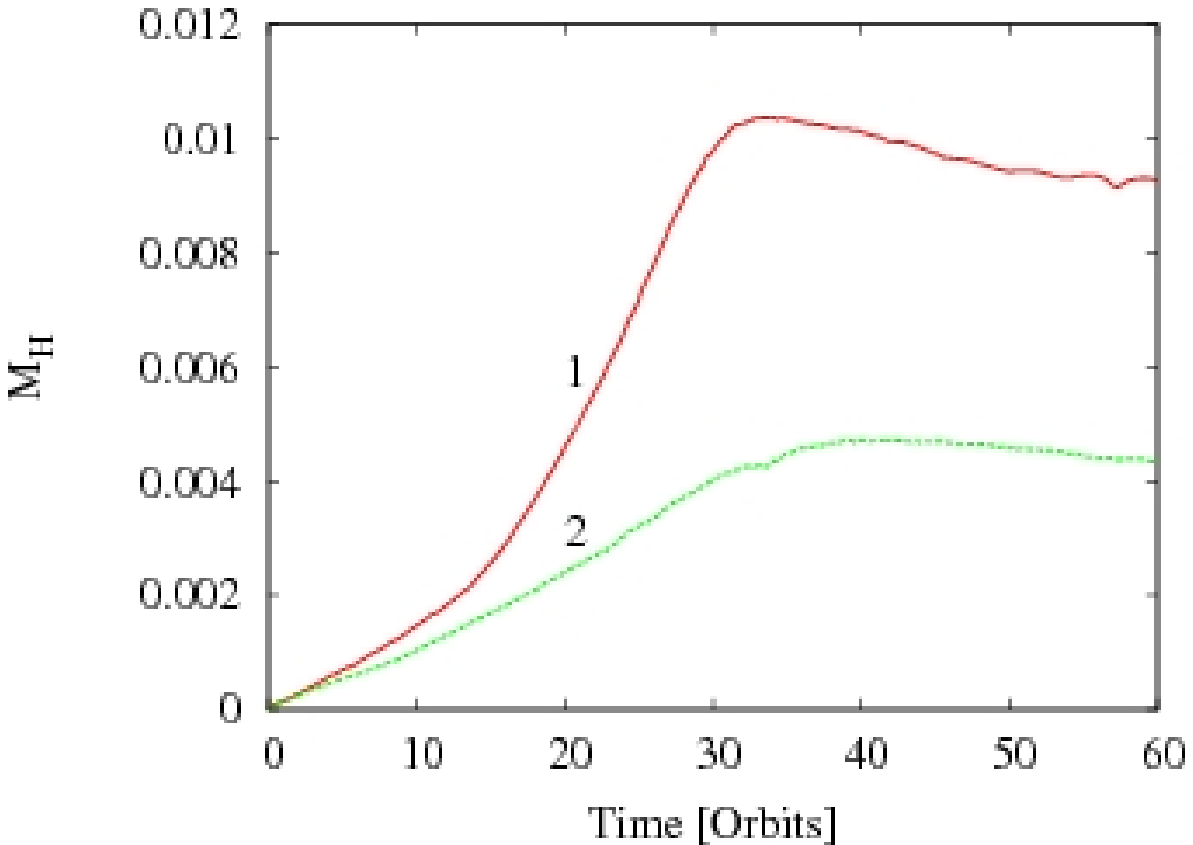}
\caption{The changes of the planet mass (left panel) and the mass of the gas inside
the Hill sphere (right panel) during the orbital evolution. Curves 1 and 2 correspond
to M1 and M2 models respectively. The initial planet mass is $M_\rmn{P} = 0.001$.}
\label{fim1_mp}
\end{figure*}

\section{Outward migration - standard cases}
\label{outward_mig}

Although the driving mechanism for inward and outward type III
migration is similar, we find that the initial conditions for each
have to differ significantly. As found in Paper~II, a relatively low
mass disc (about twice the MMSN) is sufficient to start the inward
migration of a Jupiter mass planet with an initial semi-major axis
$a=3$. To start outward migration for such a planet initially placed
at $a=0.8$, we need a three times more massive disc (with a sharp
inner edge). Another difference is that for outward migration the
planet accumulates a much larger amount of gas as it moves through
the disk. This rapid increase of the effective planet mass $\widetilde
M_\rmn{P}$ significantly modifies the planet's orbital
evolution. However, the evolution of $\widetilde M_\rmn{P}$ depends
strongly on the chosen parameters such as $h_\rmn{p}$, the scale
height of the circumplanetary disc. Thicker (warmer) circumplanetary
discs accumulate less material, and thus results in a different orbital
evolution. They also represent two different mechanisms of stopping
type~III migration.

To be able to compare the cases of high and low mass accumulation, we
will here keep $h_\rmn{p}$ constant at 0.4, but instead in one case
allow the effective planet mass to grow, by adding the mass content
within the smoothing length ($M_\rmn{P}^* = \widetilde M_\rmn{P}$,
model~M1) and in the other keep the effective planet mass constant at
the initial value ($M_\rmn{P}^* = M_\rmn{P} = 0.001$, model~M2). For
both models the initial surface density profile has 
$\mu_\rmn{D} = 0.01$ and $\alpha_\rmn{\Sigma} = -0.5$, with an
inner edge placed at $r=0.8$. At this edge the surface density drops by
a factor of $10$ (see Paper~I, Fig~2).

Note that these models were used in Paper~I, Sect.~5.2.2 in the
discussion of the dependency of the planet's orbital evolution on the
mass accumulation in the circumplanetary disc. Here we discuss them in
more detail.\footnote{Notice the difference between Fig.~14 in Paper~I
  and Fig.~\ref{fim1_mp} in the present paper. All plots in Paper~I and
  most of the ones in this paper present data averaged over 5 orbital
  periods, however in the case of model~M2 the torque and the
  migration rate oscillate strongly in time and we instead averaged
  over 10 orbits.}

\subsection{Orbital evolution}

Figure~\ref{fim1_a} shows the evolution of several key quantities for
models M1 and M2 (curve 1 and 2 respectively). The upper left panel
displays the evolution of the planet's semi-major axis. Both models
show two phases of migration. The first phase is one of rapid, outward
directed migration, lasting for about $30.7$ and $34.4$ orbits in M1
and M2 respectively. It is followed by a phase of inward migration,
which is slower for M1 than for M2. In M1 the effective planet mass
grows rapidly during the first phase of migration, and the planet,
after reaching $a \approx 1.72$, becomes locked in the disc in a
type~II like migration. In M2 this growth is artificially inhibited,
and the planet migrates further, reaching a maximum semi-major axis at
$a \approx 2.26$. At this position the planet starts an
inward-directed rapid migration phase.

In our investigation we concentrate on the first, rapid outward
directed migration phase and on the reversal of the migration
direction. We will not describe in detail the second, inward migration
phase, which corresponds to the transition between type III and type
II migration regime in M1, and to the rapid inward migration in M2.

The time evolution of the effective planet mass $\widetilde M_\rmn{P}$
and the mass of the gas inside the Hill sphere $M_\rmn{H}$ are shown
in Fig.~\ref{fim1_mp} (see also Paper~I,
Fig.~14, but notice the change of the labels). In both models the
planet accumulates gas during the outward migration phase. In M1
$M_\rmn{H}$ grows faster than in M2, and reaches $11 M_{\jupiter}$ at
$30$ orbits. This is a considerable fraction of the original mass of
the protoplanetary disc, which contained $29 M_{\jupiter}$ within
$r=2$. In the slow inward migration phase the planet's mass is almost
constant and ${\widetilde M}_\rmn{P} \approx 10 M_{\jupiter}$.
Comparing $\widetilde M_\rmn{P}$ and $M_\rmn{H}$ we can see that most
of the mass is actually accumulated inside the planet's gravitational
smoothing length. In M2 a smaller amount of gas enters the Hill sphere
and a maximum of $M_\rmn{H} \approx 5 M_{\jupiter}$ is reached at
$t=40$ orbits.

This growth of $\widetilde M_\rmn{P}$ is possible due to the use of a
locally isothermal equation of state with a constant circumplanetary
disc aspect ratio $h_\rmn{p}$. We study the effects of varying
$h_\rmn{p}$ below in Sect.~\ref{subs_dep_hp}. Note that in M1
there is no mass outflow from the Roche lobe during the slow migration
phase, such as was seen in the inward migration case described in
Paper~II, Sect.~4.3.  In this respect, M1 is more similar to the
inward migration simulation with a lower $h_\rmn{p}=0.3$ (see Paper~I,
Sect.~5.1.2).

The evolution of the migration rate $\dot a$ is presented in middle
left panel of Fig.~\ref{fim1_a}. Initially $\dot a$ grows with the
planet's semi-major axis for both models and has a maximum at $t
\approx 20$ orbits ($\dot a \approx 0.0062$) and $t \approx 23.3$
orbits ($\dot a \approx 0.011$) for M1 and M2 respectively. In M1
$\dot a$ has a lower value due to the increase of the planet's
inertia. Later $\dot a$ drops in a similar way for both models,
however in M2 the planet goes through a short period of strong
oscillations of the migration rate around $t\approx 28$ orbits.  The
physical mechanisms behind the drop in migration rates are addressed
in Sects.~\ref{in_flow_str} and~\ref{stop_mig}.

During the slow inward migration phase of model M1 $\dot a$ oscillates
around $-3 \times 10^{-4}$. The oscillations are slightly more visible
after about $60$ orbits after the planet's eccentricity $e$ has
increased. As we saw before, in M2 the planet enters a rapid inward
migration phase.

The plot of the non-dimensional migration rate $Z$ presented in middle
right panel helps us understand the difference between the two
simulations. As explained in Paper~II, the non-dimensional migration
rate is defined as a ratio of the planet's migration rate $\dot a$ and
the so-called fast migration speed $\dot a_\rmn{f}$, which is given by
the ratio of the half width of the horseshoe region $x_\rmn{s}$ and
the libration timescale $T_\rmn{lib}$ \citep{PaperIV}. It is
equivalent to $Z = T_\rmn{lib}/T_\rmn{migr}$, where $T_\rmn{migr}=
x_\rmn{s}/{\dot a}$ is the migration time-scale. Using $Z$ we can
divide type III migration into a fast $|Z|>1$ and a slow
$|Z|<1$ migration regime. Note that this division is not identical
to that between type III and type II migration. Type II like migration
corresponds to the phase of nearly constant $|Z| \ll 1$.

The initial value of $Z$ is close to $1$ for both M1 and M2. However,
due to the increase in $M_\rmn{P}^*$ and the fact that $\dot a_\rmn{f}
\sim a^{-0.5}(M_\rmn{P}^*)^{2/3}$, $Z$ never exceeds $1$ for model M1,
and the planet migrates in the slow migration regime during the entire
simulation. During the outward migration phase $Z$ oscillates around
$0.7$, although both $a$ and $\dot a$ grow during the first $20$
orbits. Later $Z$ decreases reaching its final value in the slow
inward migration phase. For M2, $|Z|>1$ during essentially the entire
simulation, so in the outward as well as in the inward migration phase
the planet migrates in the fast migration regime.

As shown in Paper II and below in Sect.~\ref{in_flow_str}, the value
of $|Z|$ correlates with the asymmetry of the co-rotation region, and
gap opening occurs for $|Z| \ll 1$. Model M1 thus experiences gap opening,
and a smooth transition to inward migration, whereas model M2 maintains
a strong asymmetry in the co-orbital region, without any gap opening.

The upper right panel in Fig.~\ref{fim1_a} shows the changes of the
eccentricity $e$. In M1 $e$ is approximately constant during the
initial outward migration phase and does not exceed 1\%. At $t\approx 53$ 
orbits, the eccentricity starts to rapidly increase. This
coincides with an increase of the eccentricity of the outer
circumstellar disc (outer with respect to the planet's orbit). At
about $48$ orbits the outer disc becomes eccentric (lower left panel
in Fig.~\ref{fim1_dens_disc}) and the torque $\Gamma$ starts to
oscillate (lower left panel in Fig.~\ref{fim1_a}). Note that the
oscillations of $\Gamma$ are driven by the gas placed outside the Hill
sphere. The oscillations of the torque generated by the gas from
within the Hill sphere are order of magnitude smaller and start to be
visible after about $60$ orbits. In the later stages $e$ oscillates
around $4\%$. In M2 the eccentricity grows during the outward
migration phase reaching $3\%$ and decreases in the inward migration
phase.

The lower left panel in Fig.~\ref{fim1_a} presents the time evolution
of the total torque exerted by the gas on the planet. $\Gamma$ grows
during the rapid outward migration phase in both models, however in M1
there is a short ($\sim 10$ orbits) period of decreasing torque. The
torque has a maximum at about 20 orbits and its value, unlike the
maximum value of the migration rate, is similar for both models. Later
on $\Gamma$ decreases rapidly and the planets enter the inward
migration phase. There is visible a short period of strong
oscillations of $\Gamma$ at about 28 orbits in M2. As noted above,
$\Gamma$ starts to oscillate in M1 at 48 orbits, when the eccentricity
of the outer disc starts to grow.  M2 also experiences a short period
of torque oscillations around $t\approx 28$ orbits.

In Paper~II Sect.~4.1 we described the relation between the averaged
torque and the non-dimensional migration rate $Z$ for the inward
migration case. $|\Gamma|$ was shown to grow linearly with $|Z|$ in
the slow migration regime, and to decrease slowly in the fast
migration regime. The trend for the outward directed migration is
similar, however there are also some differences. As we saw above,
models M1 and M2 show very different values for $Z$, and it we can
study the relation between $\Gamma$ and $Z$ in slow ($Z<1$) migration
regime by looking at M1, and in the fast ($Z>1$) migration regime
using M2.  We show this relation in the lower right panel in
Fig.~\ref{fim1_a}. In the slow migration regime (M1) the torque
evolution is strongly influenced by the mass accumulation in the
circumplanetary disc and shows two different stages. In the first one
(starting at $Z\approx 1$) $\Gamma$ grows reaching the maximum at $Z
\approx 0.6$. The second stage with $\Gamma\sim Z$ corresponds to a
slowing down of migration and agrees with the results for the inward
directed migration. Model M2 (always in the fast migration regime)
shows two stages too. The first one qualitatively agrees with the
results for the inward directed migration with $\Gamma$ oscillating
between $0.003$ and $0.0045$, and being less dependent on $Z$. In the
second stage the planet reverts rapidly its direction of migration and
$\Gamma$ decays quickly with $Z$.


\begin{figure*}
\includegraphics[width=84mm]{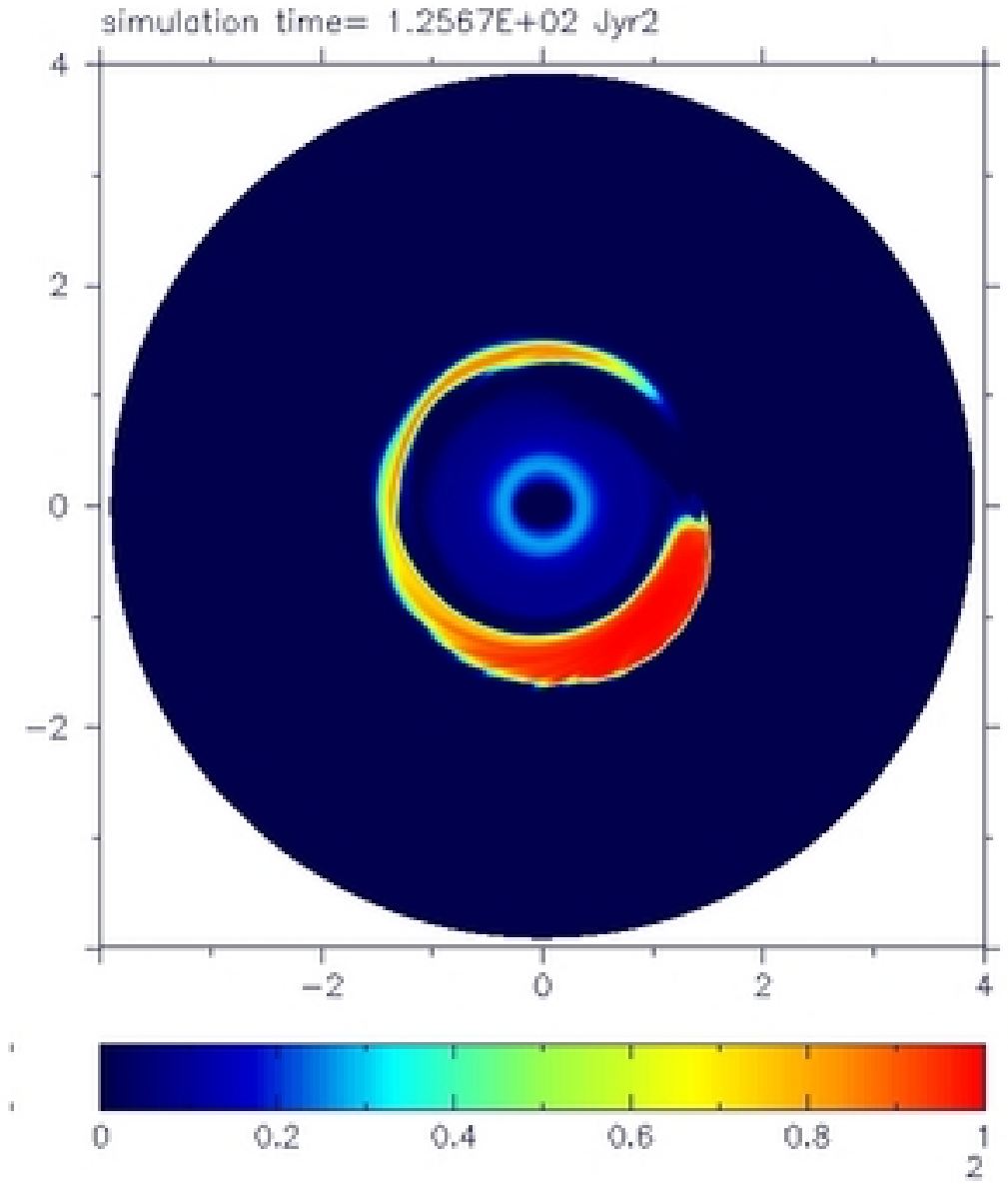}
\includegraphics[width=84mm]{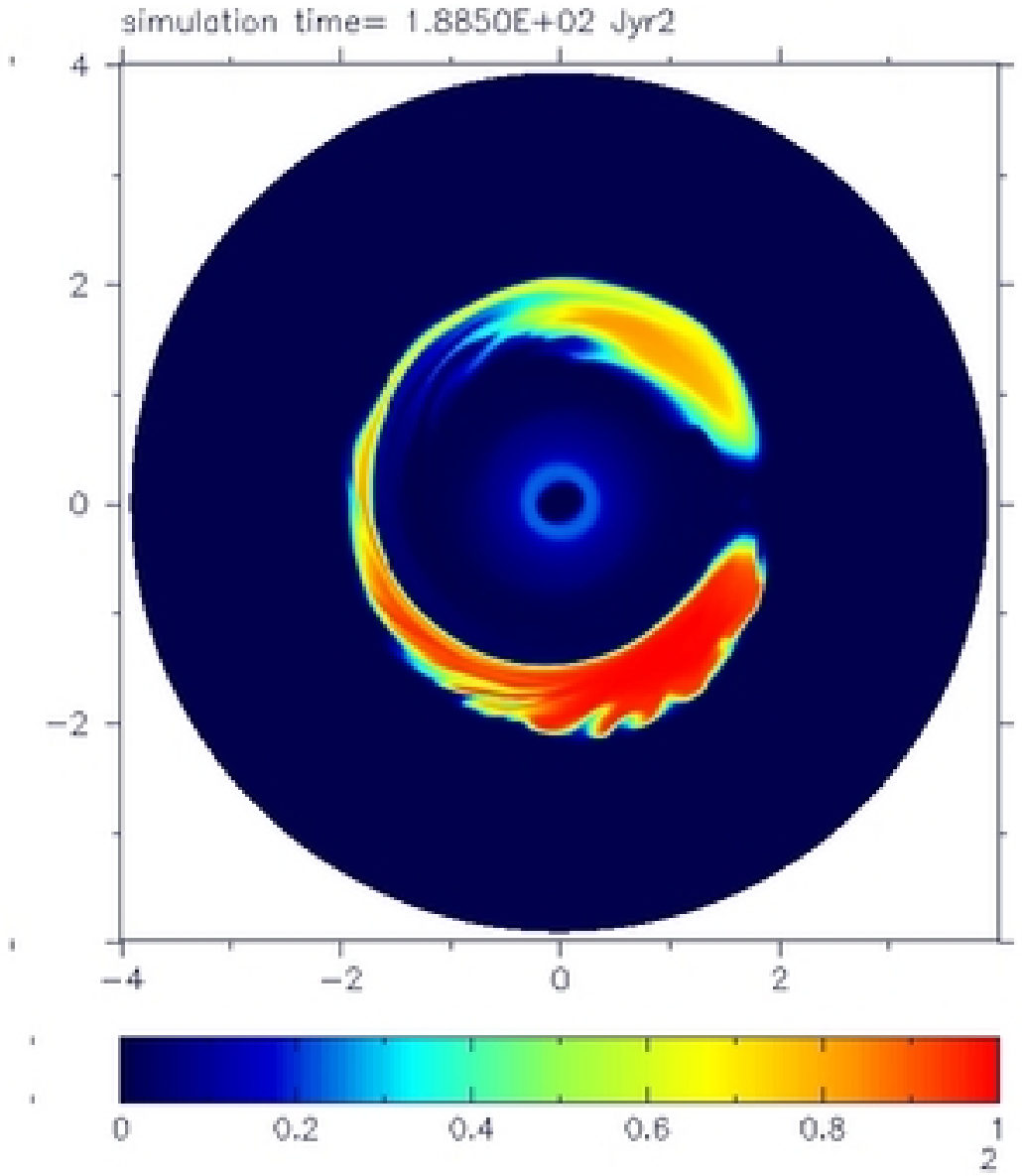}
\includegraphics[width=84mm]{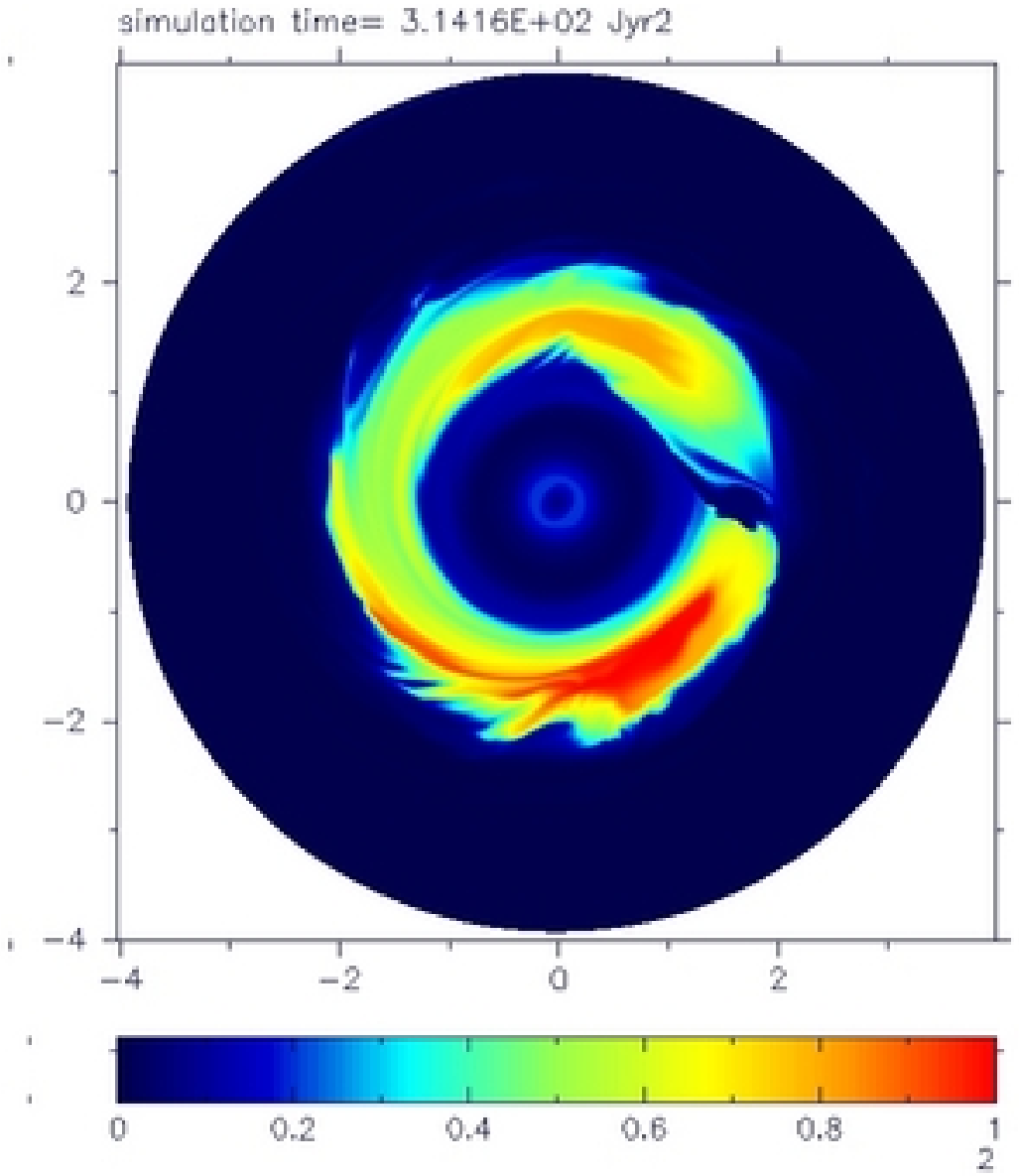}
\includegraphics[width=84mm]{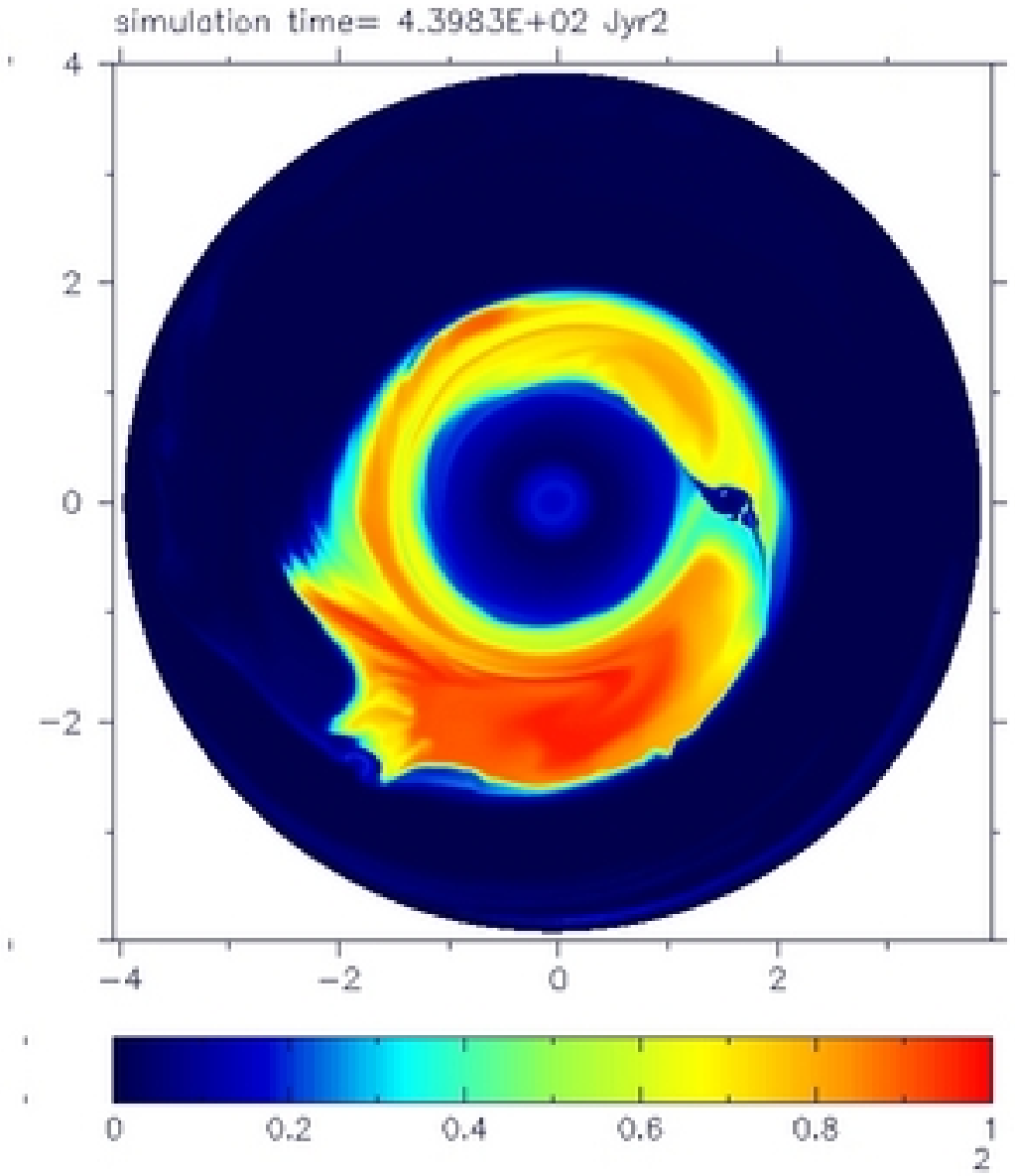}
\caption{The mass fraction of the gas that initially was placed in the
 corotational region for model M1. The mass fraction goes from $0$
 (region occupied by the gas from the inner or outer disc only) up to
 $1$ (region occupied by the gas from corotation only). The top left
 ($t=20$ orbits $Z\approx 0.6$) and a top right ($t=30$ orbits
 $Z\approx0.1$) panels present the outward migration phase. The lower left and
 lower right panels show the slow migration phase $|Z| \ll 1$ at $t=50$
 and $70$ orbits respectively. The upper row and the lower left panel
 present the disc for the planet's eccentricity $e < 0.01$. After about
 $50$ orbits both the planet's and the outer disc eccentricity rapidly
 grow giving a strongly asymmetric flow. It is presented in the lower
 right panel. All plots are done in co-moving reference frame with the
 planet at $(x,y)$ equal $(1.42,0)$, $(1.71,0)$, $(1.66,0)$ and
 $(1.62,0)$ respectively.} 
\label{fim8_comp_disc}
\end{figure*}

\begin{figure*}
\includegraphics[width=84mm]{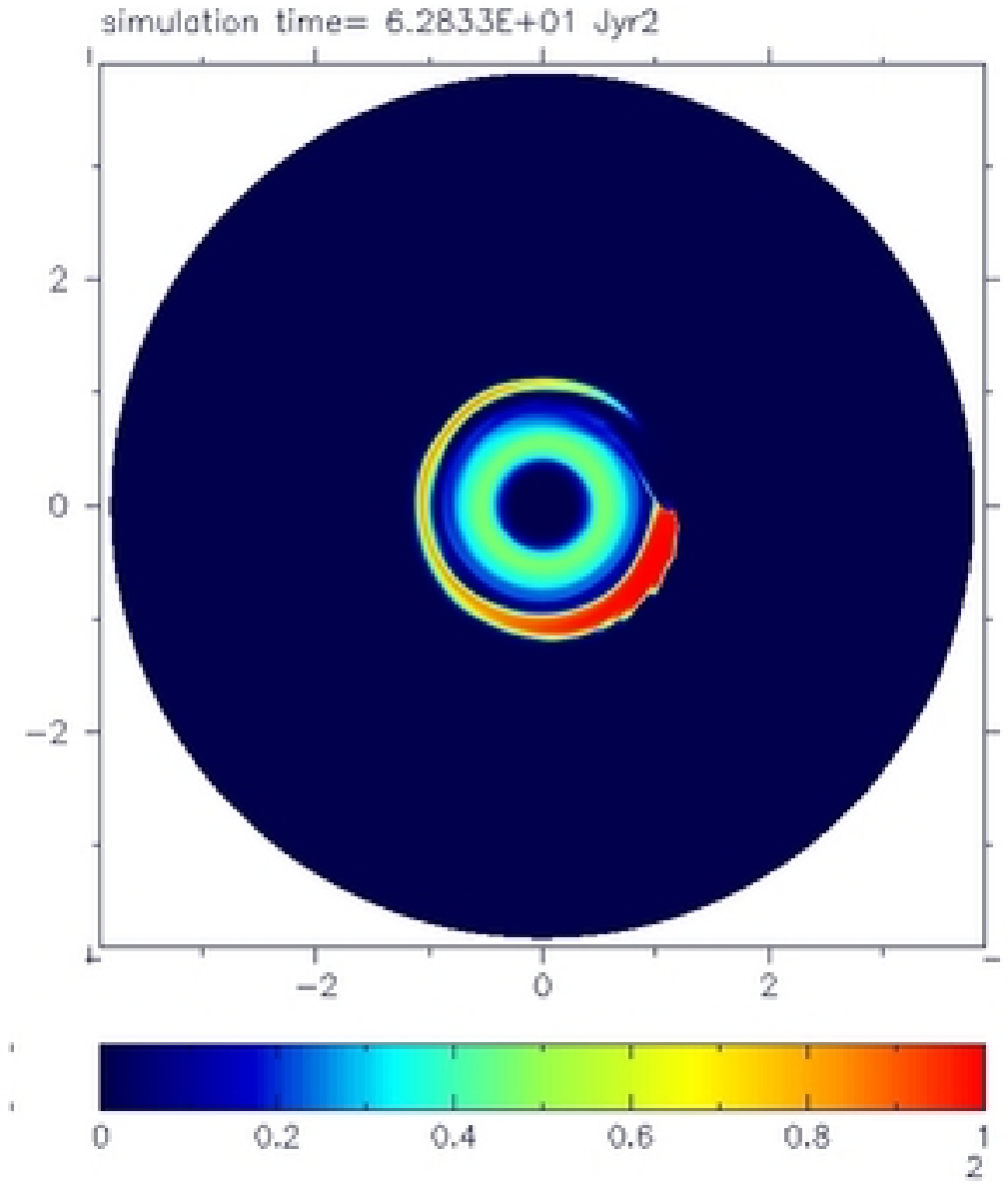}
\includegraphics[width=84mm]{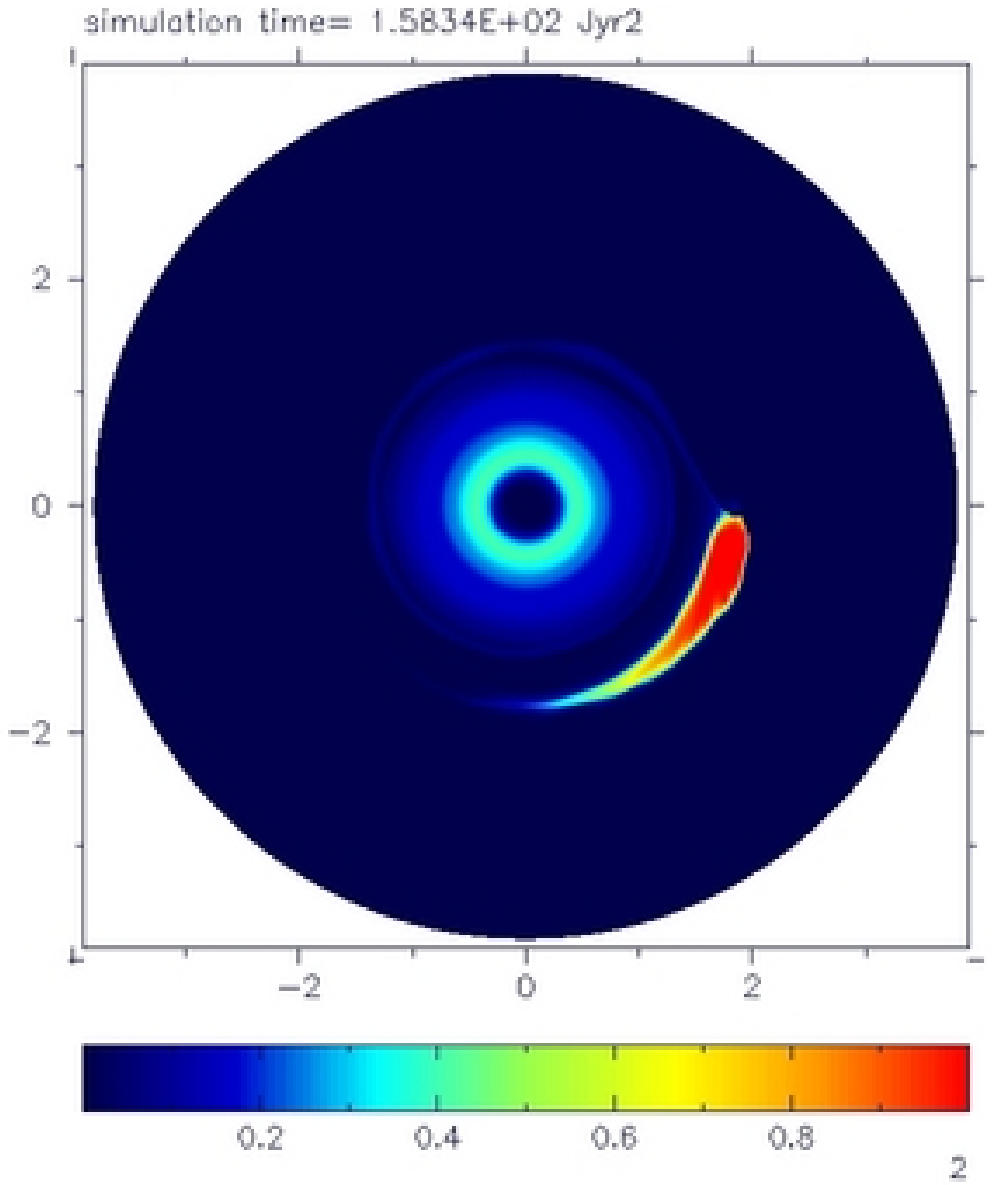}
\includegraphics[width=84mm]{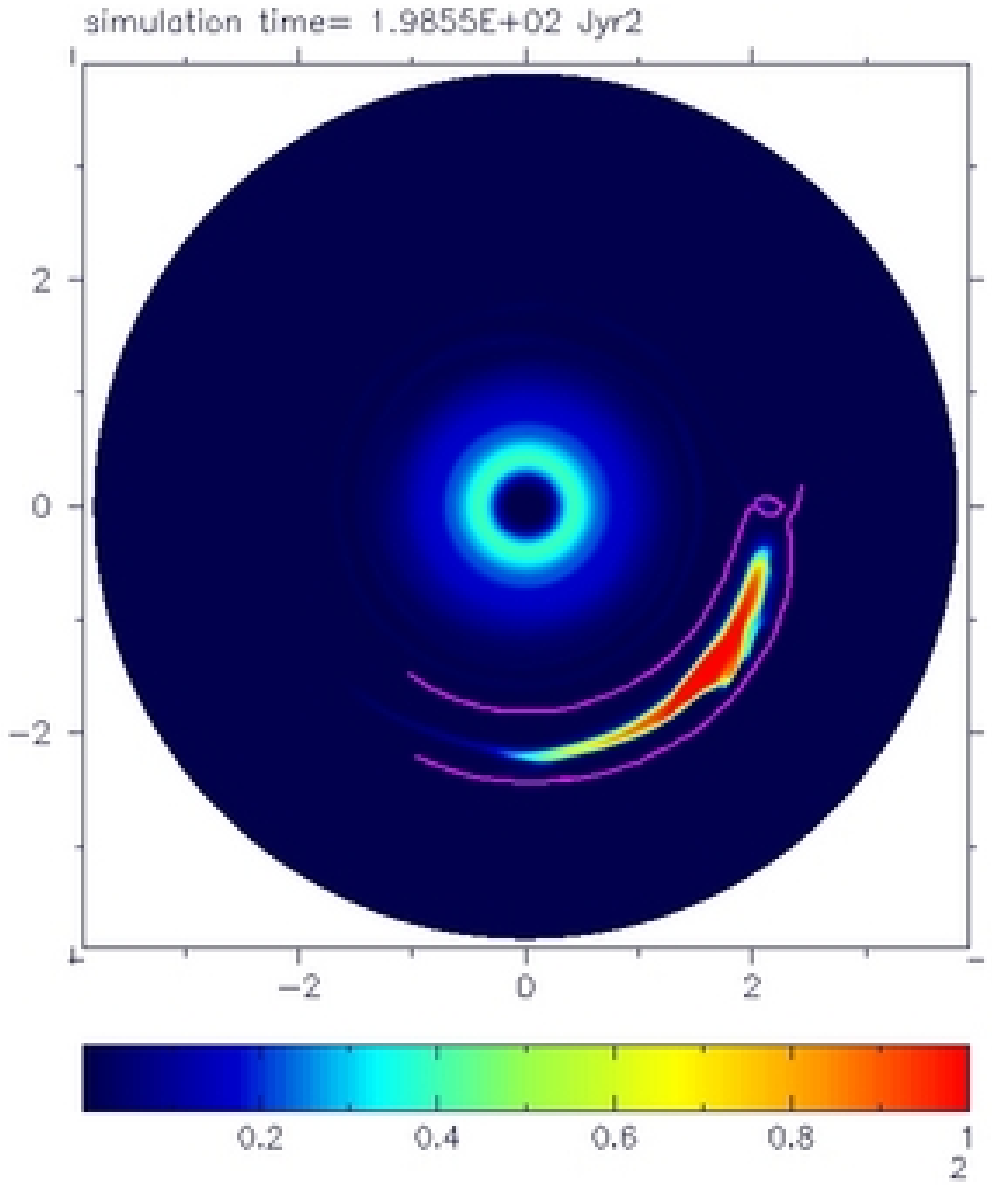}
\includegraphics[width=84mm]{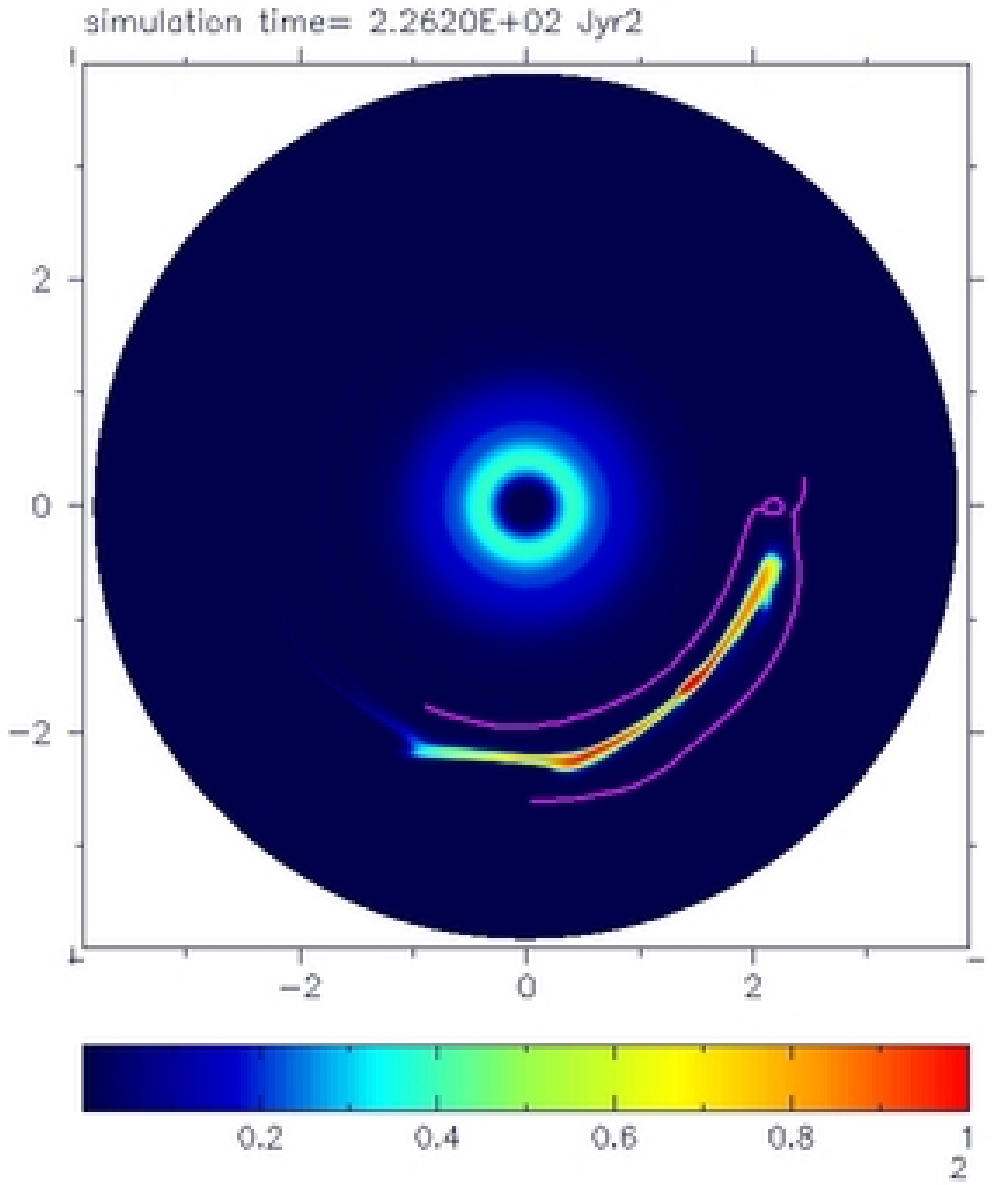}
\caption{The mass fraction of the gas that initially was placed in the
  corotational region for model M2. The top left ($t=10$ orbits
  $Z\approx 1.7$) and a top right ($t=25.3$ orbits $Z\approx 4.5$)
  panels present the migrating planet increasing the non-dimensional
  migration rate. The lower left and lower right panels show the
  planet in the stage of slowing down the migration at $t=31.6$ ($Z
  \approx 2$) and $36$ ($Z \approx 0$) orbits respectively. The pink
  lines show the flow lines at the border of the horseshoe region. All
  plots are done in co-moving reference frame with the planet at
  $(x,y)$ equal $(1.1,0)$, $(1.87,0)$, $(2.1,0)$ and $(2.25,0)$
  respectively.}
\label{fim8_comp_disc_2}
\end{figure*}

\subsection{Flow structure in the co-orbital region}

\label{in_flow_str}

In Paper~II we discussed the relation between the time-averaged torque
and the asymmetry of the horse shoe region for the inward migration
case. We have found the shape of the horseshoe region to be dependent
on the relation between the migration time scale $T_\rmn{migr}$ and
the libration timescale $T_\rmn{lib}$, as expressed by the
non-dimensional migration rate $Z$. In the slow migration regime the
horse shoe region extends almost the entire $2\pi$ angle of the
co-rotation region, and is fairly symmetric. The co-orbital flow is
limited to a small region near to the planet. In the fast migration
regime the horseshoe region shrinks to a single tadpole-like region
of limited azimuthal extent, disconnecting from the planet on one side
and the co-orbital flow takes place over a wide azimuthal range.

A similar dependency is observed for the outward migration
case. However, the fact that the volume (and mass) of the co-orbital
region grows with $a$ during outward migration produces important
differences. The relation between the shape of the horseshoe region
and the non-dimensional migration rate for the outward migration in
the slow and the fast migration regimes is illustrated by models M1
and M2, respectively. The results are presented in
Figs.~\ref{fim8_comp_disc} and~\ref{fim8_comp_disc_2}. The plots show
the mass fraction of the gas that initially was placed in the
corotational region $a_\rmn{init}-2 R_\rmn{H} < r_\rmn{s} < a+2
R_\rmn{H}$, where $r_\rmn{s}$ is here the distance to the star and
$a_\rmn{init}$ is the initial planet's semi-major axis. The mass
fraction goes from $0$ (dark blue; region occupied by the gas from the
inner or outer disc only) up to $1$ (red; region occupied by the gas
from corotation only).

We first consider the slow migration regime (M1,
Fig.~\ref{fim8_comp_disc}). In this case, unlike the inward directed
migration, most of the gas placed initially in the co-orbital region
is captured by the planet in the tadpole-shaped horseshoe region
around the generalised libration point G5\footnote{In Paper~II we
introduced the concept of {\it the generalised Lagrangian points} G4 and
G5, which for a slowly migrating planet correspond to L4 and L5
libration points. This distinction is necessary, since for a rapidly
migrating planet these libration points move in a direction of a
planet.} and migrates together with
the planet. The top left and a top right panels present the outward
migration phase for $Z\approx 0.6$ ($t=20$ orbits) and $Z\approx 0.1$
($t=30$ orbits) respectively. In the first panel the horseshoe region
is visibly asymmetric, with only one tadpole being present. The
azimuthal extent of the horseshoe region is about $1.7\pi$ and the
co-orbital flow is a relatively broad stream, giving a large positive
value for the torque $\Gamma$. In the second panel both tadpoles are
present and the asymmetry of the horseshoe region is starting to
disappear. As the azimuthal extent of the horse shoe region grows,
that of the co-orbital flow shrinks. This reduces the co-orbital
torque $\Gamma_\rmn{CR}$ (and the total torque $\Gamma$ which is
dominated by the co-orbital torque) considerably. The co-orbital flow
keeps decreasing with decreasing $Z$ and disappears in the subsequent
inward slow migration phase, when the horseshoe region covers the
entire azimuthal range (lower left panel). For $|Z| \ll 1$ the
planet's radial motion is slow and a gap has started to be cleared as
the planet makes the transition to type II migration. The last panel
displays the mass fraction at $t=70$ after the planet has almost
opened a gap in the strongly eccentric circumstellar disc.

During the outward migration phase with $Z \sim 1$ (upper left panel)
the gas captured in the horseshoe region does not mix with the co-orbital
flow. The situation changes for the slower migration. The horseshoe
region increases its volume with decreasing $Z$ and the planet captures
part of the gas from the co-orbital flow starting a strong mixing of
the gas. 

This behaviour of model M1 is similar to the scenario described in
Paper~II, where we also found a smooth transition between the fast and
slow migration regimes corresponding to an increase in the extent and
symmetry of the horse shoe region, and a shrinking of the co-orbital
flow.  Note however that the reasons are different: decrease of the
co-orbital mass for the inward migration case and increase of
$\widetilde M_\rmn{P}$ for M1.

Model M2 shows a different evolution, see Fig.~\ref{fim8_comp_disc_2}.
As in M1 most of the gas initially placed in the co-orbital region is
captured by the planet in the tadpole-shaped horseshoe region around
the G5 libration point and migrates together with the planet. The top
left and a top right panels present the migrating planet during the
phase of increasing $Z$, with $Z \approx 1.7$ ($t=10$ orbits) and
$Z\approx 4.5$ ($t=25.3$ orbits). At both times the horseshoe region
is a single tadpole-like region, and the co-orbital flow has a wide
azimuthal range. The azimuthal extent of the horseshoe region clearly
decreases with $Z$ and is about $1.7 \pi$ and $0.5 \pi$ for $Z=1.7$
and $Z=4.5$ respectively. On the other hand the radial extent of this
region increases, since $x_\rmn{s} \sim a$, and the amount of mass in
the co-orbital region grows during the migration making outward
migration potentially a self-accelerating, run-away process. In M1 the
run-away character is suppressed due to the rapid grow of the planet's
inertia.  But even in M2, where the planet mass is constant, we do not
find run-away migration. The reason is that for the given disc
structure there is a critical value of $Z$, for which the horseshoe
region disconnects from the planet, leading to a decrease of the
co-orbital torque. This process is shown in the lower left and lower
right panels. Here the planet is slowing down its migration at
$t=31.6$ ($Z\approx 2$) and $36$ ($Z \approx 0$) orbits
respectively. The pink lines are the flow lines at the border of the
horseshoe region and show that the gas marked by the red colour does
not fill the whole horseshoe region. After the initial horseshoe
region disconnects from the planet causing a drop in
$\Gamma_\rmn{CR}$, the azimuthal extent of the horseshoe region
grows rapidly as the migration rate drops. As a result a large amount
of gas from the co-orbital flow is captured by the planet in the new
horseshoe region, stopping the mass transfer between outer and inner
disc which was responsible for the rapid migration. We find the
stream-lines in the co-orbital region to change rapidly and no smooth
transition from outward to inward migration occurs. This behaviour is
similar to the stopping of inward migration at a disc edge (Paper II,
Sect.~6). We will discuss this in more detail in
Sect.~\ref{stop_mig}. However we will encounter these two stopping mechanisms
below in other simulations. We will refer to them as stopping mechanism
M1 (for stopping due to mass accumulation), and M2 (for loss of the
horseshoe region).

Model M2 does not show not much mixing between the gas
originally captured in the horseshoe region and the co-orbital
flow. Note that during the entire phase of fast migration a small
amount of the gas is leaving the horseshoe region, flowing along the
bow shock in the inner disc. This is visible in the plots as a light
blue spiral.

\begin{figure*}
\includegraphics[width=84mm]{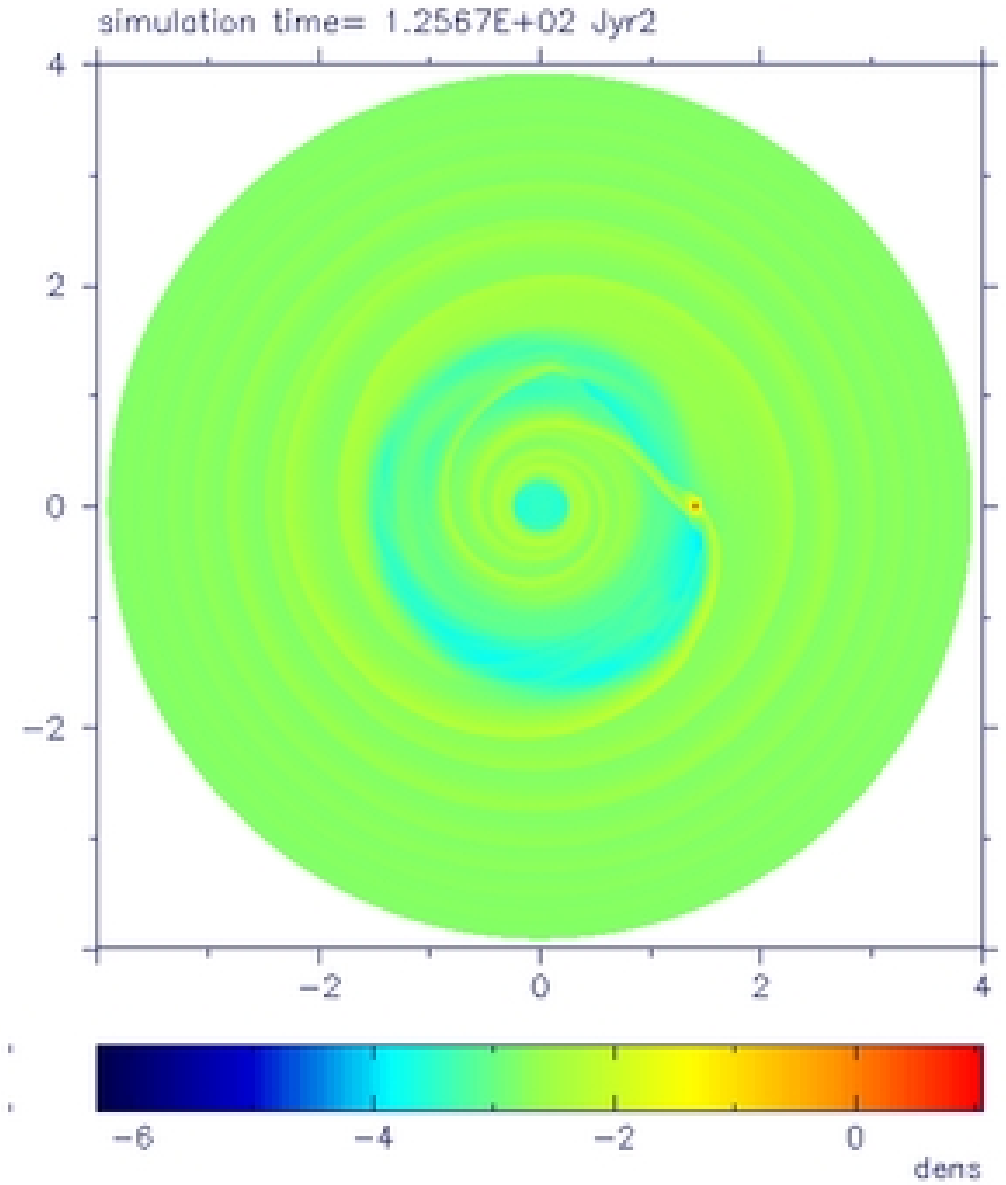}
\includegraphics[width=84mm]{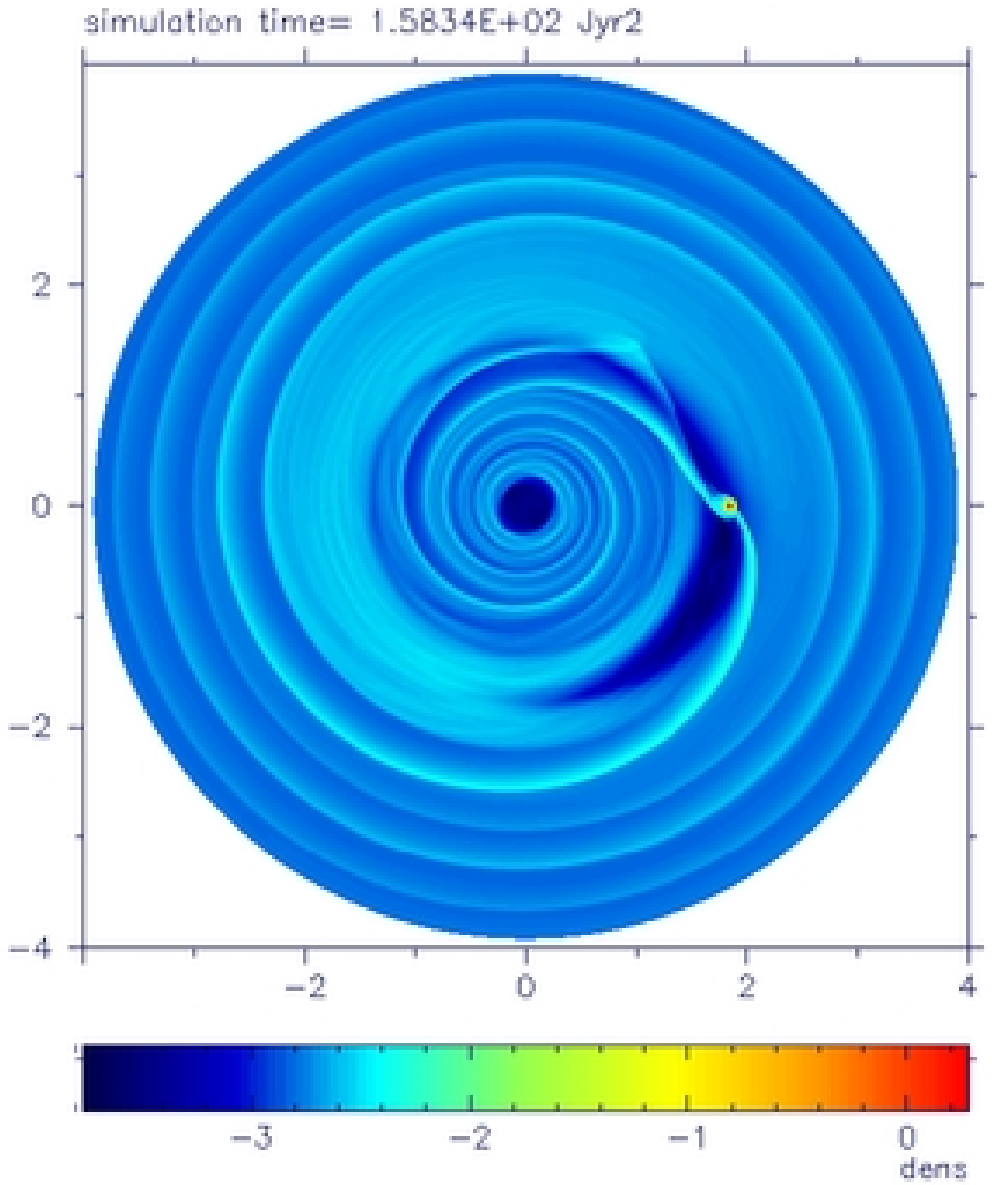}
\includegraphics[width=84mm]{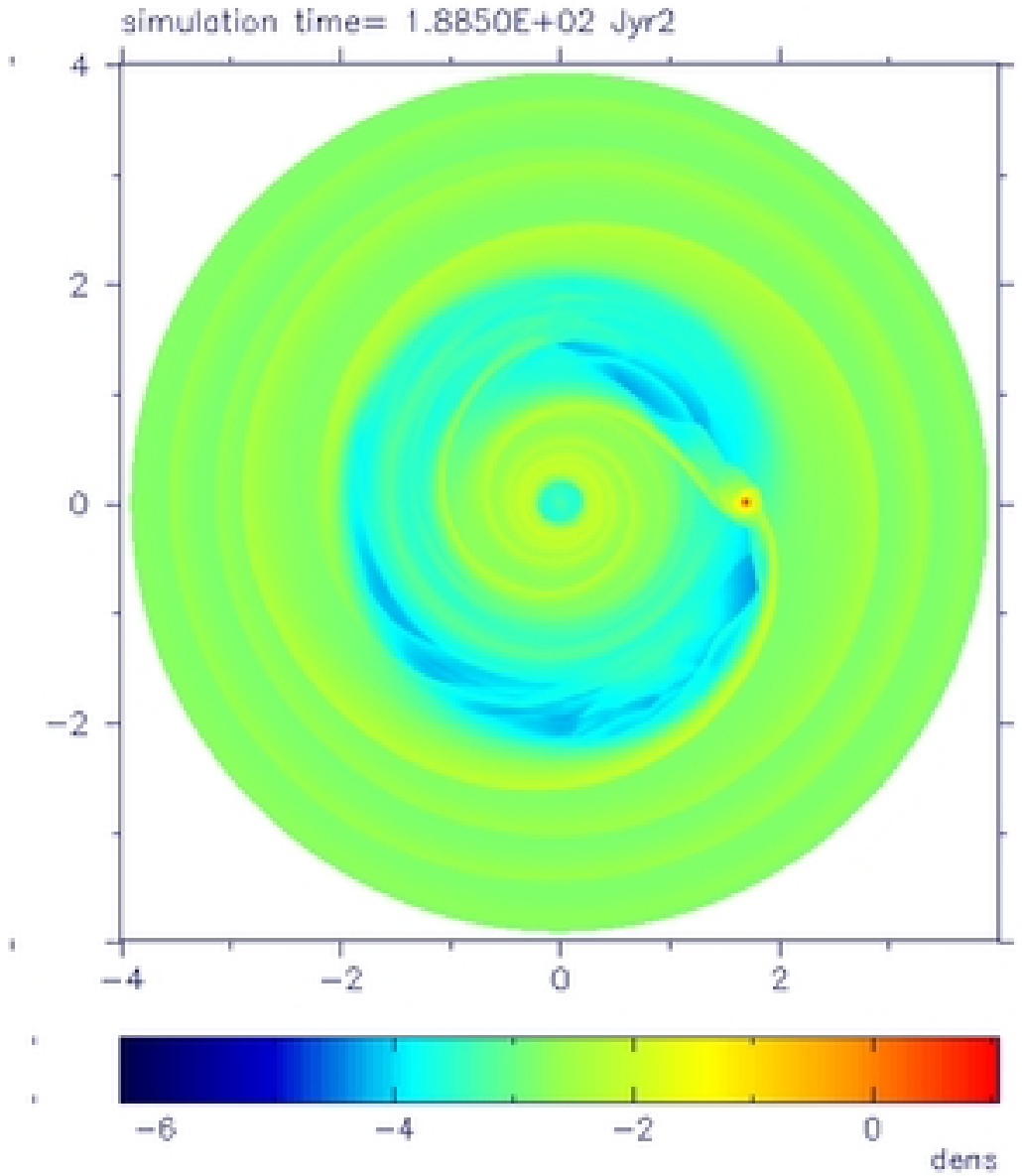}
\includegraphics[width=84mm]{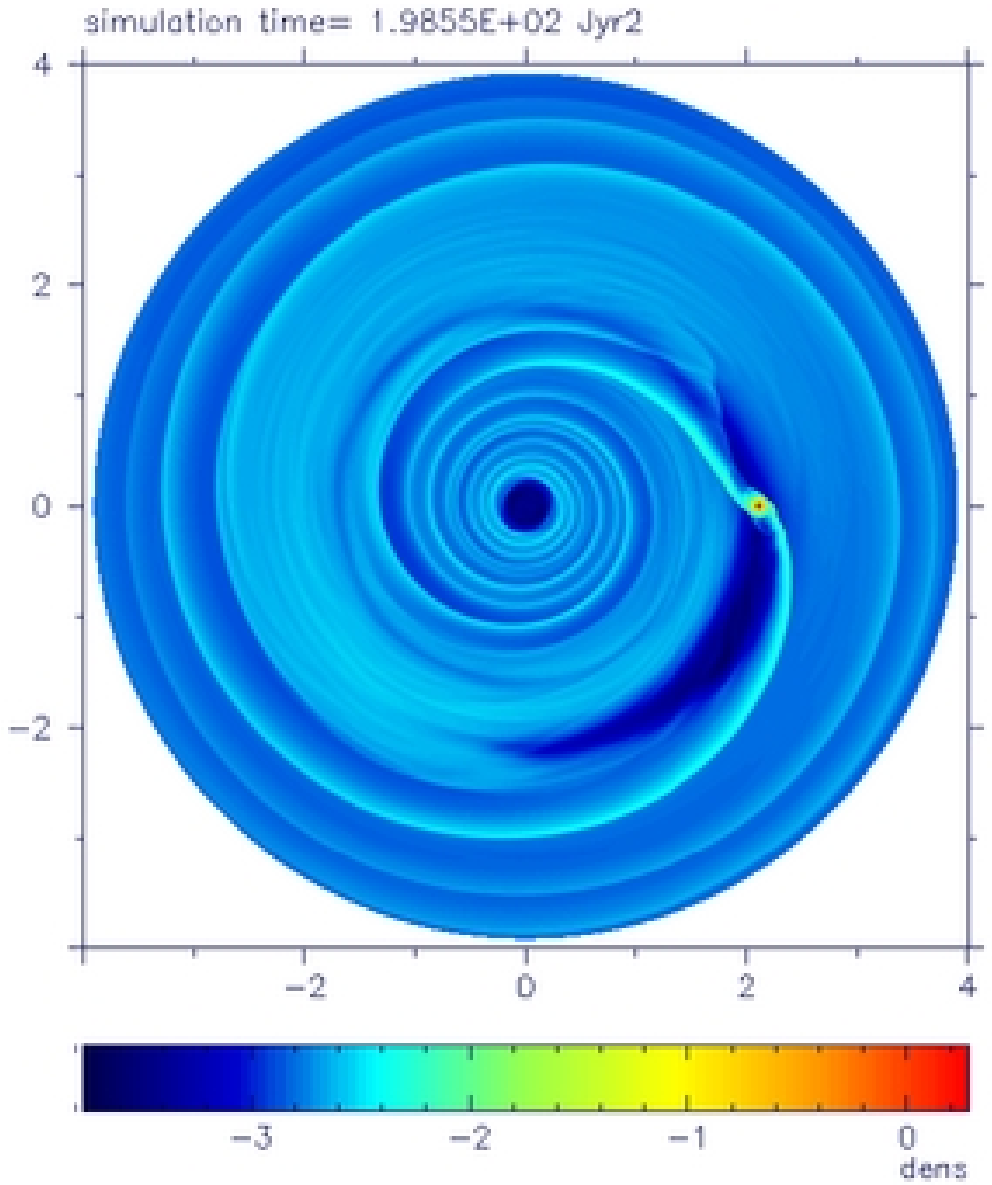}
\includegraphics[width=84mm]{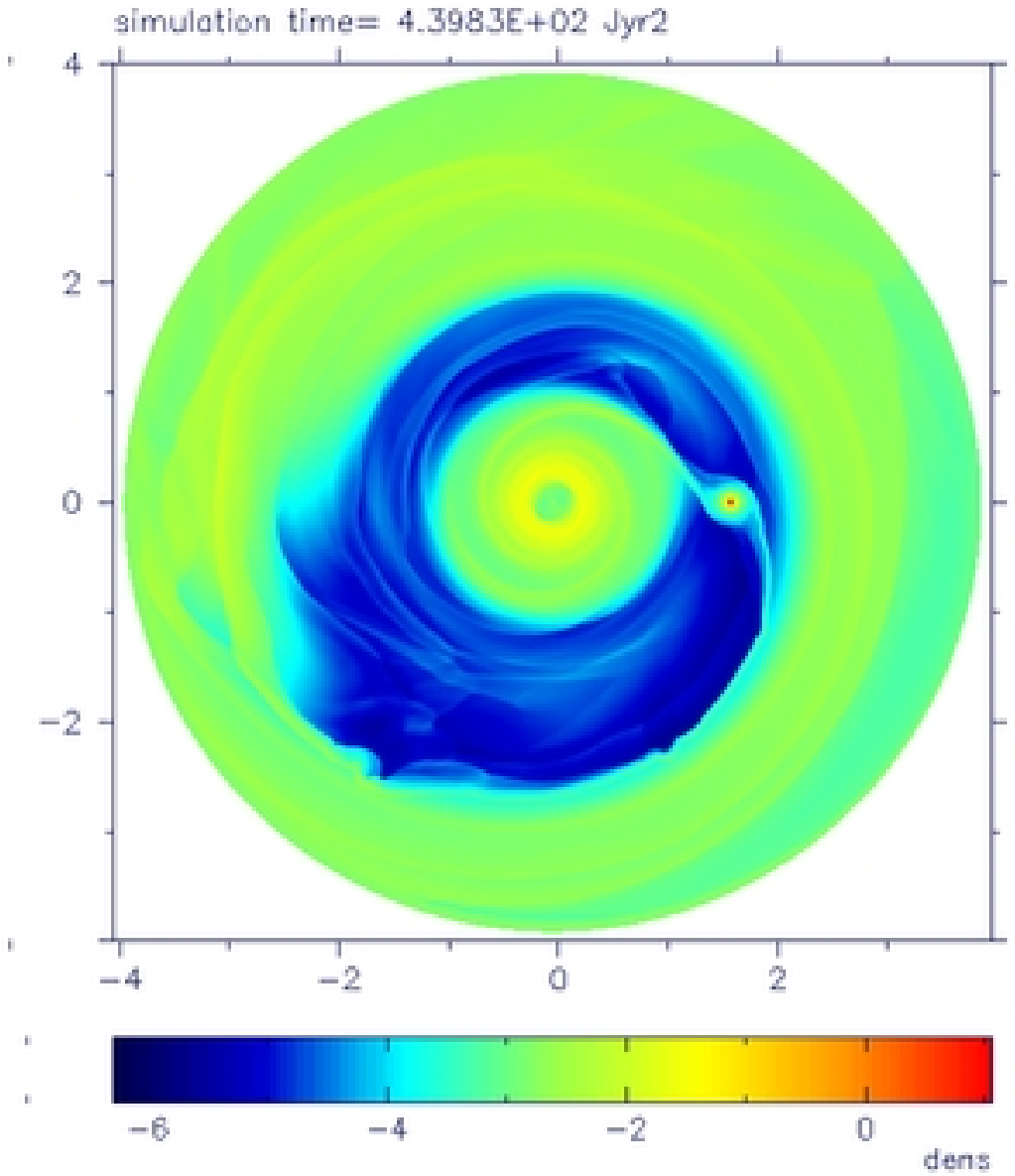}
\includegraphics[width=84mm]{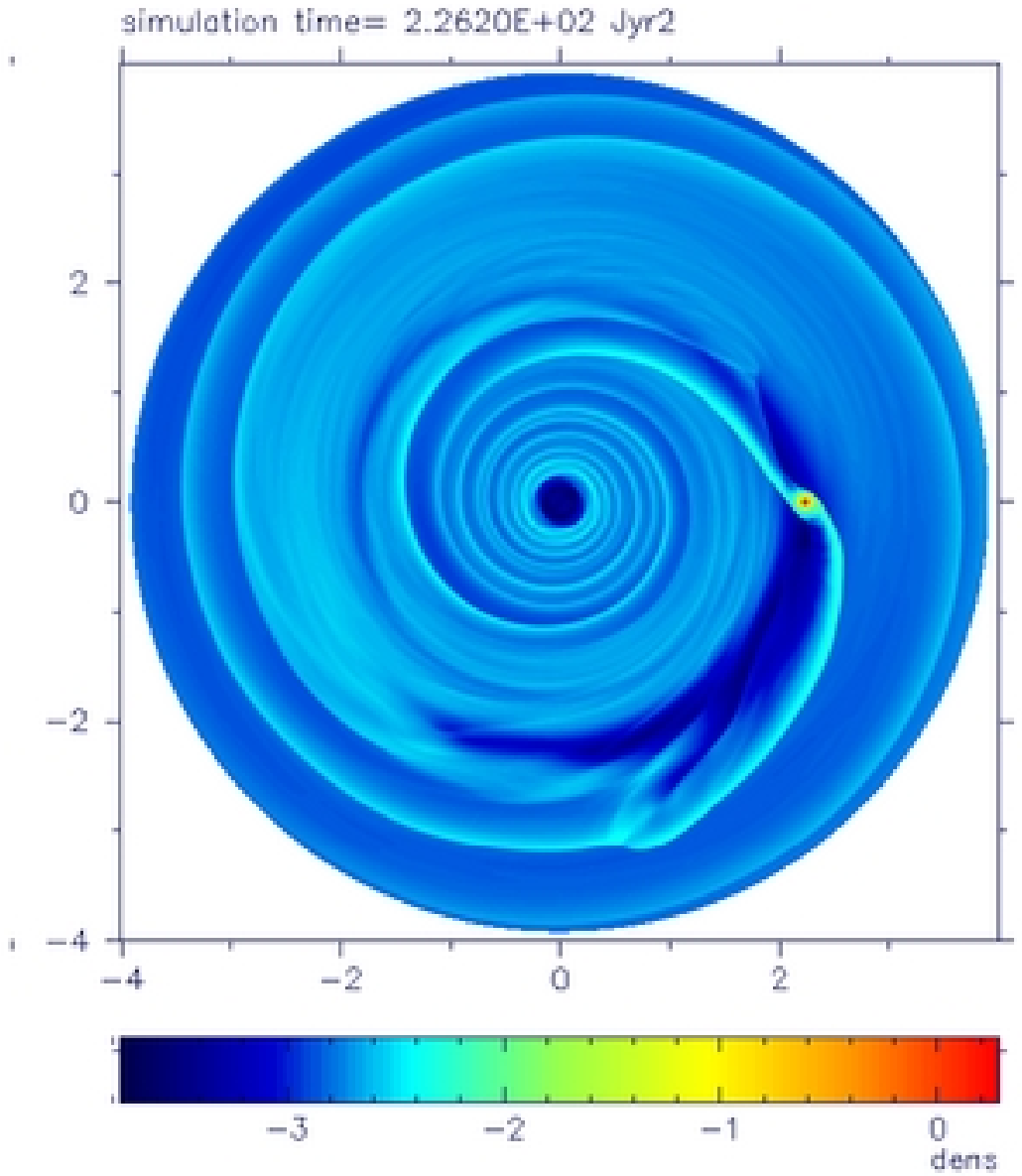}
\caption{Global surface density for the outward migration case. Left and
 right panels correspond to models M1 and M2 respectively. The upper
 left and middle left panels present the outward migration phase at
 $t=20$ ($Z\approx 0.6$)and $30$ ($Z\approx 0.1$) orbits and the
 planet's eccentricity $e < 0.01$. The lower left panel shows the inward,
 slow migration phase $|Z| \ll 1$ at $70$ orbits and 
 $e \approx 0.04$. The top right panel presents the stage of migration for the
 maximum value of the non-dimensional migration rate ($t=25.3$ orbits
 $Z\approx 4.5$). The middle right and lower right panels show the
 planet in the stage of slowing down the migration at $t=31.6$ ($Z
 \approx 2$) and $36$ ($Z \approx 0$) orbits respectively. All plots are
 done in co-moving reference frame with the planet. The colour scale is
 logarithmic and differs between M1 and M2.} 
\label{fim1_dens_disc}
\end{figure*}

\begin{figure*}
\includegraphics[width=84mm]{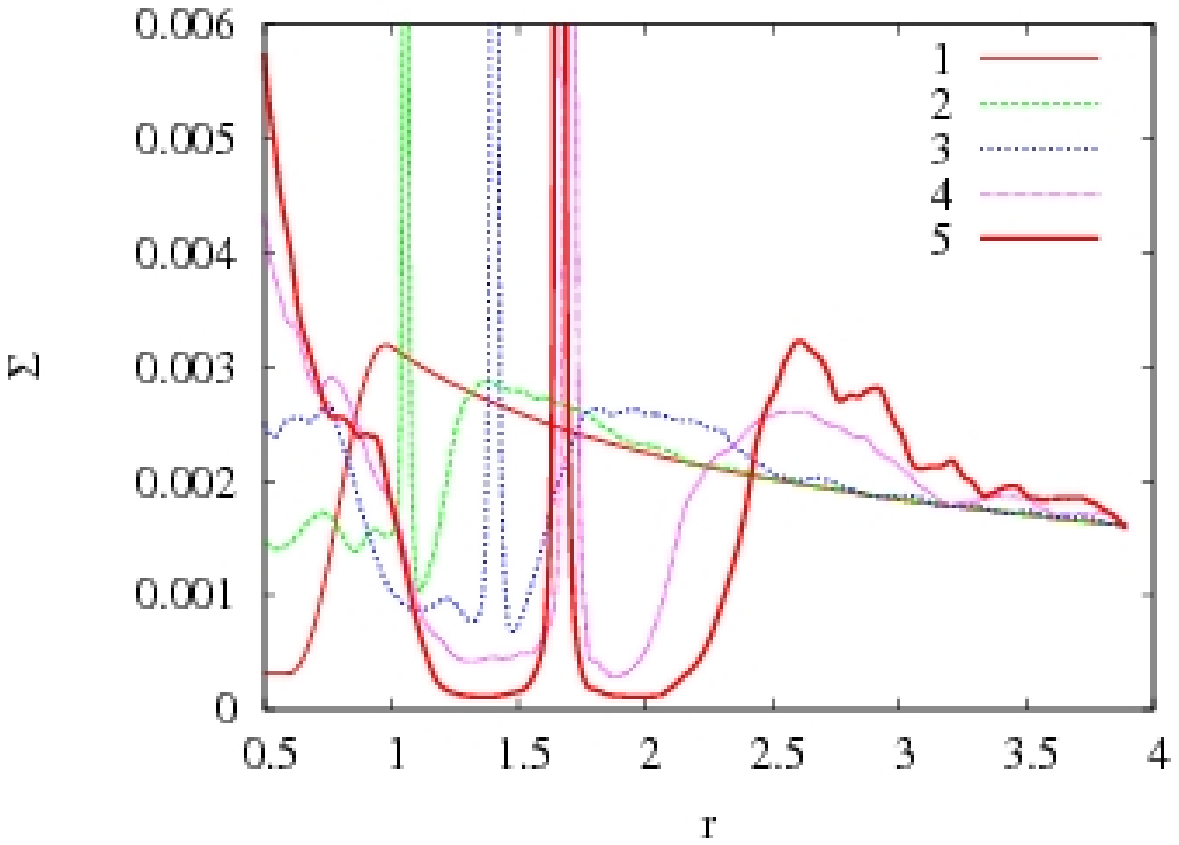}
\includegraphics[width=84mm]{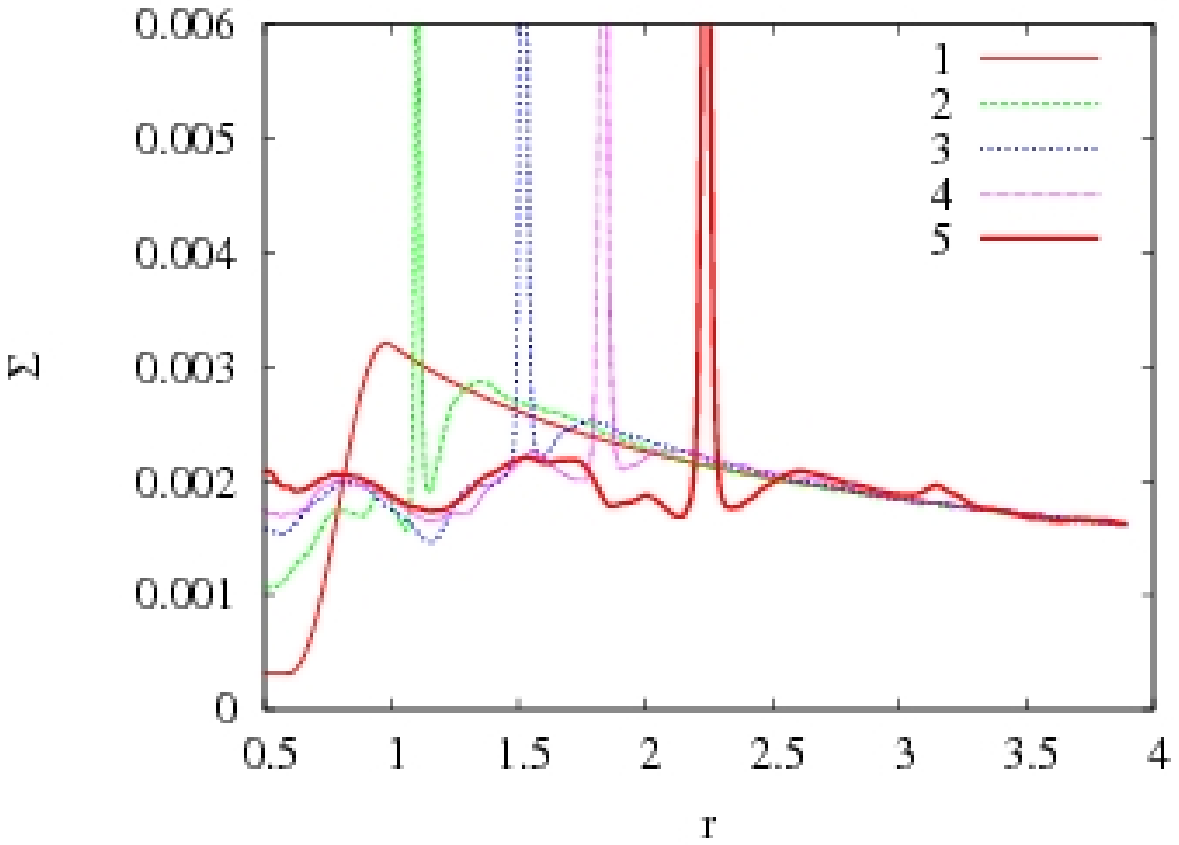}
\includegraphics[width=84mm]{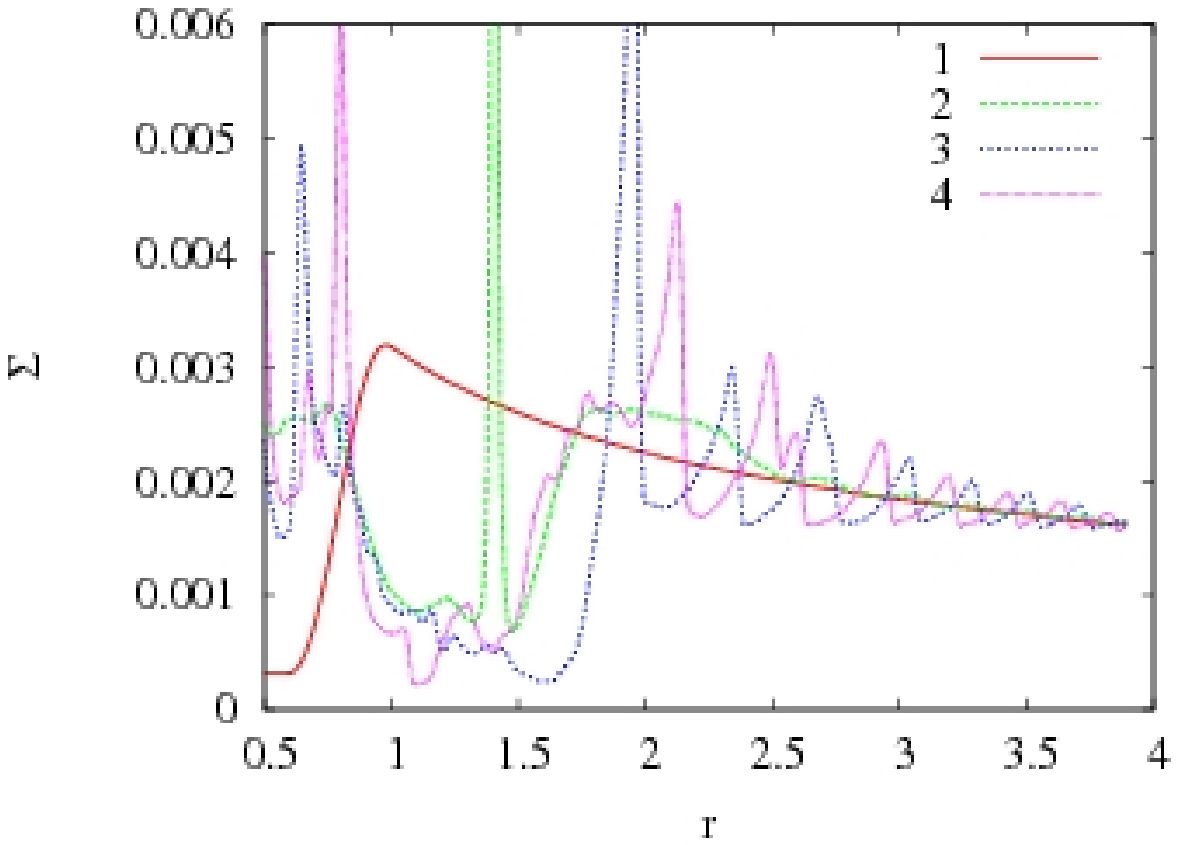}
\includegraphics[width=84mm]{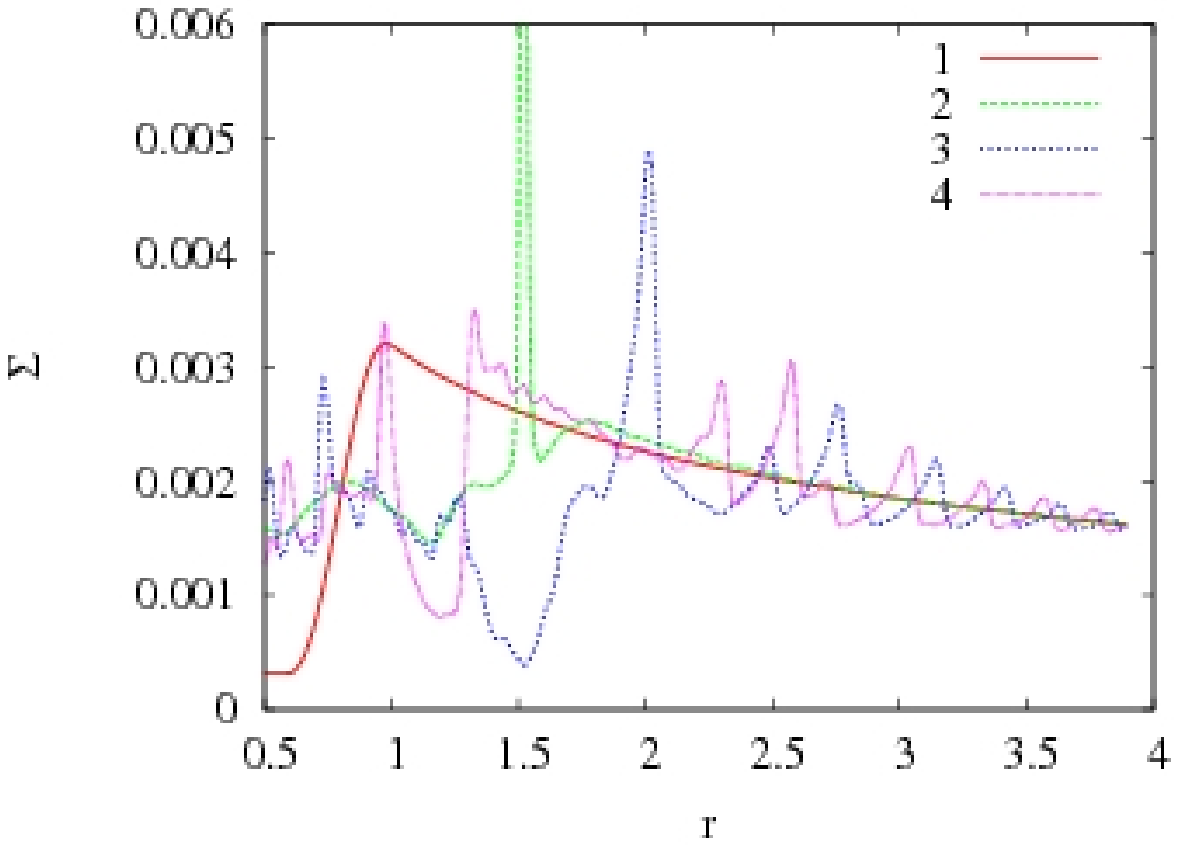}
\caption{The surface density profiles for models M1 (left column) and
  M2 (right column). The upper row shows the azimuthal average of the
  surface density $\Sigma$. Upper left panel presents it at time $t$
  equal $0$, $10$, $20$, $30$ and $40$ orbits (curves 1, 2, 3, 4 and
  5; $Z$ equal $0$, $0.7$, $0.6$, $0.1$ and $-0.02$ respectively
  ). Upper right gives the density profile at $0$, $10$, $20$, $25.2$
  and $36$ orbits (curves 1, 2, 3, 4 and 5; $Z$ equal $0$, $1.7$,
  $2.7$, $4.5$ and $0$ respectively). The planet position is visible
  as a strong spike. The lower row presents the initial density
  profile (curve 1), the azimuthal average of the surface density
  (curve 2) and the surface density cuts through the libration points
  L5 (curve 3) an L4 (curve 4). Lower left and lower right panels
  correspond to time equal $20$ ($Z \approx 0.6$) and $25.2$ ($Z
  \approx 4.5$) orbits respectively.}
\label{fim11_sigma_time}
\end{figure*}

\subsection{Surface density}
\label{sect_surf_dens}

The global surface density distribution is presented in
Fig.~\ref{fim1_dens_disc}.  The left column shows the surface density
profiles for model M1 at $t=20$ (upper), $30$ (middle) and $70$
(lower) orbits. These correspond to upper left, upper right and lower
right panels in Fig.~\ref{fim8_comp_disc}. The first plot shows the
planet migrating outward with $Z \approx 0.6$. The shape of the low
density region in the co-orbital region agrees mostly with the shape
of the horseshoe region, but fills the whole azimuthal range and the
first signs of gap creation are visible. This is more prominent at
$t=30$~orbits ($Z \approx 0.1$), when the horseshoe region fills the
whole corotation region and captures gas from the co-orbital flow. There
is a small asymmetry between the Lagrangian points L4 and L5, which
does not grow until the planet's eccentricity increases rapidly. After
about $50$ orbits both the planet's and the outer disc eccentricity
quickly grow and the interaction of the disc with the outer boundary
starts to be important. After $120$ orbits about $15\%$ ($14
M_{\jupiter}$) of the disc mass leaves the computational domain
through the outer boundary. The surface density for the eccentric
outer disc is presented in the last plot.

The right column presents the surface density for model M2 at times
$25.3$ (upper), $31.6$ (middle) and $36$ (lower) orbits, corresponding
to the upper left, upper right and lower right panels in
Fig.~\ref{fim8_comp_disc_2}. There is a visible asymmetry in the
co-orbital region in all the plots, but only in the first and second
plot is this related to the co-orbital flow. These two plots show the
planet migrating in the fast migration regime with $Z \approx 4.5$
(first plot) and $Z \approx 2$ (second plot). The first one
corresponds to the maximum value of $Z$ (and $\dot a$) and shows the
smallest extent of the horseshoe region during the entire
simulation. The low density region represents the gas that was
initially captured by the planet and covers the whole horseshoe
region. The strong co-orbital flow is visible. The second plot
represents the phase when the planet has reduced its migration
rate. Here the horseshoe region has grown and captured gas from the
co-orbital flow. The low density region does not cover the whole
horseshoe region and only represents the position of the gas that was
initially captured by the planet. It has sharp edges as no mixing has
happened. The last plot shows the planet reversing its direction of
migration with $Z \approx 0$. There is a visible asymmetry in the
corotation, even though the co-orbital flow is very weak here.  This
is caused by the fact that the transition between the outward and
inward migration takes place rapidly, and the gas does not have time
to adjust to the new configuration. For the same reason the planet
does not open a gap, even though it temporarily migrates in the slow
migration regime. At this stage the low density region starts to
disappear, and it is completely gone during the inward migration phase.

These processes are further illustrated in the upper row in
Fig.~\ref{fim11_sigma_time}. The upper left panel presents the
azimuthal average of the surface density $\Sigma$ over $2 \pi$ for
model~M1 at times $t=0$, $10$, $20$, $30$ and $40$ orbits (curves 1,
2, 3, 4 and 5). The first curve shows the initial conditions, and
curves 2, 3 and 4 present the outward migration phase. Curve 5
corresponds to the inward migrating planet inside a gap. The planet in
fact starts to clear a gap shortly after the migration starts, however
the co-orbital flow is strong enough to fill the inner region of the
disc, increasing the density there considerably. The upper right panel
presents the azimuthal average of the surface density for model
M2. Curve 1 shows the initial conditions, and the curves 2, 3 and 4
correspond to the outward migrating planet. Curve 5 shows the average
density profile for the planet at the transition between outward and
inward directed migration $Z \approx 0$. During the whole simulation
the planet is unable to open a gap, even during the time when $Z
\approx 0$. Like in model M1 the co-orbital flow is strong enough to
fill the inner disc and the final disc density profile is relatively
constant.

The lowest row in Fig.~\ref{fim11_sigma_time} illustrates the density
asymmetry in the co-orbital region. It shows the initial density
profile (curve 1), the azimuthal average of the surface density (curve
2) and the surface density cuts through the libration points L5 (curve
3) an L4 (curve 4). The cut through L5 gives the approximate width of
the co-orbital region and the density $\Sigma_\rmn{g}$ inside the
horseshoe region. Similarly the cut through L4 gives the approximate
value of the density of the gas crossing the co-orbital region
$\Sigma_\rmn{s}$. As explained in Paper~II, the difference between
$\Sigma_\rmn{s}$ and $\Sigma_\rmn{g}$ is the main reason for type III
migration.

\begin{figure*}
\includegraphics[width=84mm]{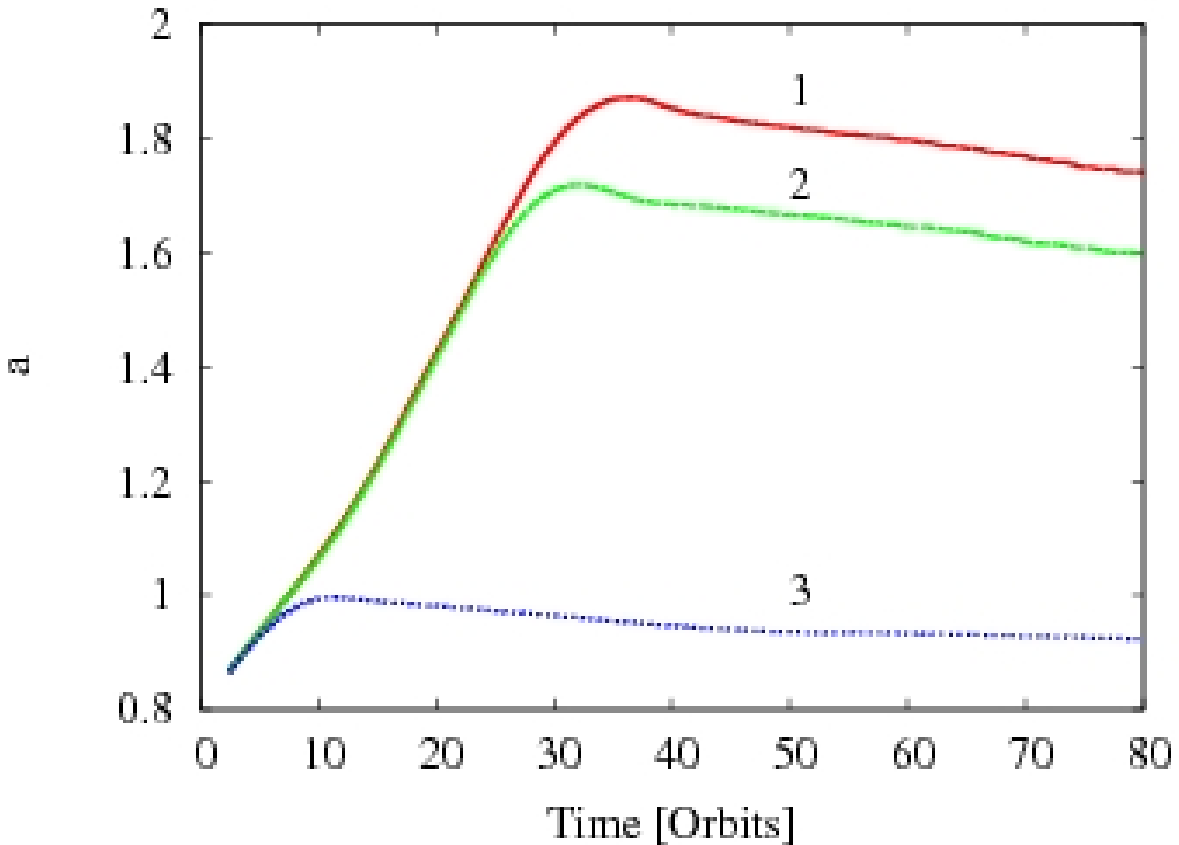}
\includegraphics[width=84mm]{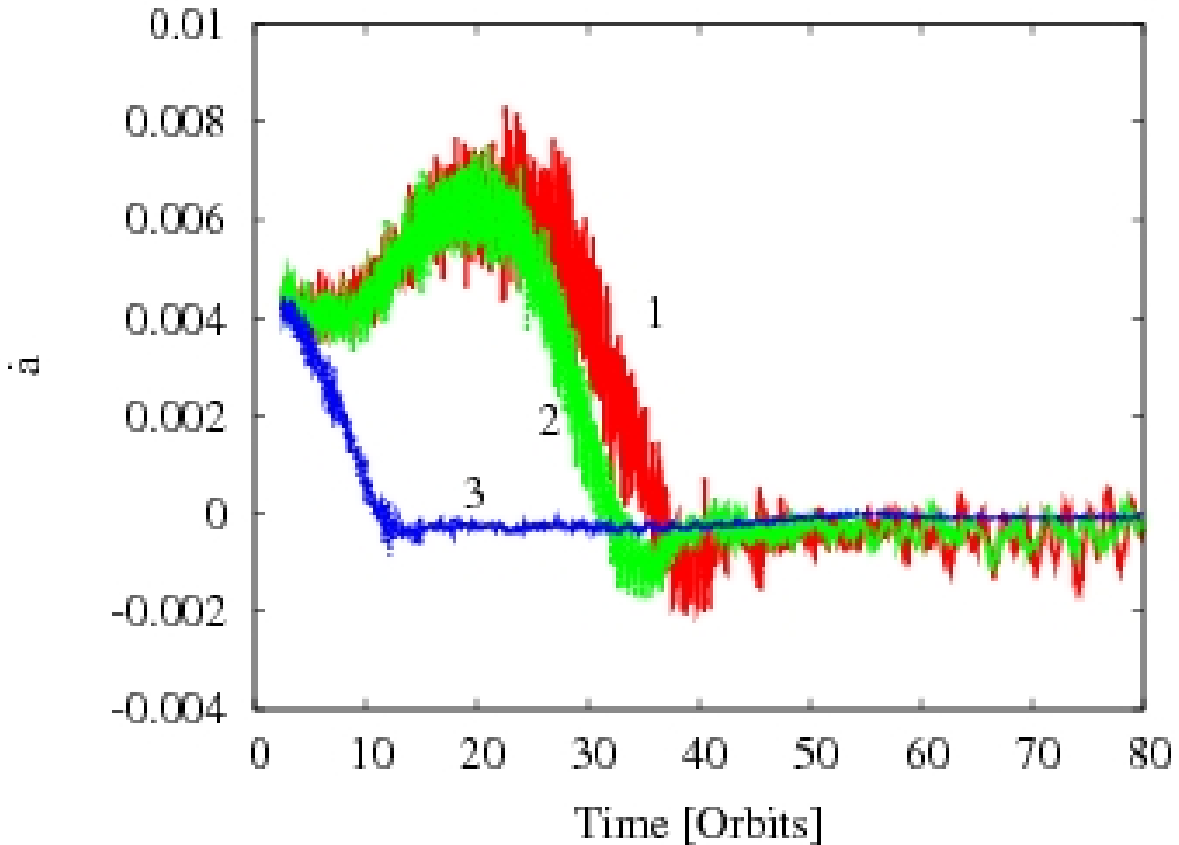}
\includegraphics[width=84mm]{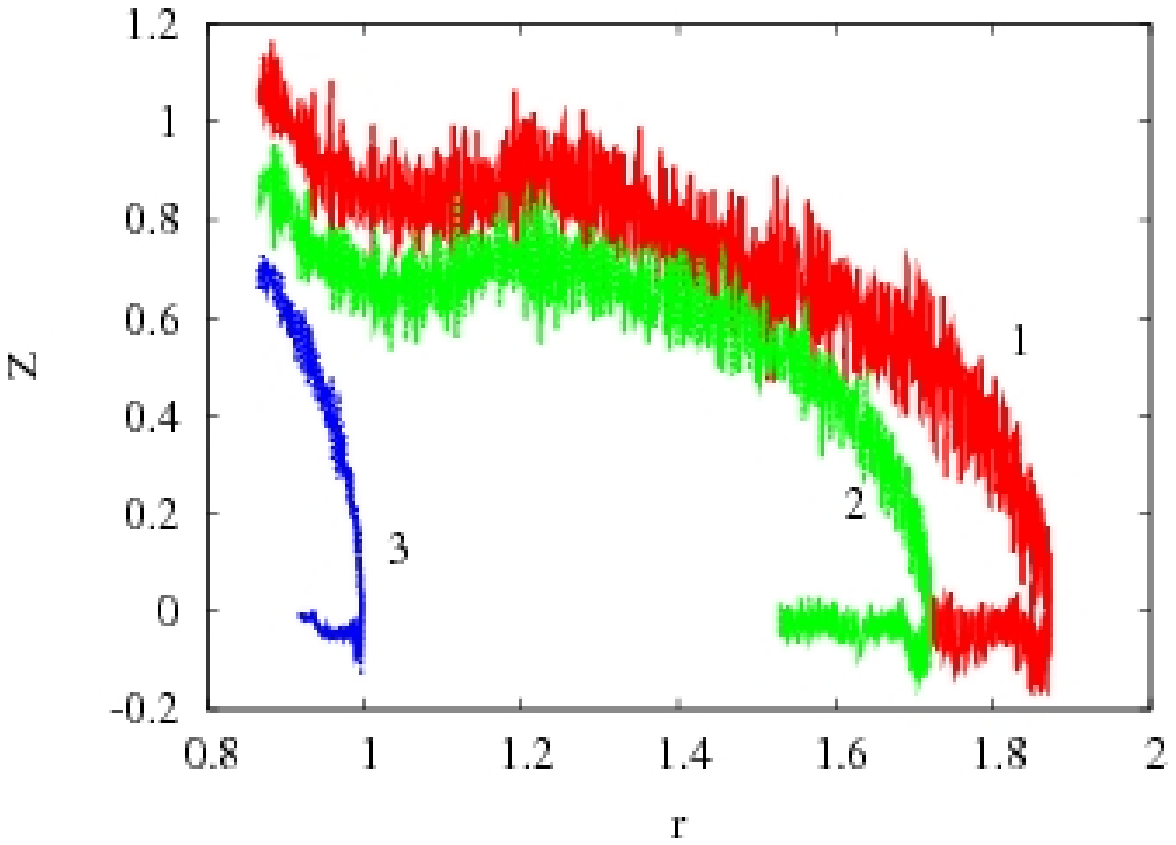}
\includegraphics[width=84mm]{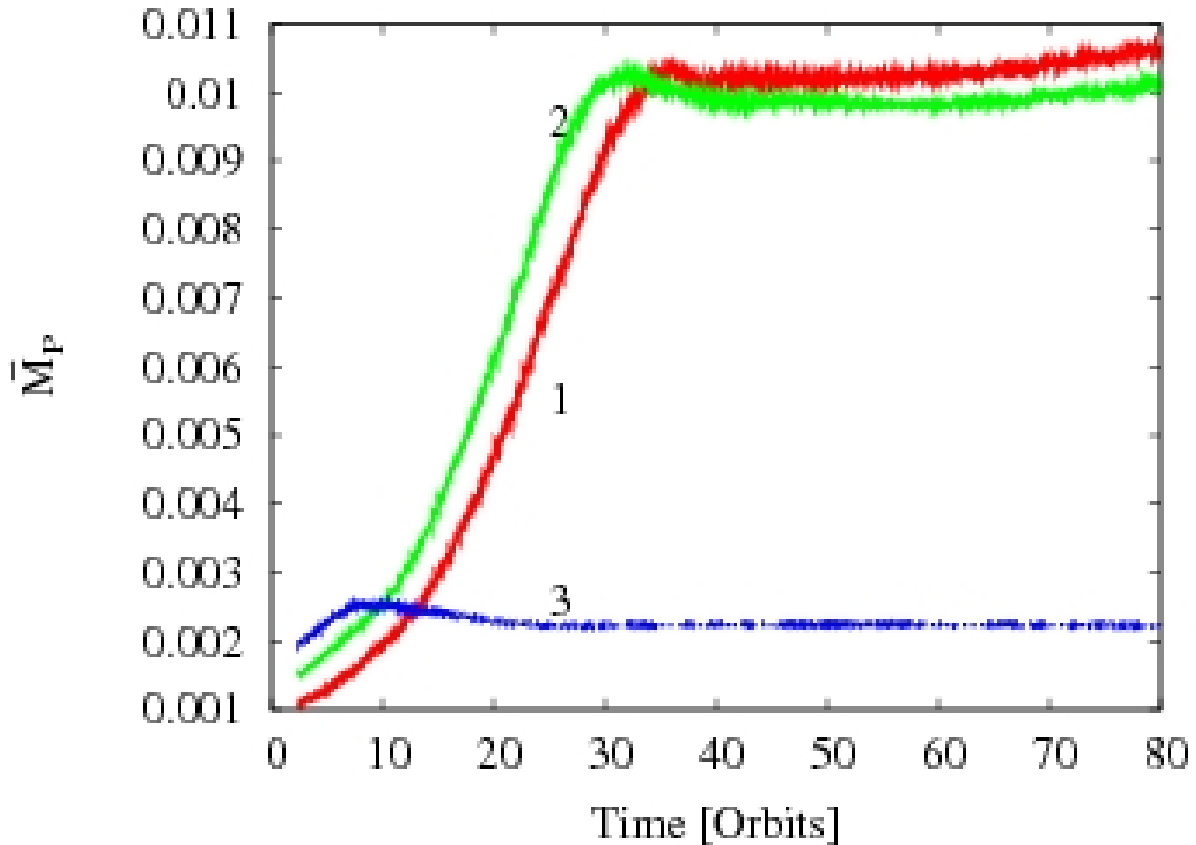}
\caption{Results of the simulations for the different planet's
 masses. Curves 1, 2 and 3 correspond to the initial planet mass
 $M_\rmn{P}$ equal $0.0007$, $0.001$ (standard case~M1) and
 $0.0013$ respectively. Upper left and upper right panels show the
 planet's semi-major axis $a$ and the migration rate $\dot a$ during the
 orbital evolution. The non-dimensional migration rate $Z$ as a function
 of the planet's position in the disc and the gravitational planet mass
 $\widetilde M_\rmn{P}$ are presented on the lower left and lower right
 panels respectively.}
\label{fmin_d_planet_mass}
\end{figure*}

The lower left panel presents M1 at $20$ orbits during the outward
migration phase ($Z \approx 0.6$). Since the planet is in the slow
migration regime and gap opening has already started, the asymmetry
between L4 and L5 is limited to the outer edge of the co-orbital
region. In contrast, model M2 (lower right panel) has the planet
migrating in the fast migration regime and the asymmetry covers the
whole co-orbital region. In this case the planet carries along the low
density horseshoe region with well defined sharp edges and an almost
static value of the surface density $\Sigma_\rmn{g} \approx
0.0005$. The value for $\Sigma_\rmn{s}$ is close to the initial
density profile at the current planet's position.

\section{Dependence on the simulation parameters}
\label{sect_dep_on_par}

After having described the two standard cases of outward migration, we
now discuss the dependence of the migration process on various
simulation parameters.

\subsection{Planet's mass}
\label{sect_diff_pl_mass}

The first important parameter is the planet mass. In our investigation
we concentrate on the orbital evolution of giant planets. In Papers~I
and~II we showed that in the fast migration regime, planet mass does
not influence the migration significantly, whereas in the slow
migration regime the migration is strongly dependent on
$M_\rmn{P}^*$. The relation between $M_\rmn{P}$ and the planet's
migration for outward migration is presented in
Fig.~\ref{fmin_d_planet_mass}. We ran models with initial masses $M_\rmn{P}$
equal $0.0007$, $0.001$ (standard case M1) and $0.0013$ (curves
1, 2 and 3 respectively). The effective planet mass was allowed to increase,
$M_\rmn{P}^* = \widetilde M_\rmn{P}$.

The upper left and upper right panels show the planet's semi-major
axis $a$ and the migration rate $\dot a$. There is a clear difference
between the evolution of the two first models and the last one,
indicating a sharp border in parameter space for which the outward
migration is allowed. In the last model the amount of mass in the
corotation is too small to support the initial impulse for outward
migration and the planet transits to an inward migration case at $a
\approx 1$. Note that even though the initial migration rate $\dot a$
is similar for all the models, the non-dimensional migration rate
(lower left panel) differs since 
$Z \sim \widetilde M_\rmn{P}^{-2/3}$. We find that to start outward
migration the initial 
average $Z$ should be bigger or close to $1$, and the last model does
not satisfy this criterion having an (extrapolated) initial value of
$Z \approx 0.9$.  Mass accumulation does not play a role here, as the
increase of $\widetilde M_\rmn{P}$ is relatively modest 
($\sim 2.5 M_{\jupiter}$ after $10$ orbits) and does not influence the
migration in a major way.

\begin{figure*}
\includegraphics[width=84mm]{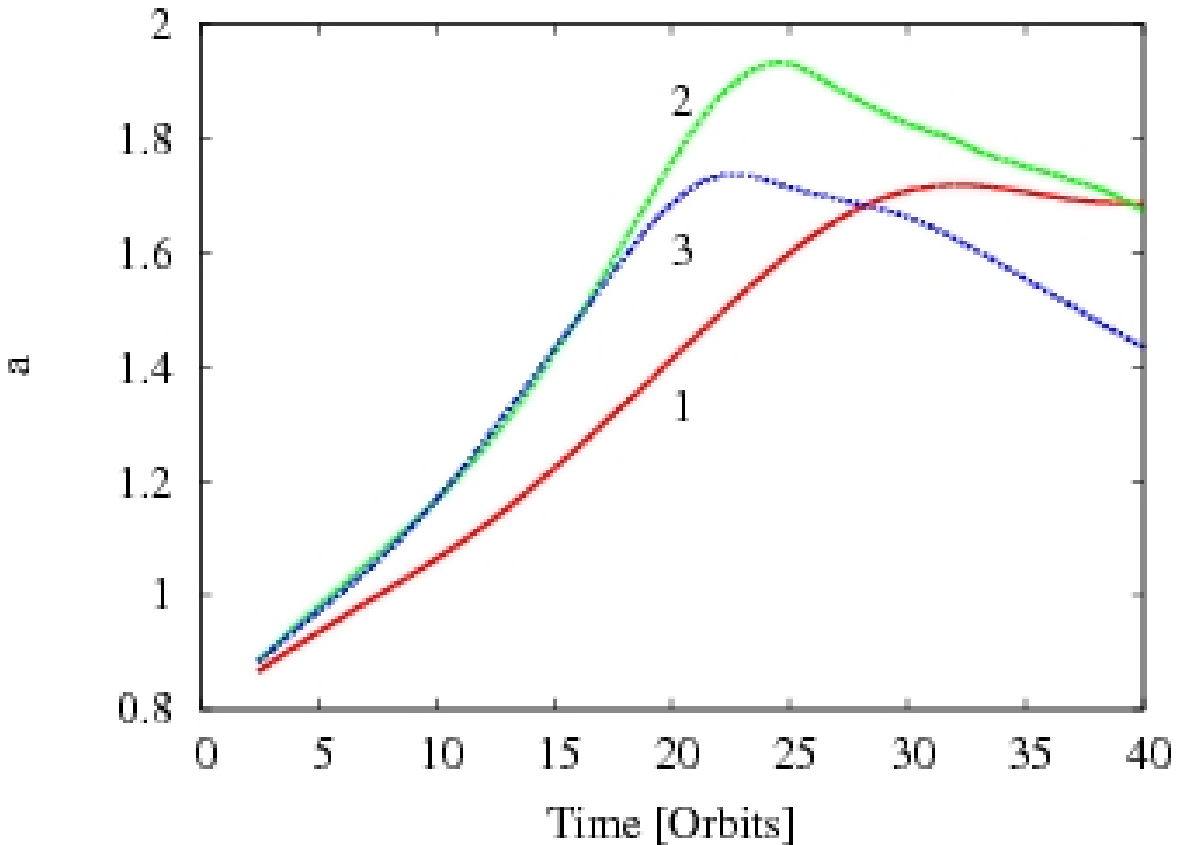}
\includegraphics[width=84mm]{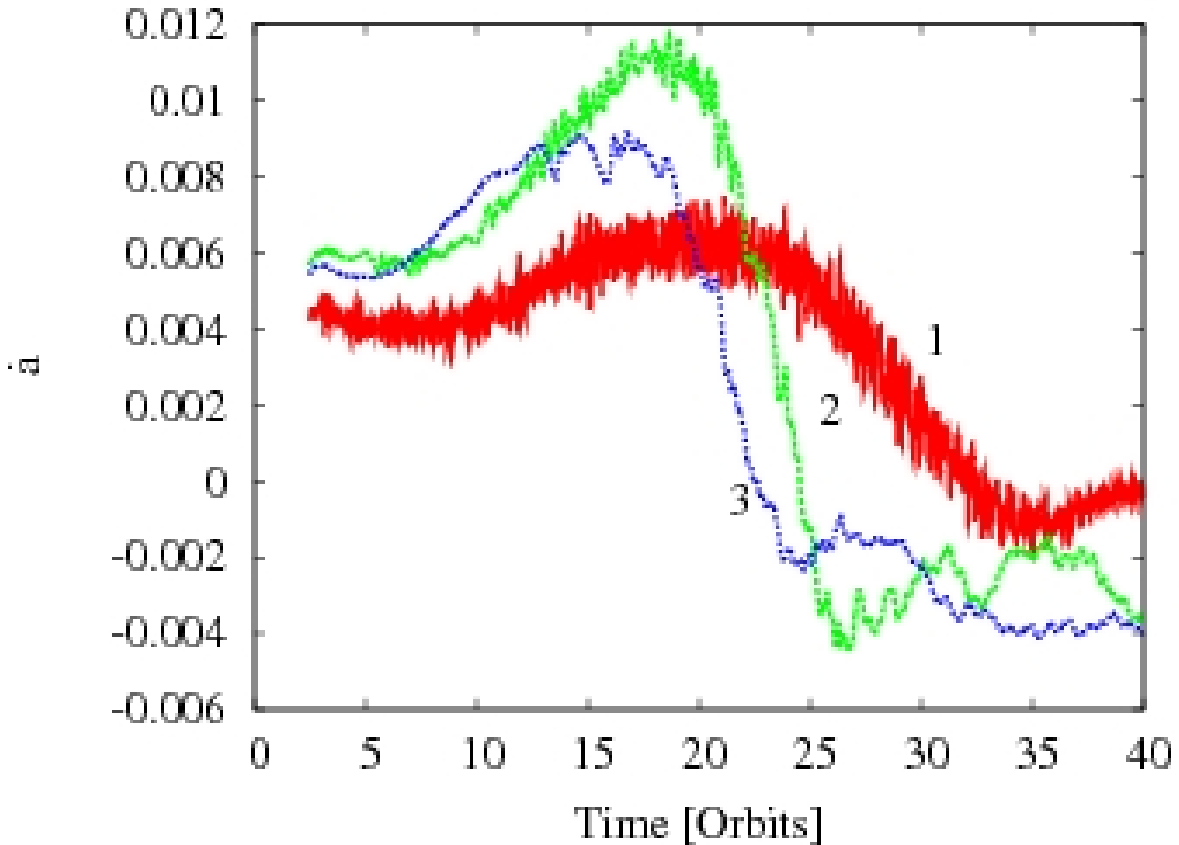}
\includegraphics[width=84mm]{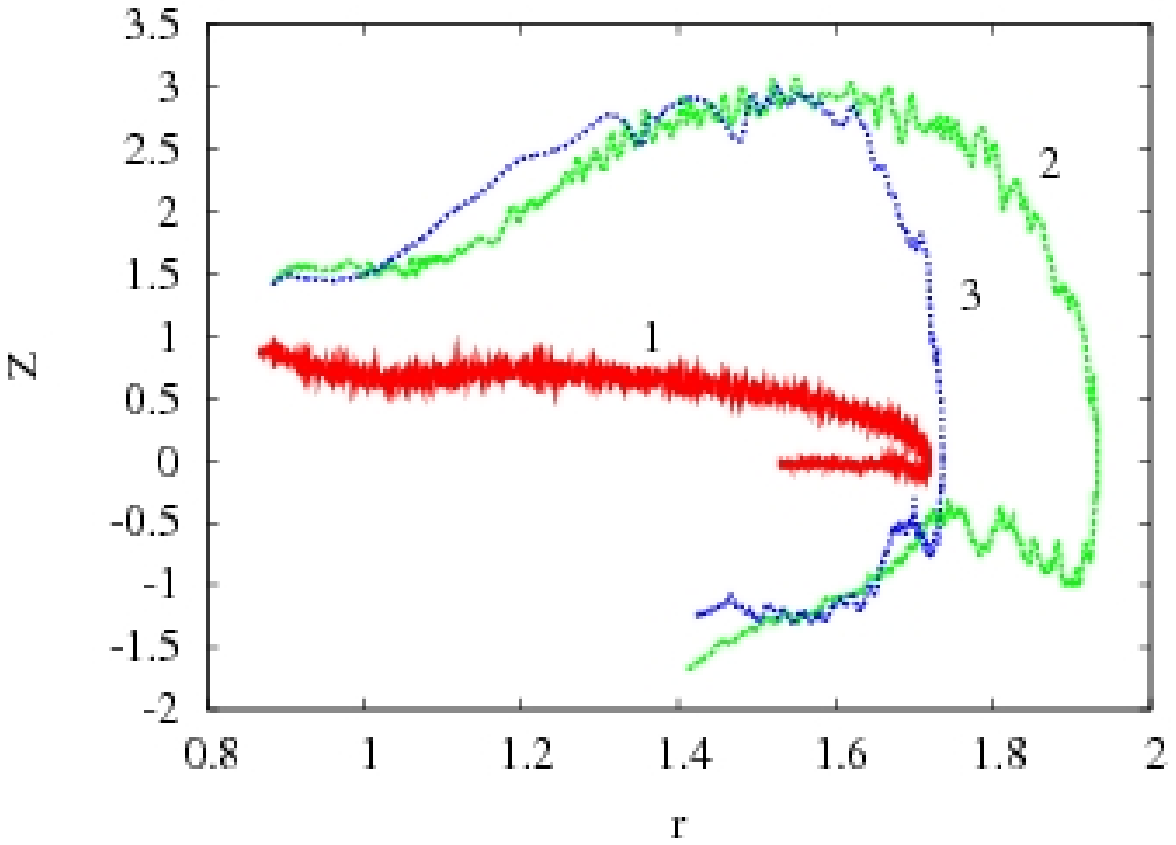}
\includegraphics[width=84mm]{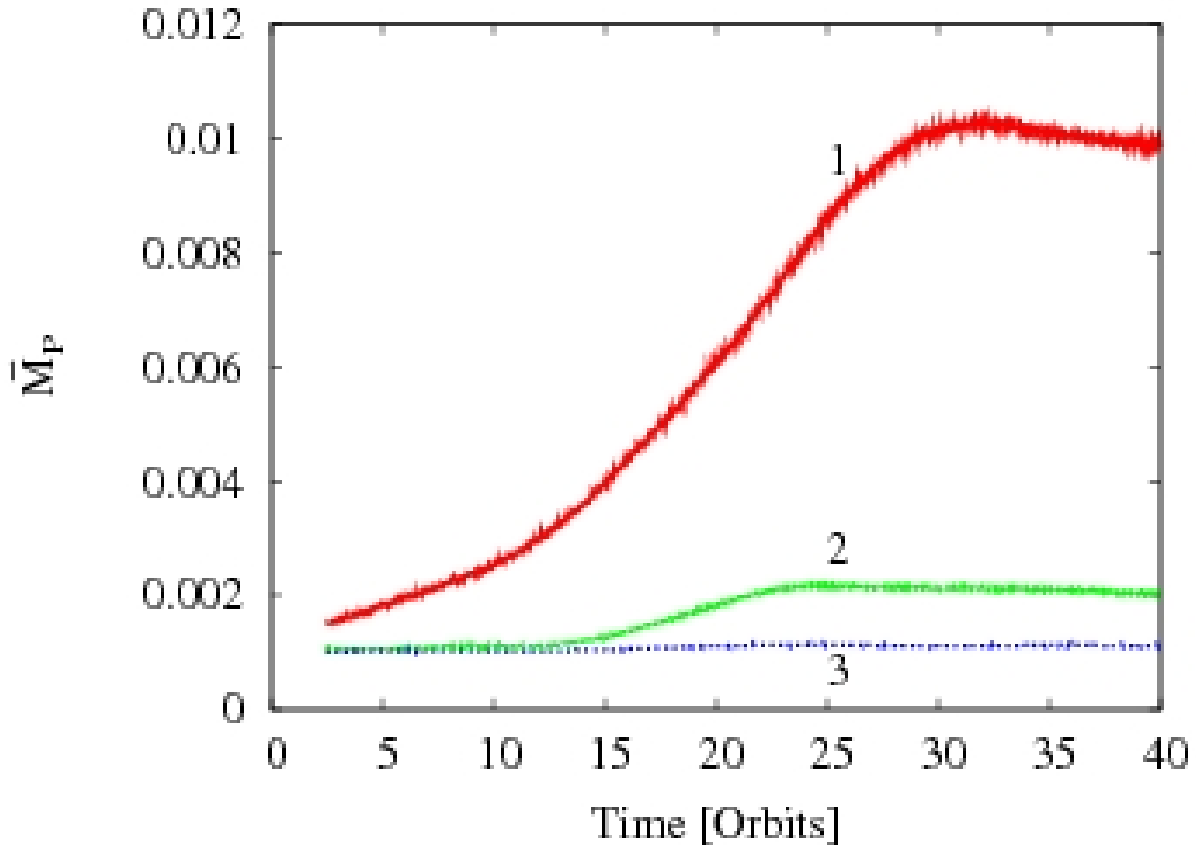}
\caption{Results of the simulations for the different circumplanetary
 disc aspect ratios $h_\rmn{p}$. Curves 1, 2 and 3 correspond to
 $h_\rmn{p}$  equal $0.4$ (standard case M1), $0.5$ and $0.6$
 respectively. The panels are the same as in
 Fig.~\ref{fmin_d_planet_mass}.}
\label{fmin_d_planet_h}
\end{figure*}

The two lower mass models satisfy the criterion stated above, with
extrapolated values of $Z \approx 1.15$ and $Z \approx 1.05$,
respectively. During the first $25$ orbits both systems evolve almost
exactly the same way, even though the migration takes place in the
slow migration regime. This can be understood from the fact that the
non-dimensional migration rate is relatively close to 1 ($Z > 0.65$)
and the migration is mostly determined by the increase of the
gravitational planet's mass, which grows in a similar way in both
models (reaching about $10 M_{\jupiter}$). Unlike the inward migration
case (Paper~II Sect.~5.1), the amount of mass accumulated by these
planets is almost independent of its initial mass. The difference 
is relatively small (of order of $0.5 M_{\jupiter}$) and does not
play the role while $\dot a$ is increasing.

The evolution starts to differ after $Z$ drops below $0.6$ and $\dot a$
starts to decrease. The migration rate drops faster for the model
with the larger initial planet mass and the planet moves from the outward
to inward migration at about $30$ orbits and $a \approx 1.7$. For the
lighter planet this transition takes place at about $35$ orbits and 
$a \approx 1.9$. 

The relation between planet mass and orbital evolution is visible on
the lower left panel, where the non-dimensional migration rate $Z$ is
plotted as a function of the planet's position in the disc. The
increase of planet mass, until $M_\rmn{P}$ reaches the critical mass
for which outward migration is no longer possible, causes $Z$ to
diminish smoothly and moves the curve to the right. In this case the
region of the disc where the rapid migration can take place shrinks
with increasing $M_\rmn{P}$ and the planet stops closer to the star.

\subsection{Effects of circumplanetary disc aspect ratio}
\label{subs_dep_hp}

The next parameter we investigate is the circumplanetary disc
aspect ratio $h_\rmn{p}$. We performed a set of three simulations with
different values of $h_\rmn{p}$ and $M_\rmn{P}^* = \widetilde
M_\rmn{P}$. The results are presented in Fig.~\ref{fmin_d_planet_h}. We use
the same set of plots as in Fig.~\ref{fmin_d_planet_mass}: the evolution
of the planet's semi-major axis $a$ and the migration rate $\dot a$ in
the upper panels, and the non-dimensional migration rate $Z$ as a
function of the planet's position in the disc together with the
evolution of the effective planet mass in the lower panels. Curves 1, 2
and 3 correspond to $h_\rmn{p}=0.4$ (standard case M1),
$0.5$ and $0.6$ respectively. 

The planet's mass $\widetilde M_\rmn{P}$ is strongly dependent on the
circumplanetary disc aspect ratio and decreases with increasing
$h_\rmn{p}$. In the first model with
$h_\rmn{p}=0.4$, the gravitational planet's mass grows up to $10
M_{\jupiter}$ causing the planet to migrate in the slow migration
regime. In the second model with $h_\rmn{p}=0.5$, $\widetilde M_\rmn{P}$
grows during the whole outward migration stage reaching 
$2 M_{\jupiter}$, however it stays very low during the first $10$ orbits
allowing the planet to migrate in the fast migration regime. A similar
behaviour is seen for $h_\rmn{p}=0.6$, but in this case
$\widetilde M_\rmn{P}$ is almost constant.

There is an important difference between the first ($h_\rmn{p}=0.4$)
and the second ($h_\rmn{p}=0.5$) model. In the first model the outward
migration stops due due to the rapid increase of the planet's inertia,
but in the second one $\widetilde M_\rmn{P}$ does not grow enough to
reduce $Z$ considerably and $\dot a$ increases significantly. This
gives a rapid increase of the non-dimensional migration rate up to 
$Z \approx 3$ at which point changes in the flow structure in the
co-orbital region similar to those found for standard case M2, slow
down the migration. The increase of the circumplanetary disc aspect
ratio allows the planet to migrate faster and to larger radii. The planet
reaches $a \approx 1.72$ at $t\approx 21$ orbits and $a \approx 1.9$ at
$t\approx 24$ orbits in the first and the second model respectively.

\begin{figure*}
\includegraphics[width=84mm]{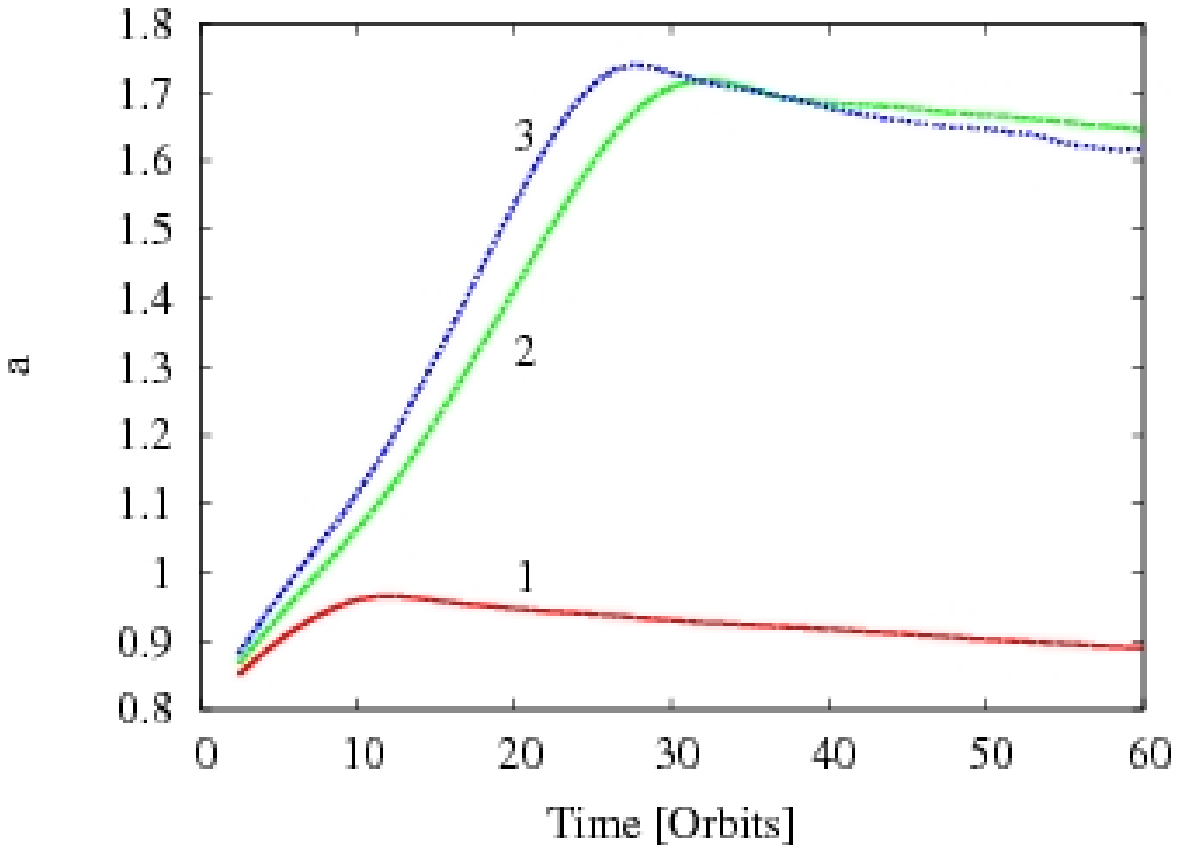}
\includegraphics[width=84mm]{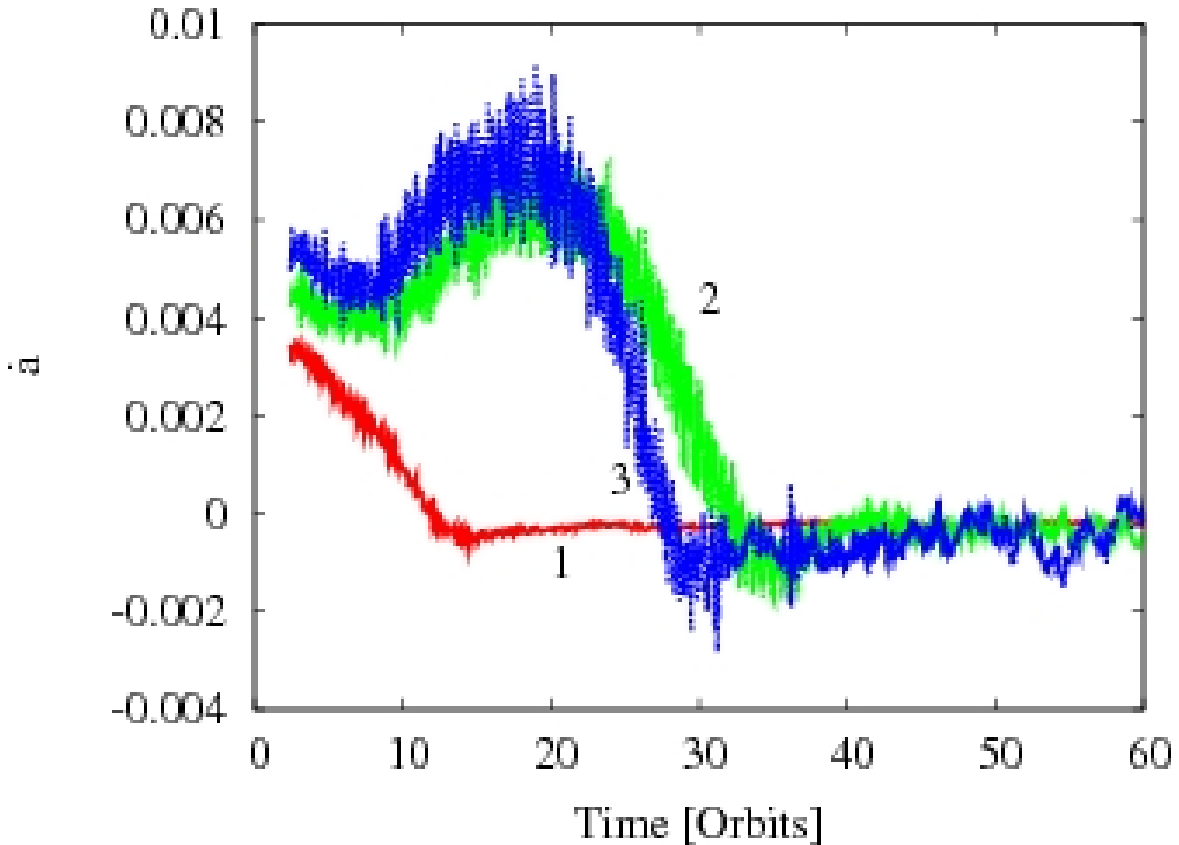}
\includegraphics[width=84mm]{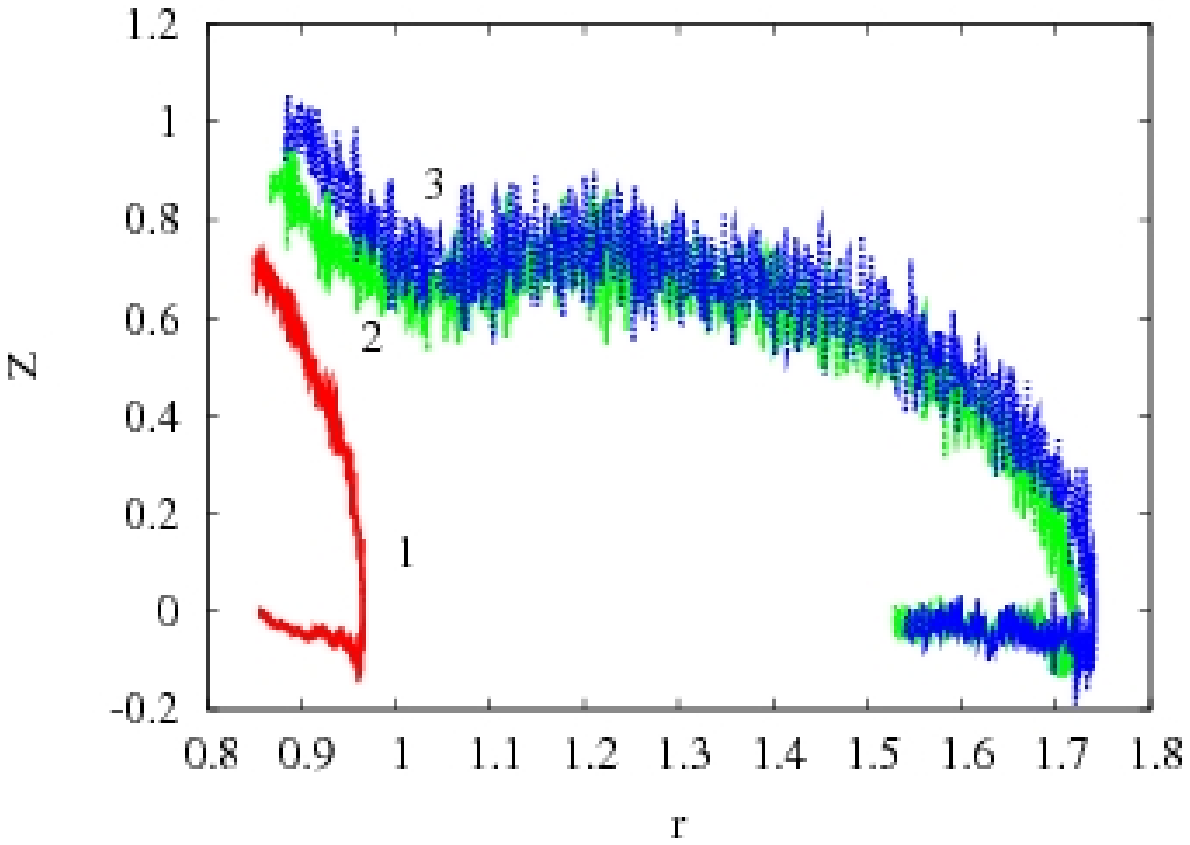}
\includegraphics[width=84mm]{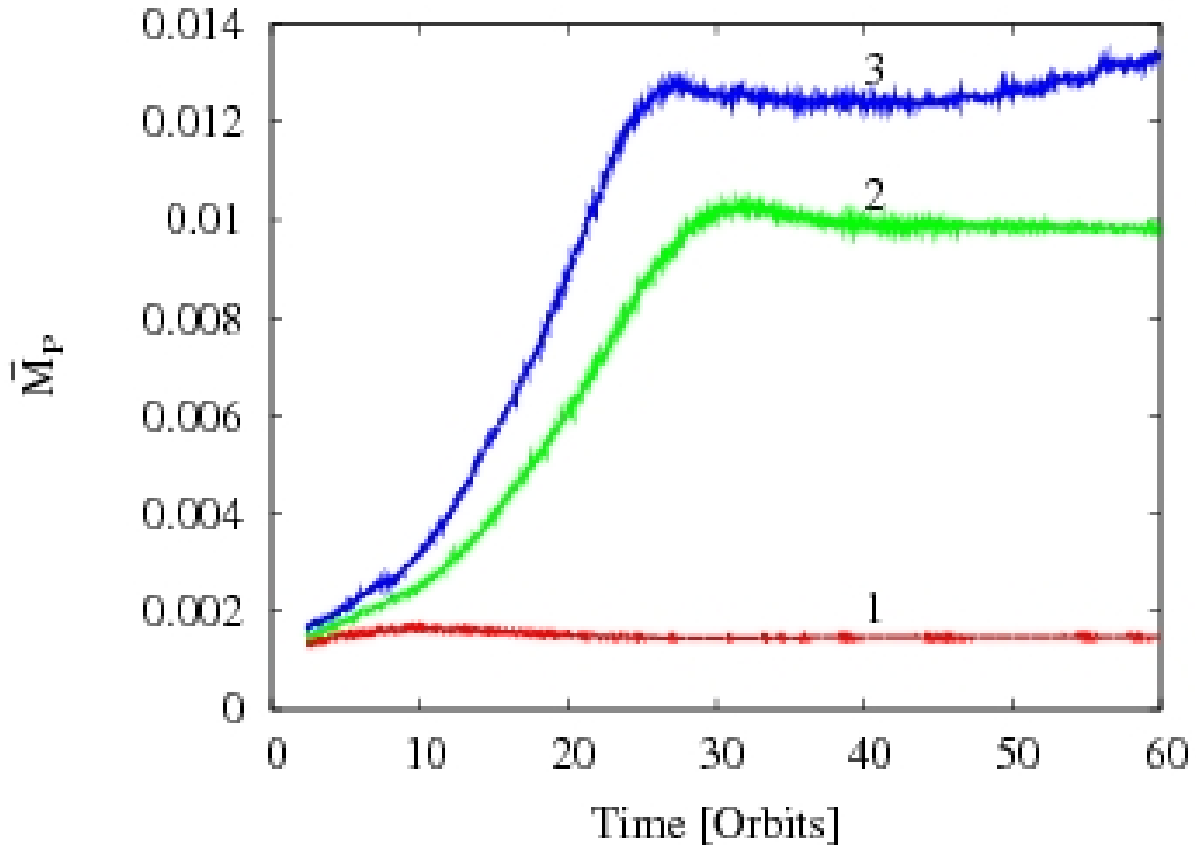}
\caption{Results of the simulations for the different total disc masses
 for $M_\rmn{P}^* = \widetilde M_\rmn{P}$. Curves 1, 2 and 3 correspond
 to the disc to the primary mass ratio $\mu_\rmn{D}$ equal $0.0075$,
 $0.01$ (standard case M1) and $0.0125$ respectively. The panels
 are the same as in Fig.~\ref{fmin_d_planet_mass}.}
\label{fmin_d_disc_mass}
\end{figure*}

A further increase of $h_\rmn{p}$ does not influence the migration
significantly, and the evolution of the second and the third model is
similar. However, increasing $h_\rmn{p}$ implies lowering 
$\widetilde M_\rmn{P}$, giving a faster increase of $Z$. That is why in
the model with $h_\rmn{p}=0.6$ the flow inside the Roche lobe changes its
structure for a lower value of $\dot a$ and the planet changes to
inward migration at $t\approx 23$ orbits reaching a maximum orbit 
$a \approx 1.7$. In the last model the migration rate grows up to $0.09$,
but in the second model $\dot a$ reaches $0.11$.

Summarising, as in the inward migration case the temperature profile
influences the planet's orbital evolution mostly by limiting the amount
of mass in the Hill sphere. $h_\rmn{p} \le 0.4$ gives a rapid grow of
$\widetilde M_\rmn{P}$ and the planet stops due to the increase of its
inertia. Higher values for $h_\rmn{p}$ reduce the effective planet mass
and lead to the other stopping mechanism (M2), where the flow in the planet's
vicinity changes structure for some critical value of $Z$.
In this case the lower gravitational
planet mass (corresponding to higher $h_\rmn{p}$) allows to reach
this critical value of $Z$ faster and stop outward migration closer to
the star.

The reader has to keep in mind that we present the results of
two-dimensional simulations with $h_\rmn{p}$ constant in time,
an important simplification. In the case of strong mass inflow into
the planet's vicinity, $h_\rmn{p}$ should be a function of the gas
density in the circumplanetary disc, and the possible vertical motions of
the gas should be taken into account. 
\begin{figure*}
\includegraphics[width=84mm]{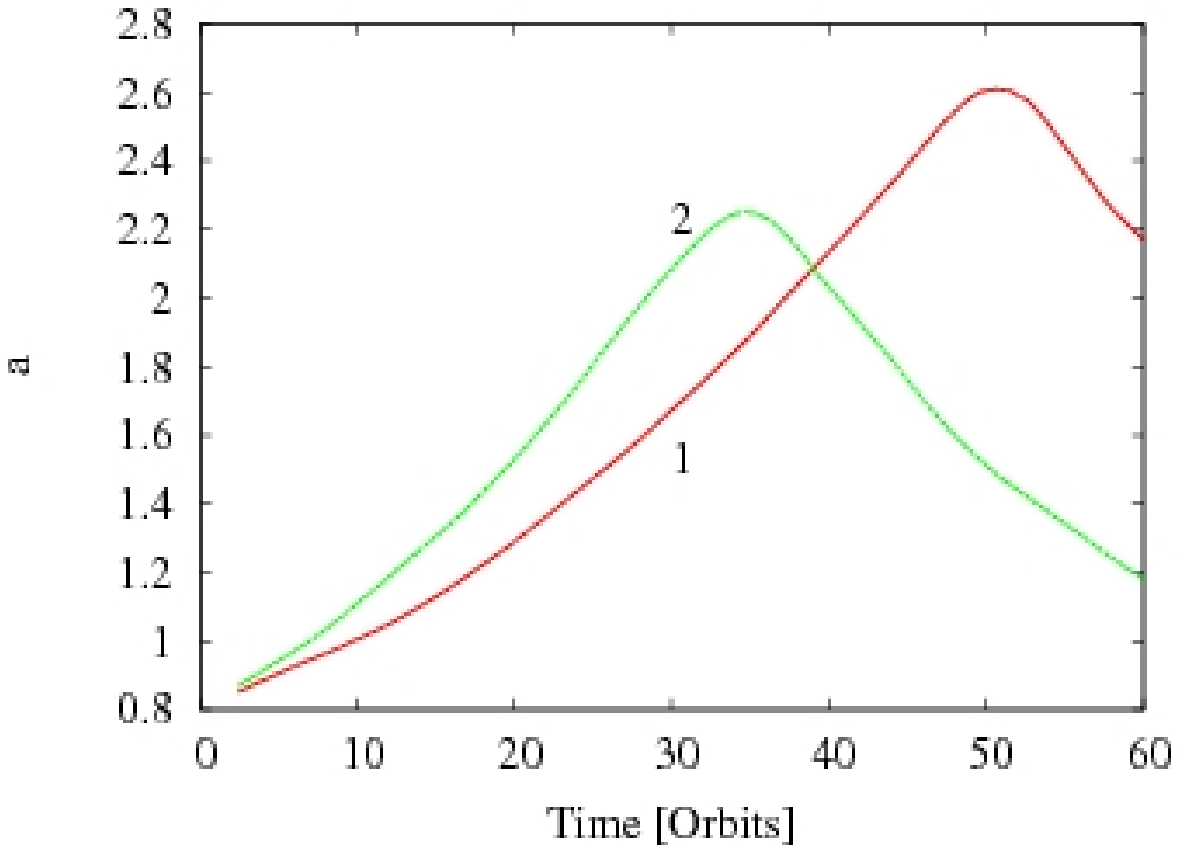}
\includegraphics[width=84mm]{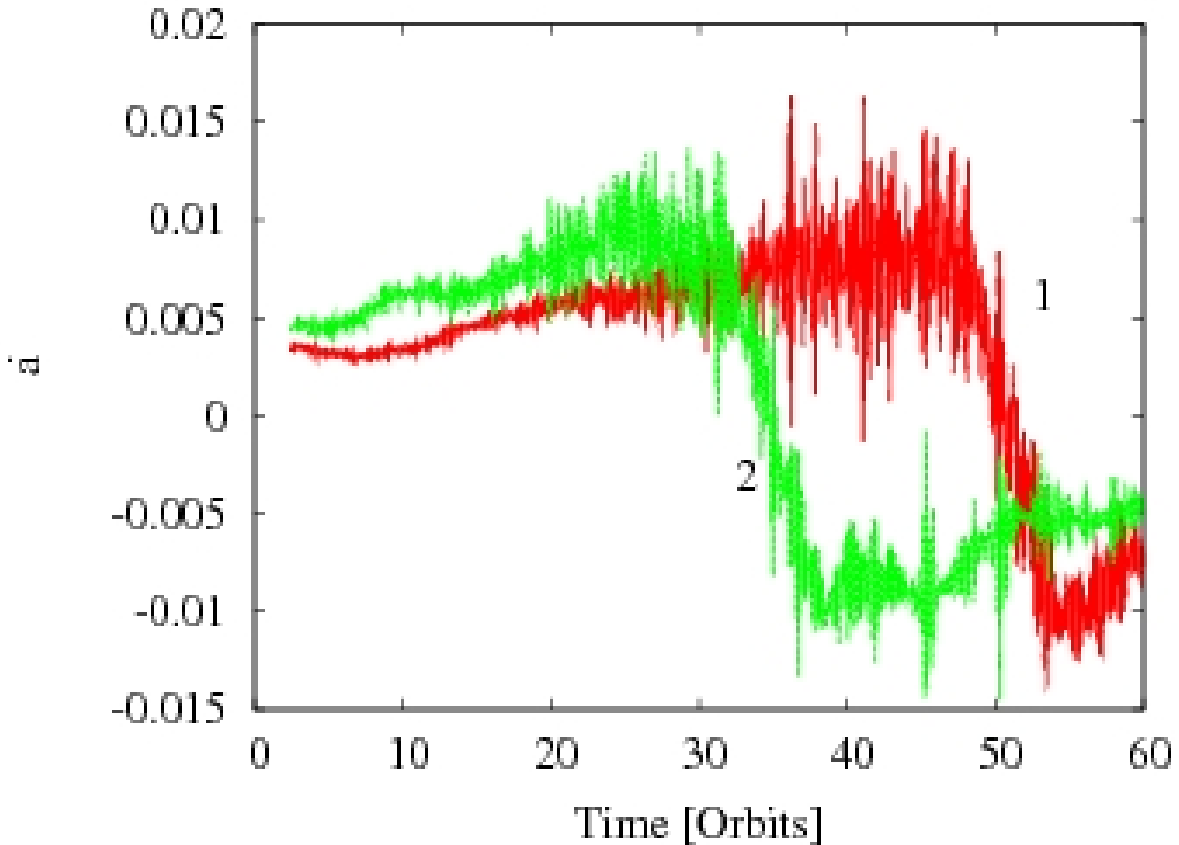}
\includegraphics[width=84mm]{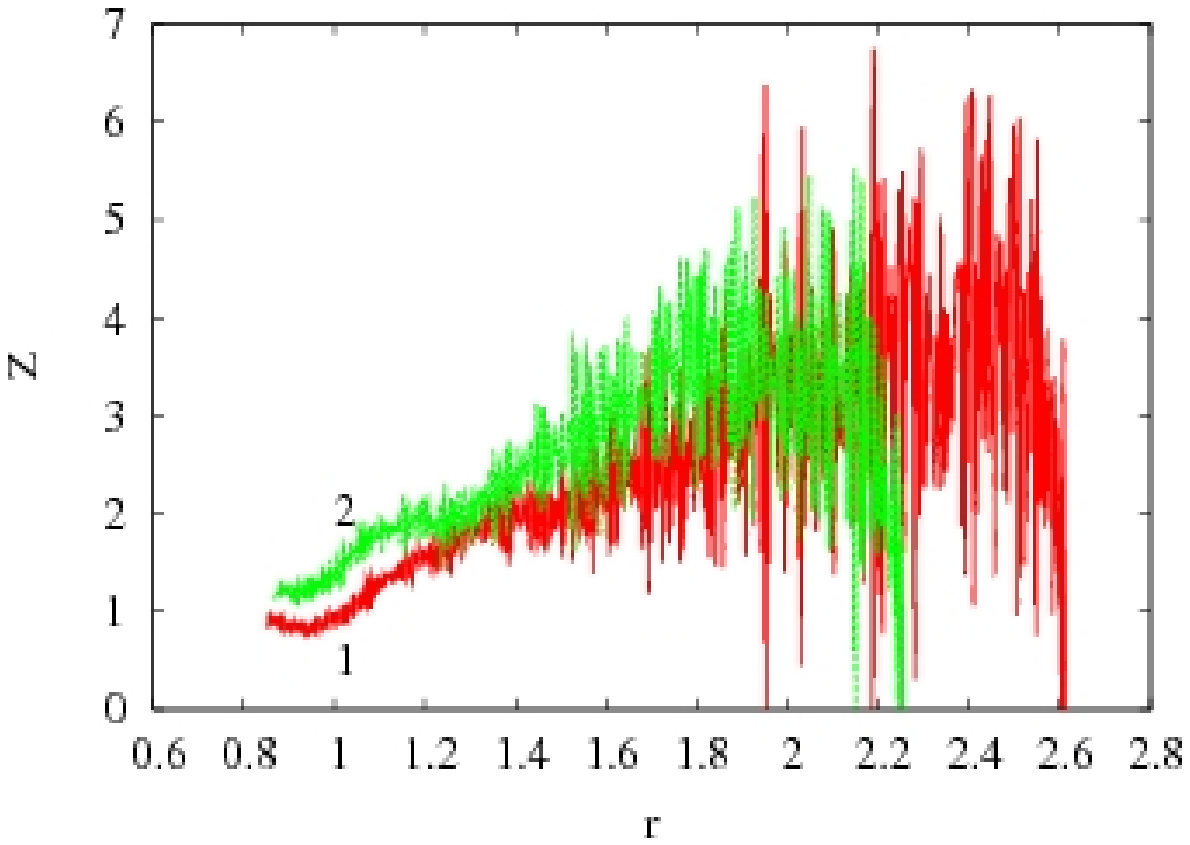}
\includegraphics[width=84mm]{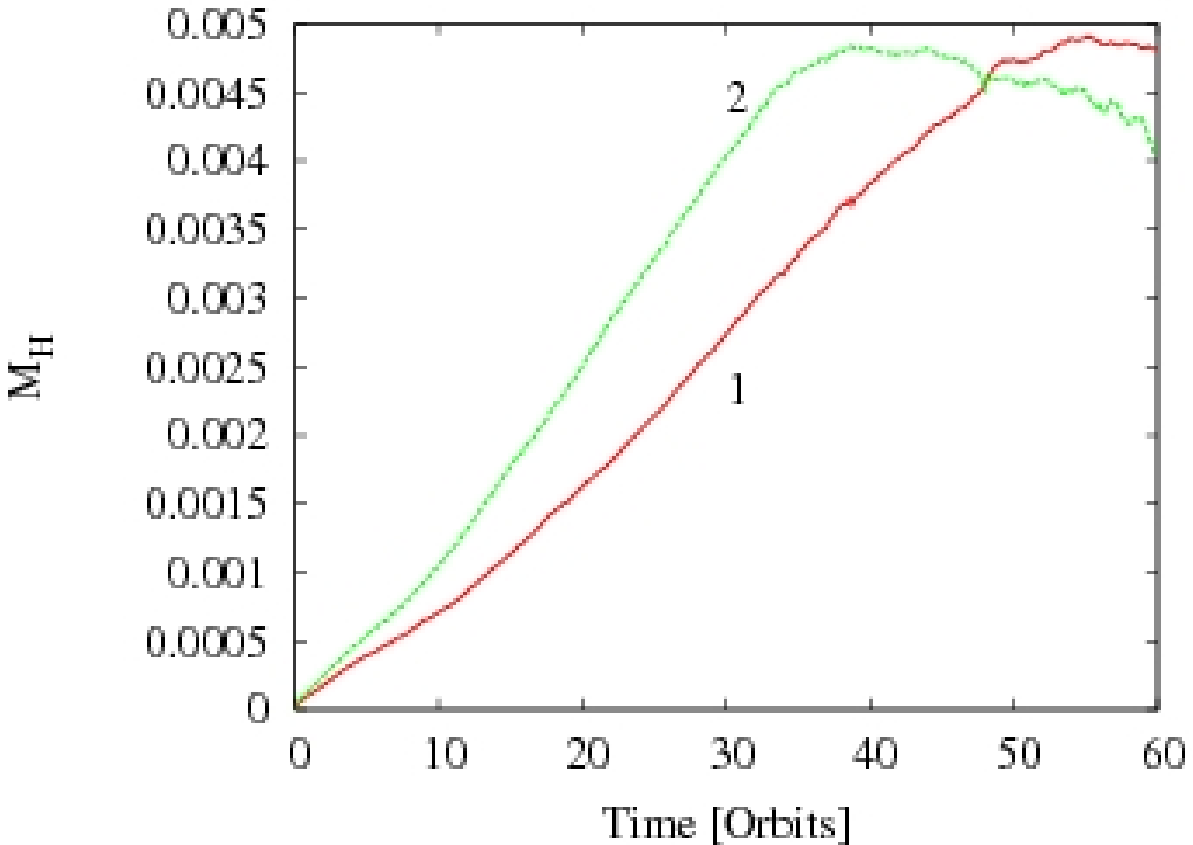}
\caption{Results of the simulations for the different total disc masses
 for $M_\rmn{P}^* = M_\rmn{P} = const$. Curves 1 and 2 correspond to the
 disc to the primary mass ratio $\mu_\rmn{D}$ equal $0.0075$ and $0.01$
 (standard case M2), respectively. The panels are the same as in
 Fig.~\ref{fmin_d_planet_mass}, however the lower right panel shows the
 changes of the mass inside the Hill sphere during evolution instead of
 effective planet's mass.} 
\label{fmin_d_disc_mass_2}
\end{figure*}

\subsection{Dependence on the total disc mass and the initial density profile}

\label{sub_dep_dens_prof_tot_mass}

Above we discussed the effects of parameters connected to the planet and its
circumplanetary disc. In this section we consider parameters describing
the global circumstellar disc: the total disc
mass given by the disc to the primary mass ratio $\mu_\rmn{D}$, and the
initial density profile described by the exponent
$\alpha_\rmn{\Sigma}$. 

\subsubsection{Total disc mass}

To describe the effect of changing the total disc mass we performed
two sets of simulations: one set for the gravitational planet mass
increased by the mass within the planet's gravitational softening
$M_\rmn{P}^* = \widetilde M_\rmn{P}$ (similar to M1) and one set for
the constant planet's mass $M_\rmn{P}^* = M_\rmn{P}$ (similar to M2).
In both sets $M_\rmn{P} = 0.001$ and $\alpha_\rmn{\Sigma} = -0.5$.

The first set contains three models with $\mu_\rmn{D}$ equal $0.0075$,
$0.01$ (standard case M1) and $0.0125$ (curves 1, 2 and 3 in
Fig.~\ref{fmin_d_disc_mass}). We see that the results are similar to
those of the simulations performed for different planet masses
(Sect~\ref{sect_diff_pl_mass} and Fig.~\ref{fmin_d_planet_mass}). As
in the case of inward migration (Paper~II Sect.~5.3), an increase of
the disc mass acts in a similar way as decreasing the planet mass.
There are however also some differences.

In the simulation with the lowest disc mass the amount of the mass
within corotation is not sufficient to support type III migration, and
the planets stops outward migration after $14$ orbits reaching $a
\approx 0.96$. In this case $Z \approx 0.85$ at the start of the
simulation, and the $Z>1$ criterion for outward migration is not
satisfied. This illustrates again the sharp border in parameter space
for which outward type III migration is allowed. The combination of
planet and disc masses gives a single parameter defining the disc
region where the type III migration is allowed.

In the second and the third model the condition for the outward
migration is satisfied with the initial averaged $Z$ equal $1.05$ and
$1.1$ respectively. The evolution of these models is similar, however the
migration rate for the most massive disc is slightly higher and the
planet reaches $a \approx 1.73$ at $27$ orbits. In the second model the
outward migration stops at $32$ orbits and $a\approx 1.7$. The last model shows
the fastest increase of the gravitational planet mass too, giving
$\widetilde M_\rmn{P} \approx 13 M_{\jupiter}$ after $27$ orbits. In the
second model $\widetilde M_\rmn{P} \approx 10 M_{\jupiter}$ at $30$
orbits. This reduces the difference between the migration rate 
$\dot a$, and the plot of the non-dimensional migration rate as a function of
radius (lower right panel) shows a very similar evolution for both
systems. This is different from the models with different planet masses
(where the lighter planet can travel further away), and the inward
migration case (where $\mu_\rmn{D}$ modifies the planet's migration in a
significant way). 

In the second set of simulations ($M_\rmn{P}^* = M_\rmn{P}$) we have
two models with $\mu_\rmn{D}$ equal $0.0075$ and $0.01$. The results
are presented in Fig.~\ref{fmin_d_disc_mass_2}. Curves 1 and 2
correspond to the first and the second model respectively. In a more
massive disc the planet migrates faster and has a higher mass
accumulation rate, but the maximum value of the mass within the Hill
sphere $M_\rmn{H}$ is similar for both models and equals about $5
M_{\jupiter}$. Similarly the migration rate increases faster for the
more massive disc, but both models reach the same maximum value, $\dot
a \approx 0.01$ at $43$ and $25$ orbits for model 1 and 2
respectively. After the migration rate starts to decrease, $M_\rmn{H}$
is still growing and reaches its maximum at $55$ and $40$
orbits. However, the transition between outward and inward directed
migration takes place at $50$ ($a \approx 2.6$) and $35$ 
($a \approx 2.25$) orbits. So, although the more massive disc allows the
planet to 
travel faster, the outward migration phase lasts shorter and the
planet actually stops closer to the star. This is caused by the nature
of the stopping mechanism which here is M2-like.  In both models the
rapid changes start upon reaching $Z\sim 4$, a value only weakly
dependent on $\mu_\rmn{D}$. The more massive disc shows a stronger
increase of $Z$ and reaches this critical value sooner.


\begin{figure*}
\includegraphics[width=84mm]{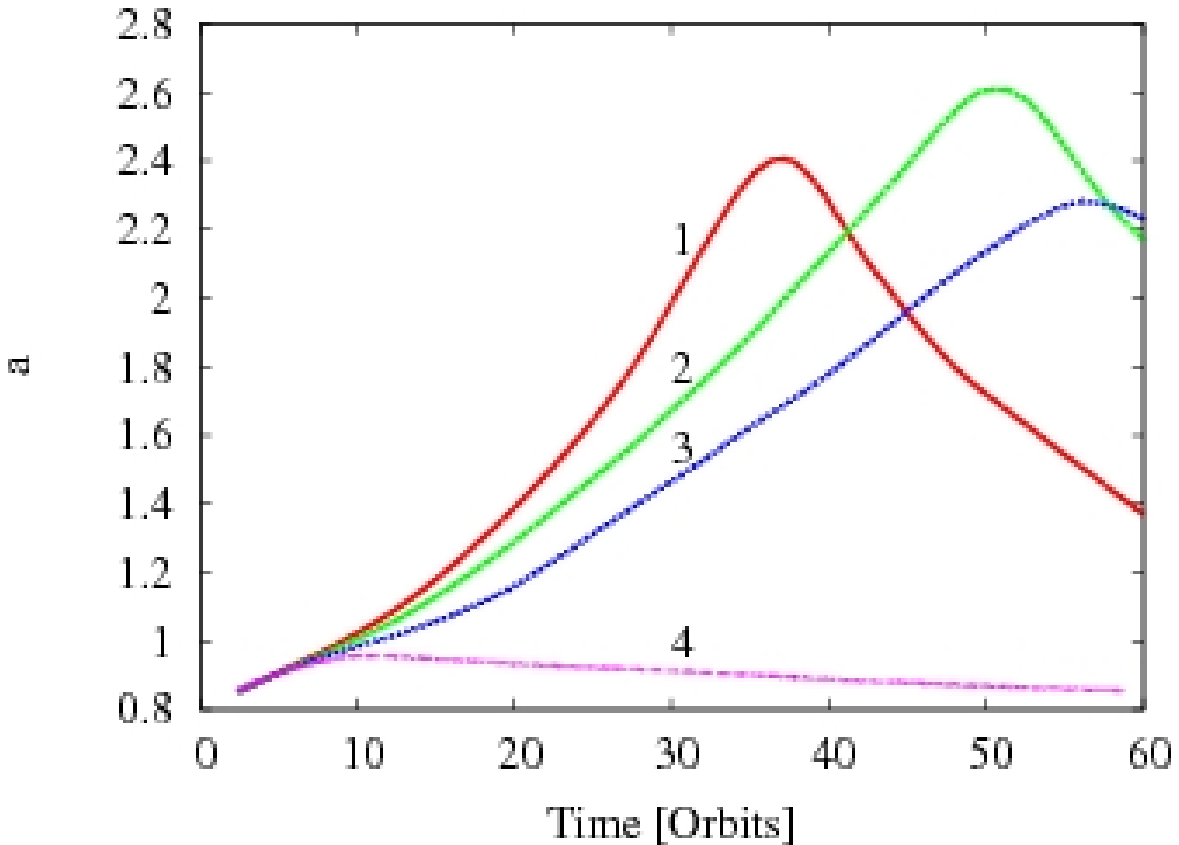}
\includegraphics[width=84mm]{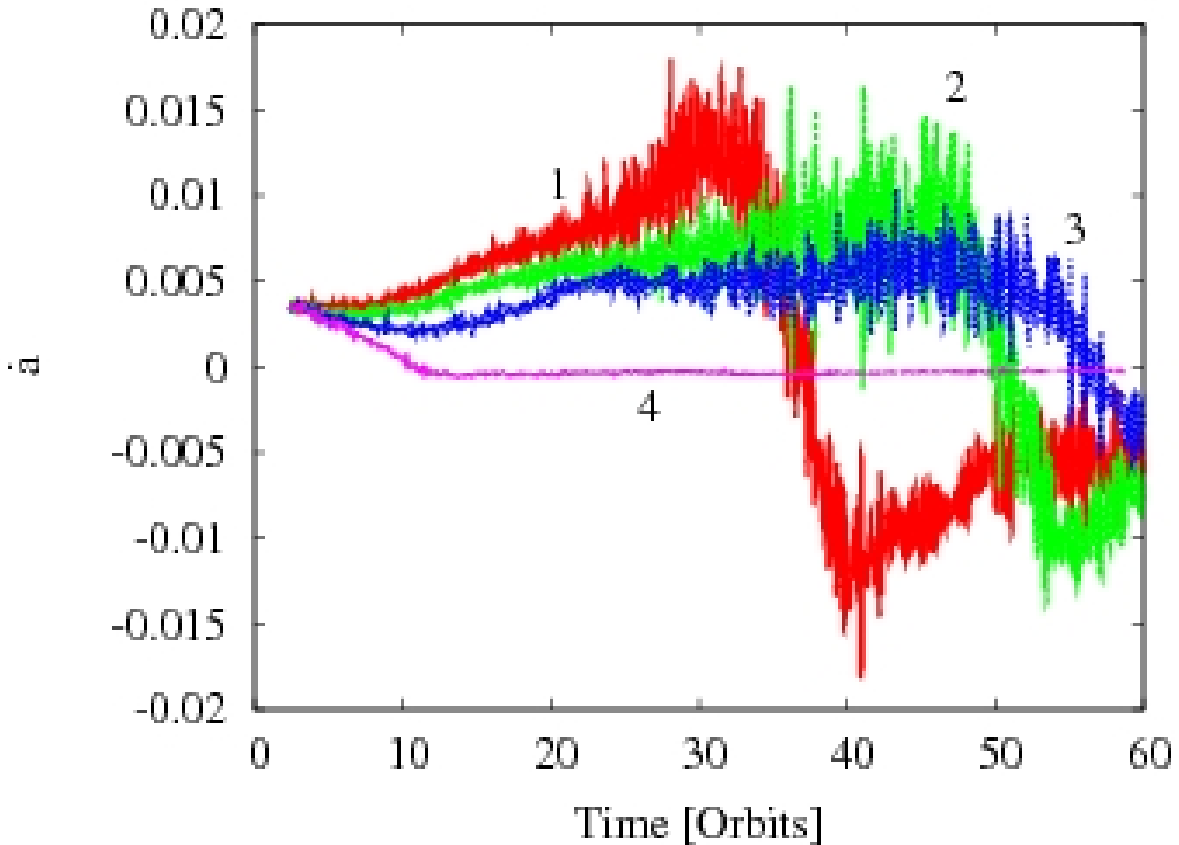}
\includegraphics[width=84mm]{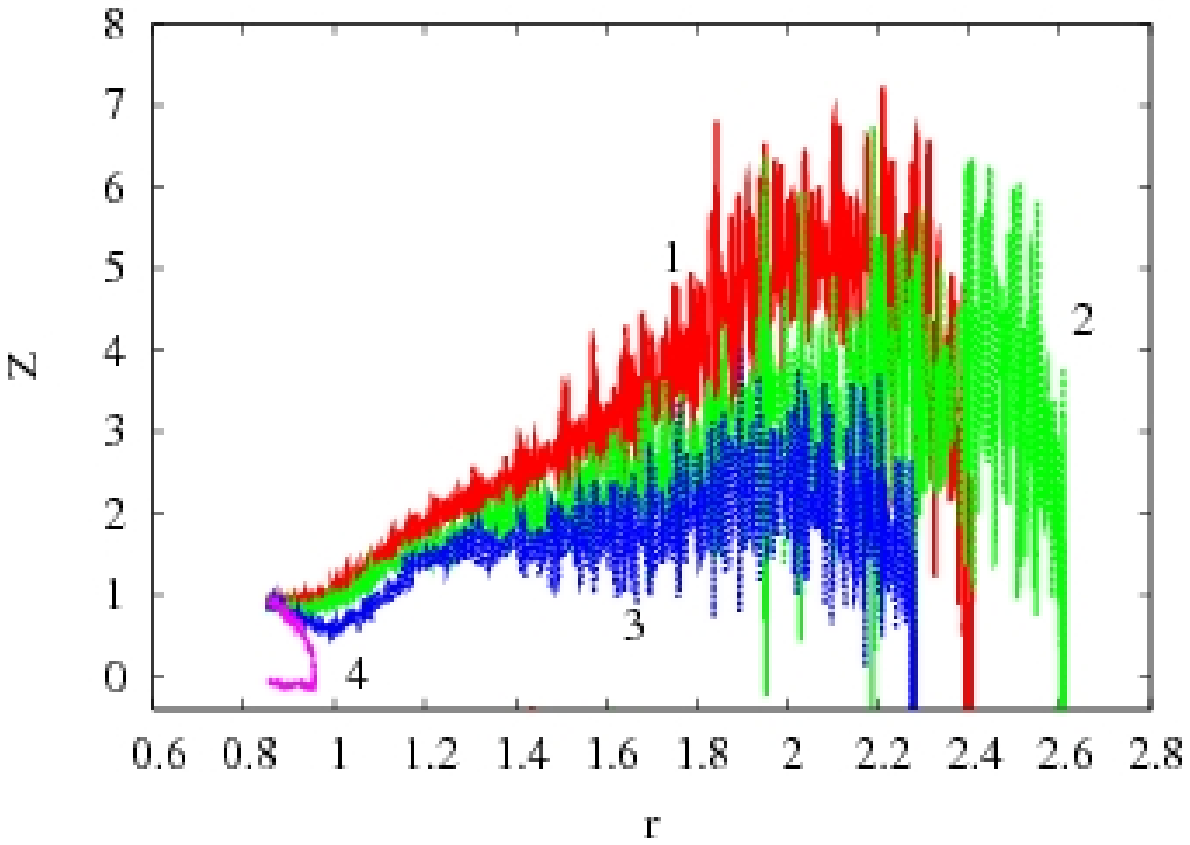}
\includegraphics[width=84mm]{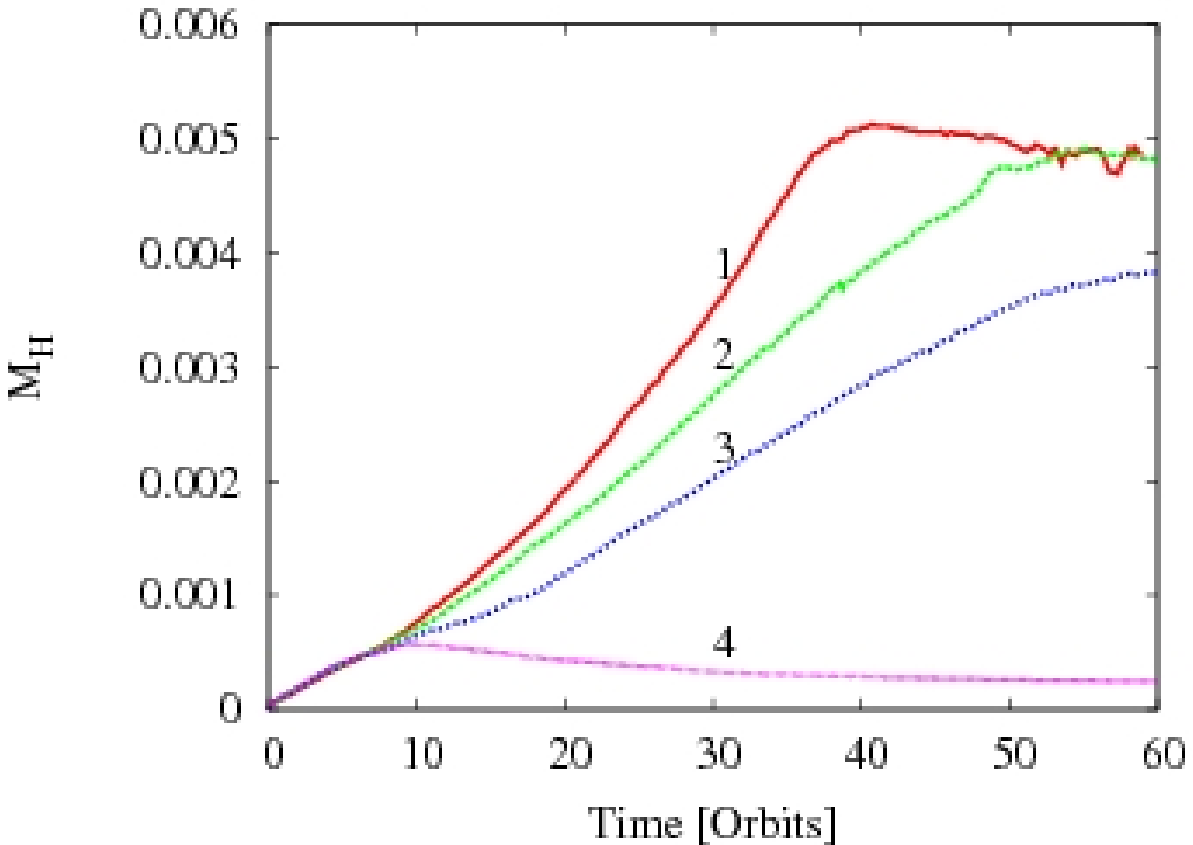}
\caption{Results of the simulations for the different initial disc
 profiles with $M_\rmn{P}^* = M_\rmn{P}= 0.001$. Curves 1, 2, 3 and 4
 correspond to $\alpha_\rmn{\Sigma}$ equal $0.0$, $-0.5$, $-1$ and
 $-1.5$ respectively. The panels are the same as in
 Fig.~\ref{fmin_d_disc_mass_2}.} 
\label{fmin_d_disc_prof}
\end{figure*}

\subsubsection{Initial density profile}

The final parameter we consider is the density exponent
$\alpha_\rmn{\Sigma}$. Since the evolution of the models with
$M_\rmn{P}^* = \widetilde M_\rmn{P}$ is dominated by the mass
accumulation and only weakly depends on $\mu_\rmn{D}$, we performed a
series of simulations with constant $M_\rmn{P}^* = M_\rmn{P}$,
$\mu_\rmn{D}=0.0075$ and $\alpha_\rmn{\Sigma}$ equal $0.0$, $-0.5$,
$-1$ and $-1.5$ (curves 1, 2, 3 and 4 in
Fig.~\ref{fmin_d_disc_prof}). The density
gradient at the initial planet position is the same in all presented
models, giving the same initial value of the migration rate $\dot
a$. 

Once again this set of simulations shows the sharp boundary in
parameter space for which outward migration is allowed. Although the
initial condition for outward migration $Z > 1$ is satisfied for all
the models, in the last simulation ($\alpha_\rmn{\Sigma}=-1.5$, curve
4) the mass within corotation is too small to support the migration
and the planet switches to inward migration after about $10$ orbits
reaching $a \approx 1$. The mass within its Hill sphere does not
exceed $0.7 M_{\jupiter}$.

The other models migrate in the fast migration regime, however the
migration rate depends strongly on $\alpha_\rmn{\Sigma}$. The smallest
gradient in the density profile corresponds to the biggest mass in the
corotation region and allows for the fastest increase of $\dot a$,
reaching $0.015$ after $30$ orbits. This gives a maximum value of 
$Z \approx 5.5$. Similarly this model shows the quickest increase of the
mass within the Hill sphere giving $M_\rmn{H} \approx 5 M_{\jupiter}$
at about $40$ orbits. The transition between outward and inward
migration takes place at about $37$ orbits and $a \approx 2.4$.

Lower $\alpha_\rmn{\Sigma}$ means lower mass within the corotation
region, and thus slower migration. That is why the migration rate
grows slower in the second model reaching $0.01$ after about $46$
orbits, and remains approximately constant with $\dot a \approx 0.005$
between $20$ and $50$ orbits in the third model. Although $\dot a$ is
almost constant in model 3, $Z$ grows in both models reaching about
$4$ and $2.5$ in the second and the third model respectively. The
lower value of $Z$ allows the planets to travel outward for a longer
time, and only switch to the inward migration phase after about $51$
($a \approx 2.6$) and $56$ ($a \approx 2.3$) in model 2 and 3
respectively. The rate of the mass accumulation in the Hill sphere
decreases with decreasing $\alpha_\rmn{\Sigma}$, however the final
mass in the first and the second model is quite similar. In the third
model $M_\rmn{H}$ reaches almost $4 M_{\jupiter}$.

One thing that this series of simulation shows is that the critical
value of $Z$ for stopping mechanism M2, is not universal, but actually
depends on the exponent in the density profile and decreases with
decreasing $\alpha_\rmn{\Sigma}$.

\begin{figure*}
\includegraphics[width=84mm]{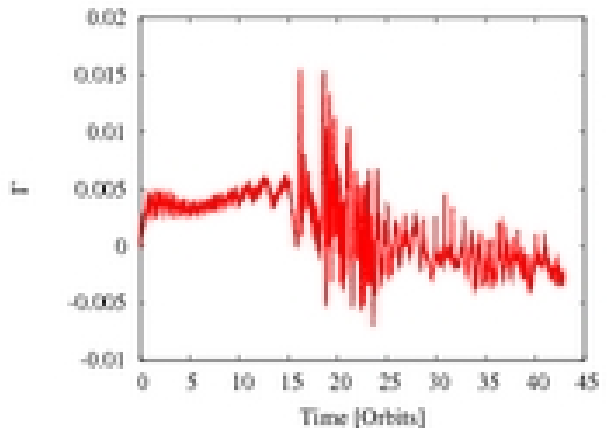}
\includegraphics[width=84mm]{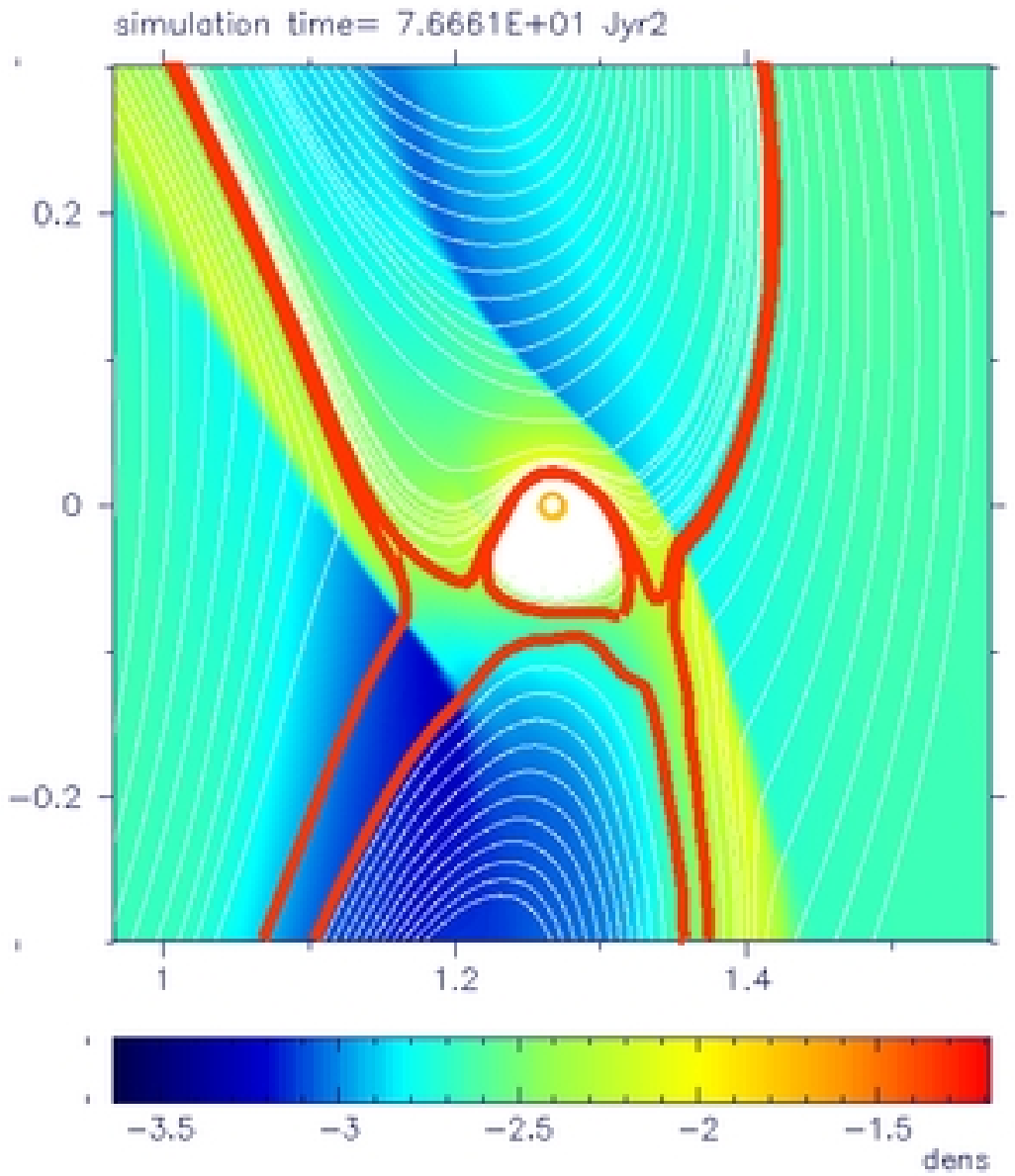}
\includegraphics[width=84mm]{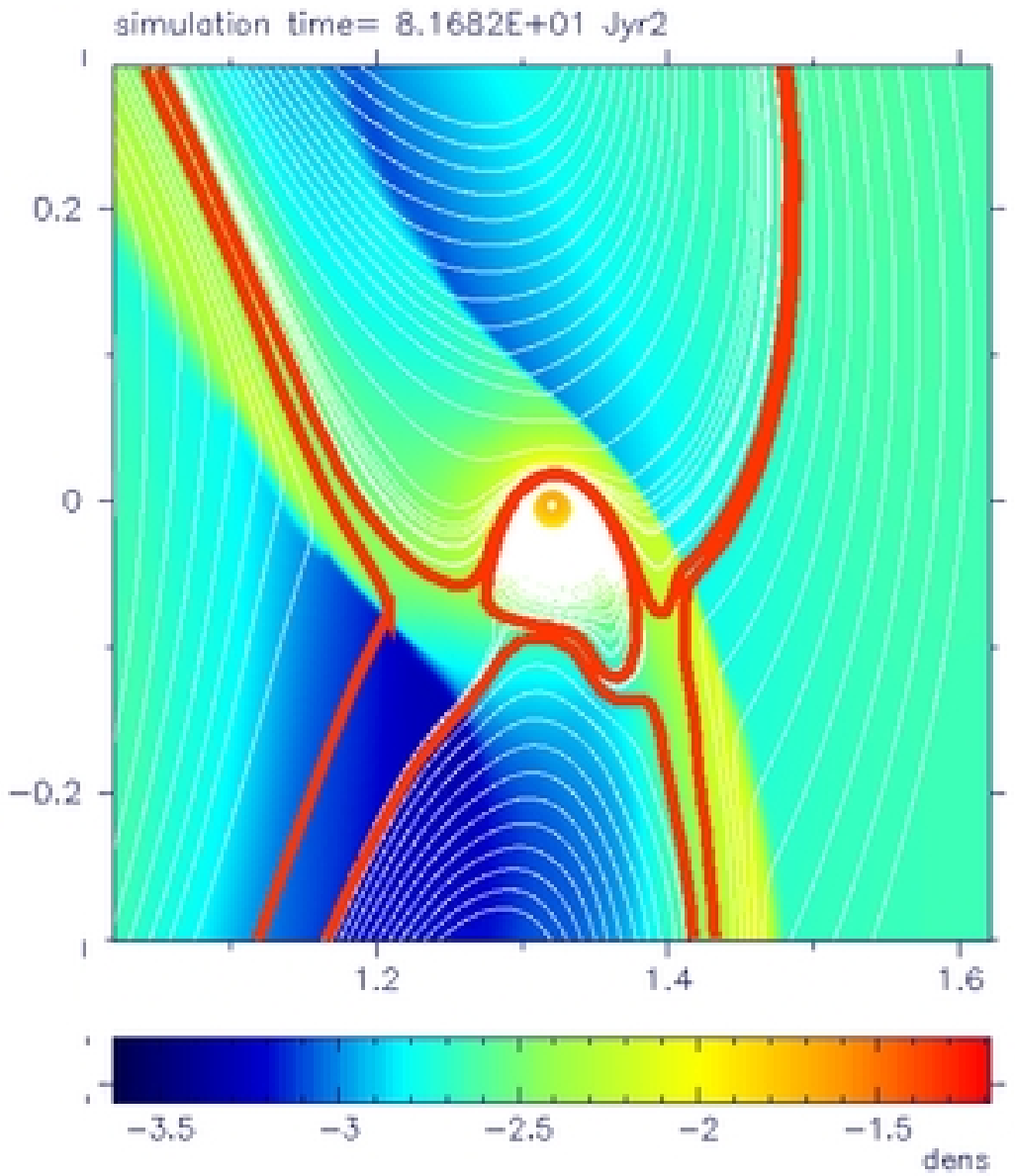}
\includegraphics[width=84mm]{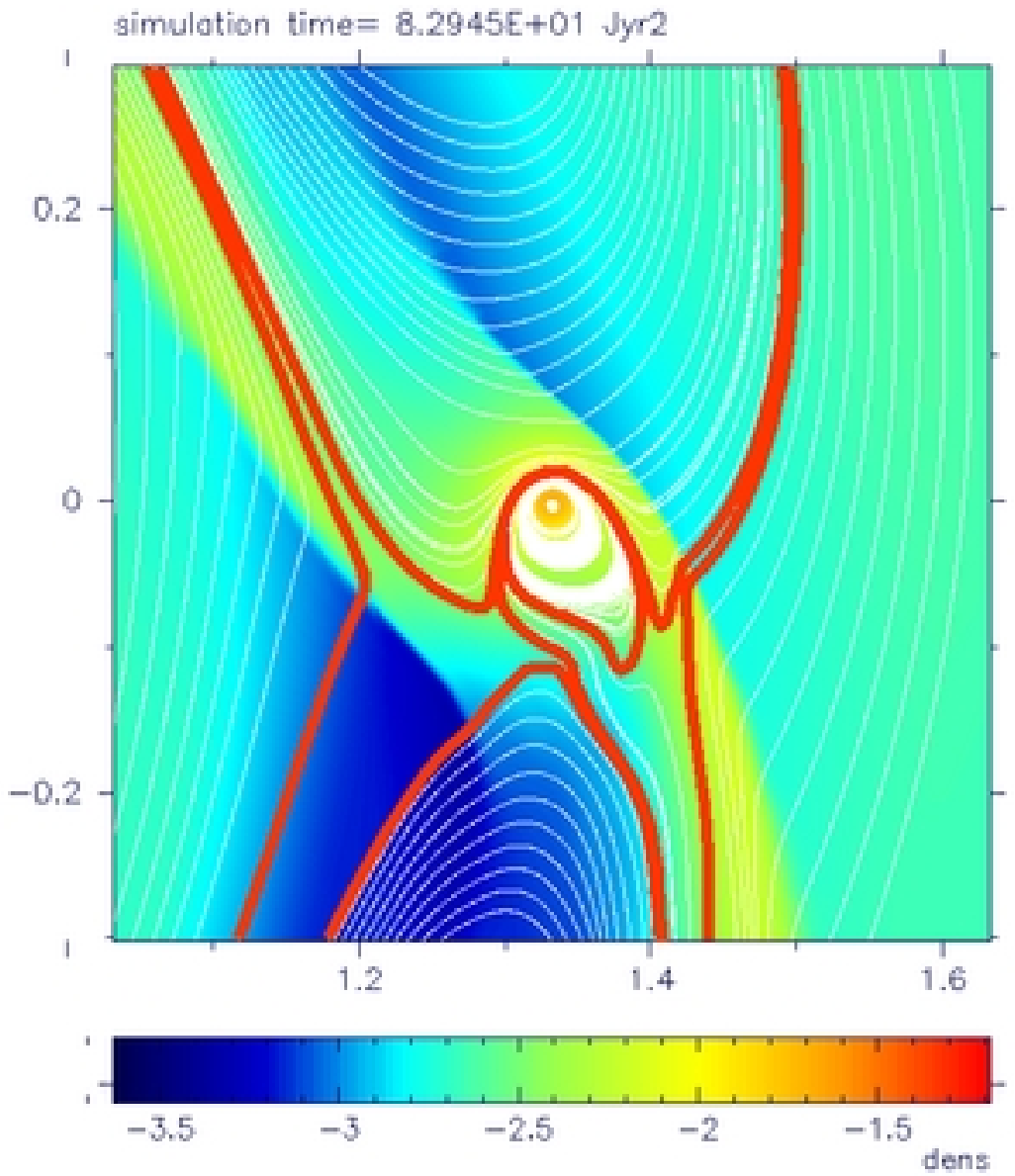}
\caption{The non-averaged torque exerted by the gas on the planet (upper
 left panel) and the details of the gas flow in the Roche lobe for model
 with $h_\rmn{p} = 0.6$. Upper right, lower left and right panels show
 the surface density and the flow lines in the planet's vicinity after
 $12.2$, $13$ and $13.2$ orbits. The colour scale is logarithmic. The
 plotted domain is square region of the size of $6 R_\rmn{H}$. The flow
 lines close to the border of different regions are made more visible.} 
\label{fi_stop_torque}
\end{figure*}


\section{Stopping type III migration}
\label{stop_mig}

In Paper~II we described two possible mechanisms of stopping type~III
migration (or more precisely, switching it to type~II
migration). Inward migration can be slowed down by the simple
geometrical effect of the shrinking of the co-orbital region with
decreasing semi-major axis, or by the interaction with a steep density
gradient, such as a disc edge. The second mechanism acts in the same
way for outward directed migration, but the first one would make
outward type~III migration a self-accelerating process. In this case
an additional mechanism is necessary to limit the non-dimensional
migration rate $Z$.\footnote{In most of the simulations $\dot a$ grows
  during the outward migration phase even though $Z$ decreases; see middle
  row in Fig.~\ref{fim1_a}.} We found two possible mechanisms to
achieve this.

The first one is the rapid increase of the planet's inertia, which
lowers $Z$ and forces the migration to proceed in the slow migration
regime ($Z<1$). The example is our standard case M1, in which the
effective planet mass grows from $1$ up to about $10 M_{\jupiter}$
during about $30$ orbits ($\sim 350$ years). Such a rapid mass
accumulation seems unrealistic and could be expected to lead to an
increase of the disc aspect ratio $h_\rmn{p}$, which would reduce the
mass accumulation rate. Our disc model does not allow for time-dependent
values of  $h_\rmn{p}$, but as the growth of the rapidly migrating planet 
is a highly dynamical process in which a large amount of mass may enter 
the Roche lobe and interact with the gaseous planet's envelope, a much
more sophisticated method would be needed to model this realistically.

Note that even though such a rapid increase of the effective planet's
mass is more likely to happen in outward migration, we even found it
in some cases of inward migration (see the model with $h_\rmn{p}=0.3$
described in Paper~I Sect.~5.1.2).

The second stopping mechanism is encountered in models in which the
growth of the effective planet mass remains limited\footnote{In the
  model with $h_\rmn{p}=0.5$ the final effective planet's mass is of
  the order of $2 M_{\jupiter}$; see Fig.~\ref{fmin_d_planet_h}.}, and
the example is our standard case M2. Here both $\dot a$ and $Z$ grow
quickly during the migration until a critical value of $Z$ is reached
and the flow structure in the co-orbital region changes
dramatically. The reason for such a rapid change are strong
oscillations of the flow in the planet's vicinity. They are clearly
visible in the simulation with $h_\rmn{p}=0.6$ (see
Fig.~\ref{fmin_d_planet_h}). In this model the outward directed rapid
migration lasts for about $23$ orbits and the planet reaches $a\approx
1.7$. This simulation has the important advantage that the
relatively high value of the circumplanetary disc aspect ratio
suppresses the highest frequency oscillations, and consequently it is
easier to study the non-time-averaged torque $\Gamma$.

\begin{figure*}
\includegraphics[width=84mm]{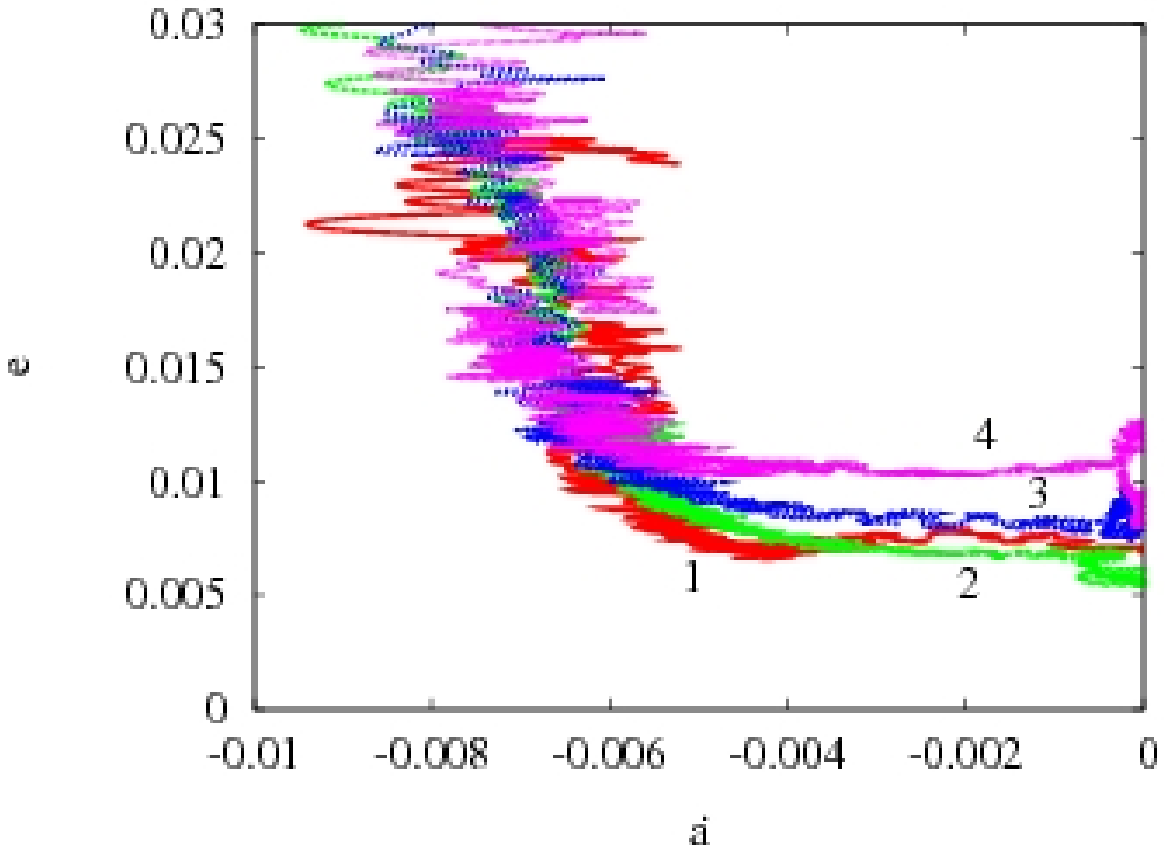}
\includegraphics[width=84mm]{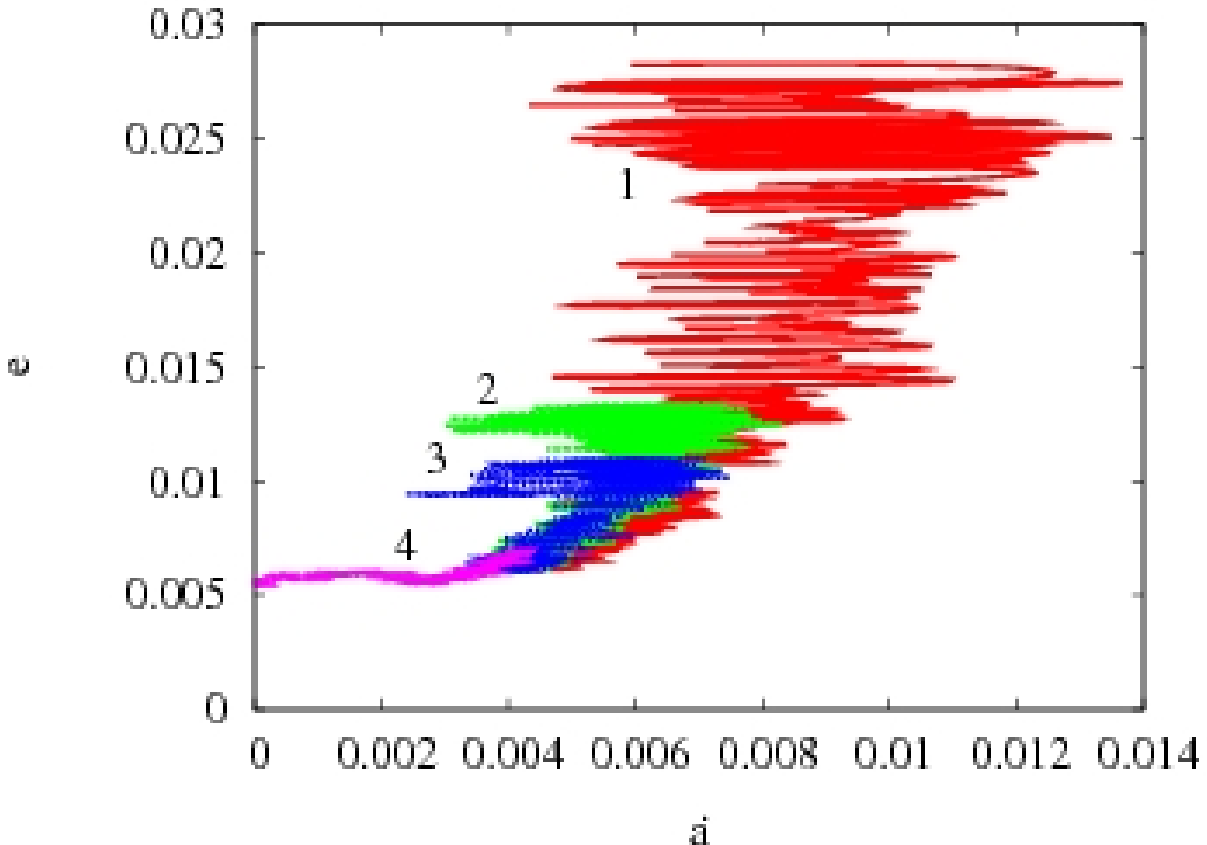}
\caption{The eccentricity $e$ as a function of the migration rate 
 $\dot a$. Left and right panels present the results of the inward and 
 outward directed migration respectively. All curves present models with
 $h_\rmn{p}=0.4$ and different planet's mass. Curve~1 gives the model
 with ${M_\rmn{P}}^* = M_\rmn{P}=0.001$. Curves 2, 3 and 4 show the
 models with ${M_\rmn{P}}^*=\widetilde M_\rmn{P}$, and $M_\rmn{P}$ equal
 $0.0007$, $0.001$ and $0.0013$ respectively.} 
\label{f_e_dota}
\end{figure*}

The results are presented in Fig.~\ref{fi_stop_torque}. The upper left
plot shows the time evolution of the non-averaged torque. There are
oscillations of $\Gamma$ that start at about $10$ orbits 
($Z\approx 2$) and grow in time. These oscillations are caused by cyclic changes
of the shape of the circumplanetary disc. One of these cycles starting
at about $12$ orbits, is presented in the upper right and lower
panels. In Paper~II we described the flow structure in the planet's
vicinity for different stages of migration, showing that in the fast
migration regime the circumplanetary disc is strongly asymmetric
(Paper~II, upper left panel in Fig.~4) and gives a significant
contribution to the total torque (Paper~II, Fig.~5). A similar
situation can be seen in the upper right panel in
Fig.~\ref{fi_stop_torque}, corresponding to a value of $\Gamma$ close
to the maximum of the cycle. The circumplanetary disc has a triangular
shape compressed at the side of the co-orbital flow and stretched at
the side of the horseshoe region. In a later phase, the orbits in the
circumplanetary disc have stretched even more at the side of the bow
shock in the outer disc (lower left panel) giving a small increase of
$\Gamma$. In the final phase, some mass from the planet's vicinity
leaves the Roche lobe flowing along the outer bow shock (lower right
panel), reducing $\Gamma$ significantly. Using the tracer fluid
(marking the position of gas placed in the initial corotation region)
we can see that this process involves a big portion of the gas from
the co-orbital flow entering the Roche lobe and being redirected back
to the outer disc and not accumulating in the planet's proximity. This
strongly perturbs the outer bow shock and giving the oscillations in
$\Gamma$.

We find that these oscillations grow in time, until they become so
strong that they destroy the regular co-orbital flow, starting a
period of strong oscillations in $\Gamma$ and $\dot a$. These strong
oscillations of the migration rate open the flow-lines in the
co-orbital region and a large amount of gas from the former
co-orbital flow is captured into the new horseshoe region. Such a short
phase of strong oscillations of averaged $\Gamma$ and $\dot a$ is
also visible in the standard case M2 (Fig.~\ref{fim1_a}). 

We note that we found a similar process (the corrugation of the inner
bow shock by the gas from the co-orbital flow being redirected back to
the inner disc) in models for inward migration.  However there the
oscillations do not appear to grow in time. This difference in
behaviour may be due to the inherently self-decelerating and
self-accelerating character of inward and outward migration.

One may wonder whether this rearrangement of the co-orbital region is
not caused by numerical effects. Although we cannot fully rule out
that our disc model influences this process, we still think it is a
rather robust mechanism. We find it in models with different
parameters and envelope masses, indicating that it is not our
self-gravity correction that plays a role here. We also point to the
results of \citet{2003ApJ...588..494M} which showed an outward directed
migration to revert its direction after a relatively short time.

\begin{figure*}
\includegraphics[width=84mm]{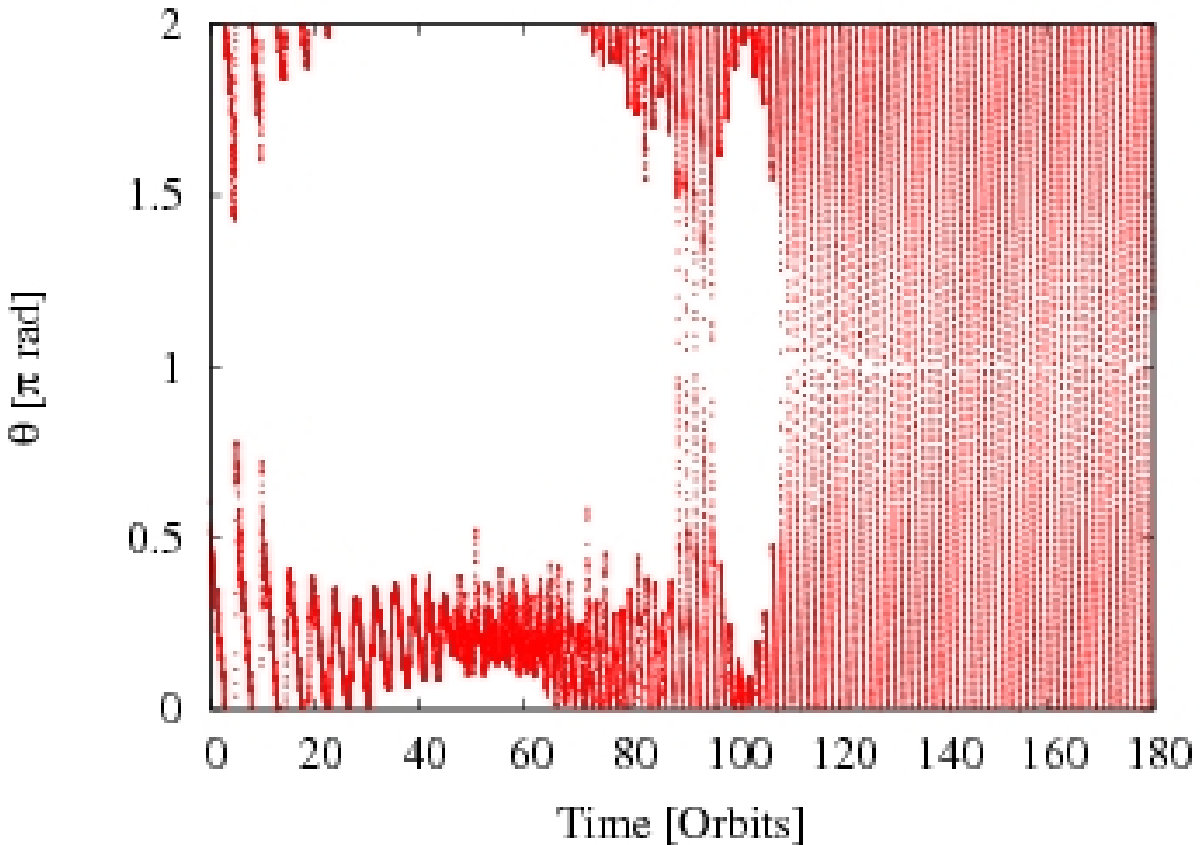}
\includegraphics[width=84mm]{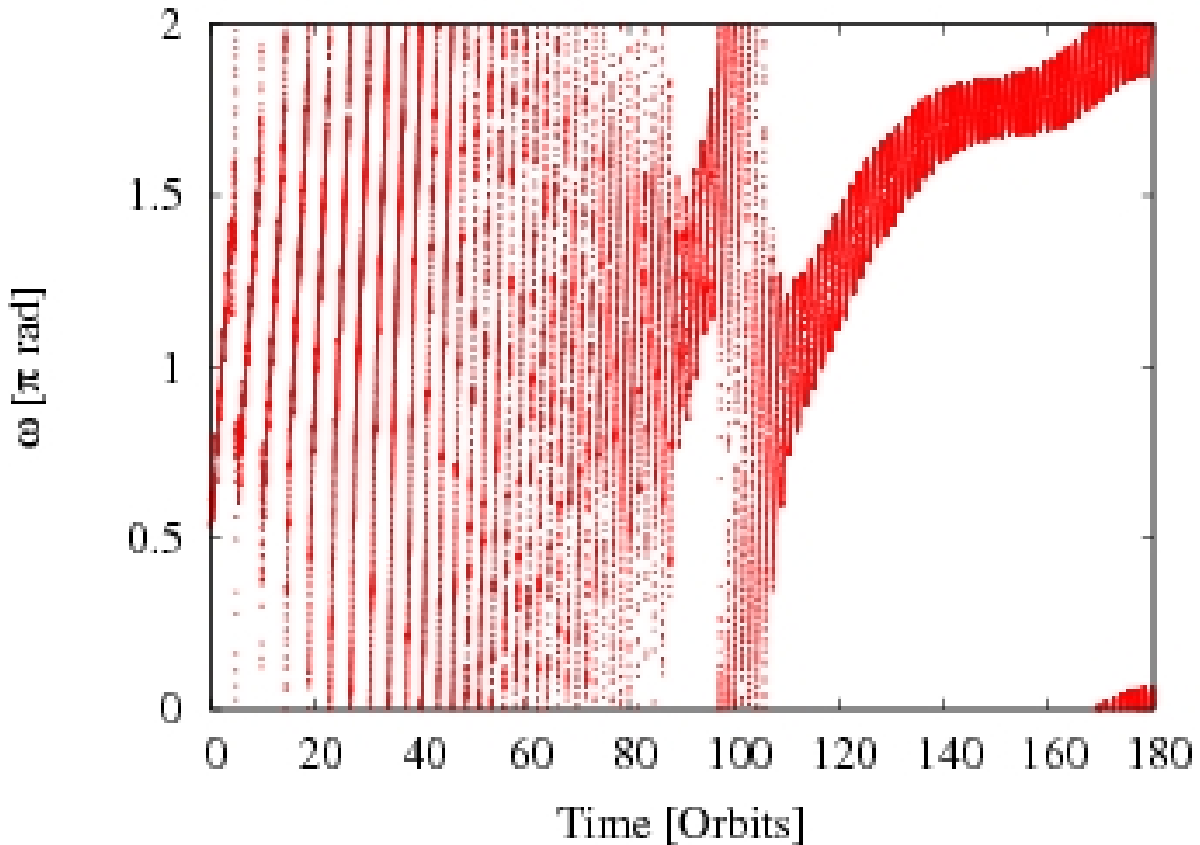}
\caption{Evolution of a true anomaly $\theta$ (left panel) and an
 argument of periastron $\omega$ (right panel) for an inward migrating
 planet. The rapid migration phase lasts here for about $85$ orbits and
 is followed by the gap formation phase. During the rapid migration
 phase $\theta$ oscillates in a relatively narrow range of angles and
 $\omega$  ranges from $0$ up to $2\pi$. The situation reverses in the
 gap formation phase.} 
\label{f_ang}
\end{figure*}


\section{Eccentricity evolution}

\label{sect_ecc_ev}

In Paper~II we discussed the eccentricity evolution for inward
migrating planets. In this case the eccentricity is damped during the
rapid migration phase on a time-scale similar to that of migration.
Moreover we found the damping time-scale to be independent of
the planet mass. In the case of outward migration we find that the
eccentricity grows with the migration rate $\dot a$ during the
whole outward migration phase. These results suggest a correlation between
$e$ and  $\dot a$, as shown in Fig.~\ref{f_e_dota}. This plot
shows the results of the inward (left panel) and outward (right panel)
directed migration models for different prescriptions of the effective
planet's mass (${M_\rmn{P}}^* = M_\rmn{P}=0.001$ for curve 1 and
${M_\rmn{P}}^*=\widetilde M_\rmn{P}$ for curves 2, 3 and 4) and
different initial planet's masses ($M_\rmn{P}$  equal $0.0007$, $0.001$
and $0.0013$ for curves 2, 3 and 4 respectively). For the outward
directed migration only the rapid migration phase is plotted.

The eccentricity is approximately constant for $|\dot a|< 0.005$ and grows
rapidly for bigger absolute values of the migration rate. This relation
is similar for inward and outward directed migration and is weakly
dependent on the planet mass. The origin of this relation is unclear,
but it may be caused by the fact that the spiral orbit of a 
rapidly migrating planet cannot be properly represented by Keplerian elements.
This conclusion is supported by the fact that during the rapid migration
phase, unlike in the gap formation phase, the true anomaly $\theta$ for
the migrating planet oscillates in a relatively narrow range of angles,
but the argument of periastron $\omega$ ranges from $0$ up to $2\pi$
(Fig.~\ref{f_ang}). 


\section{Conclusions}

\label{conclusions}

In this paper we investigated the outward directed type~III migration
of the high-mass planet embedded in a disc.  Using two dimensional
numerical simulations, performed in a Cartesian coordinate system and
the inertial reference frame, we studied the orbital evolution of
planets and the dependency on various numerical parameters. The
adaptive mesh refinement allowed us to achieve high resolution inside
the Roche lobe and study in detail the flow structure in the planet's
vicinity. To avoid problems with numerical convergence we used a
modified version of the usual local-isothermal approximation, where
the temperature depends on the distance to both the star and the
planet, and also added a correction for the gas self-gravity near the
planet.

Both inward and outward type~III migration are driven by the
co-orbital torque. However, the existence of the (inward directed)
differential Lindblad torque can be expected to break the symmetry
between the two cases. Indeed, the simulations show that outward
migration always reverts to inward migration. In fact in some cases
this reversal happens almost instantly, and the initial value of the
non-dimensional migration rate $Z$ is found to be crucial here.
Placing the planet at an initial density gradient we force the planet
to start rapid migration, however the strength of the initial impulse
depends on the planet mass and the local density profile, and it can
be measured through the initial averaged value of $Z$. We found that
the disc supports outward migration if the initial $Z >1$. Since 
$Z \sim \dot a {M_\rmn{P}}^{-2/3}$ the initial $Z$ decreases with the
planet mass and grows with the disc mass. It means that outward
migration is easier to start for less massive planets in more massive
discs. Note here that although initially $Z>1$ is required, this does
not mean that outward type~III migration can not later proceed in the
slow migration regime $Z<1$.

The second important difference between inward and outward migration
is caused by the relation between the planet's semi-major axis $a$ and
the volume of the co-orbital region. In Paper~II we showed that the
migration rate can be approximated to be dependent on the mass of the
co-orbital region, and for $\alpha_\rmn{\Sigma}>-2$ the inward rapid
migration is a self-decelerating process (the mass of the co-orbital
region decreases during migration). For the same reason outward
directed migration is a self-accelerating process. The migration rate
$\dot a$ grows with $a$ during the outward migration phase, however
the non-dimensional migration rate $Z$ can grow or decrease depending
on the effective planet mass.

We found two different stopping mechanisms for outward type III
migration.  Because of the presence of the Lindblad torques, stopping
here always implies a reversal of the migration direction.  The first
stopping mechanism occurs when the effective planet mass increases
enough to make $Z$ fall below 1 so that the migration proceeds in the
slow migration regime. In this case some aspects of the evolution of
the system resemble the evolution of the standard case of inward
migration described in Paper~II. In both cases the transition between
$Z\approx 1$ and $Z \ll 1$ is smooth and proceed in a similar way,
however the reason for slowing down migration is different (decrease
of the co-orbital mass and increase of $\widetilde M_\rmn{P}$ for the
inward and outward migration respectively). The horseshoe region is
initially restricted to a single tadpole-like region and gradually
grows filling the whole co-orbital region and decreasing the
co-orbital flow. As in the inward migration case the torque is a
linear function of $Z$. In this model the migration rate is relatively
small and the planet starts clearing a gap shortly after the
simulation starts.

The other mechanism of stopping outward type~III migration is seen in
models where the first mechanism does not operate, and the planet
keeps migrating in the fast migration regime $Z>1$.  The migration is
accelerated due to the growth of the co-orbital mass and the asymmetry
of the flow in the planet's vicinity grows too, leading to
corrugations of the external bow shock and oscillations in the
torque. These oscillations grow in time and finally destroy the
co-orbital flow, opening the flow lines in the co-orbital region. Due
to this the planet rapidly stops its outward migration and starts
migrating inward (also in the fast migration regime). The simulations
show that there is a critical value of $Z$ at which the strong
oscillations of $\dot a$ start and the planet reverts its migration
direction. We found this critical value of $Z$ to be weakly dependent
on the total disc mass $\mu_\rmn{D}$, but sensitive to
$\alpha_\rmn{\Sigma}$. For our set of simulations this critical value
for $Z$ fell in the range 3--6.

For the second stopping mechanism the whole process of direction
reversion is very brief and the planet does not have time to initiate
gap clearing. For the same reason the gas from the former (before the
slowing down of the migration started) horseshoe region does not mix
with the gas captured in the new horseshoe region and can be seen as a
low density region in the corotation region. Compared to the slow
migration regime, the averaged torque has a weaker dependency on
$Z$.

Our simulations show a correlation between eccentricity and migration
rate, which is almost independent on the planet mass. We found
the eccentricity to be approximately constant for $|\dot a|< 0.005$ and to grow
rapidly for bigger absolute values of the migration rate. This
correlation may be caused by the fact that the spiral orbit of a
rapidly migrating planet is not well represented by Keplerian elements.

Outward directed type III migration is an extreme process in
terms of mass accumulation by the planet, and clearly to make progress
a more sophisticated model for how the planet deals with such a
massive accretion rate is needed. This would involve both improvements
in the thermal model used, as well as the inclusion of the
self-gravity of the gas. However, outward type III migration can
clearly add interesting aspects to planet formation scenarios,
allowing both rapid growth and a (perhaps only temporary) increase of
its orbital radius.


\section*{Acknowledgements}

The software used in this work was in part developed by the
DOE-supported ASC / Alliance Center for Astrophysical Thermonuclear
Flashes at the University of Chicago. We thank F.\ Masset for
interesting and useful comments. Calculations reported in this paper
were performed at High Performance Computing Centre North (HPC2N) and
National Supercomputing Centre, Link\"oping, and on the Antares
cluster funded at Stockholm Observatory purchased using funding
provided to PA by Vetenskapsradet, Sweden. The authors acknowledge
support of the European Community's Human Potential Programme under
contract HPRN-CT-2002-00308 (PLANETS), as well as the NSERC, Canada,
Discovery grant (2005-2008).

%
%
\bibliographystyle{mn2e}
\bibliography{articles_astro}

\label{lastpage}

\end{document}